\documentclass[twoside]{article}

\usepackage{auxiliary/PRIMEarxiv}
\usepackage{auxiliary/arxivpackages}
\usepackage{auxiliary/mymacros}
\usepackage{auxiliary/commands}
\usepackage{auxiliary/filecommands}
\renewcommand*{\backrefalt}[4]{%
    \ifcase #1 \footnotesize{(Not cited.)}%
    \or        \footnotesize{(Cited on page~#2.)}%
    \else      \footnotesize{(Cited on pages~#2.)}%
    \fi}

\renewcommand{\cite}[1]{\citep{#1}}
\setcitestyle{authoryear,round,citesep={;},aysep={,},yysep={;}}

\newcommand{\mytitle}[1][]{Infinite Horizon Markov Economies}
\newcommand{\mysubtitle}[1][]{}
\newcommand{\myauthors}[1][]{Denizalp Goktas}
\newcommand{\myauthorsshort}[1][]{Goktas}

\title{\mytitle
}

\author{
 Denizalp Goktas\thanks{Both authors contributed equally to the paper.}\\
  Department of Computer Science\\
  Brown University \\
  Providence, RI\\
  \texttt{ denizalp\_goktas@brown.edu} \\
   \And
 Sadie Zhao\footnotemark[1]\\
 SEAS\\
 Harvard University \\
  Cambridge, MA\\
  \texttt{sadie\_zhao@g.harvard.edu} \\
  \AND
  Yiling Chen \\
  SEAS \\
  Harvard University \\
  Cambridge, MA\\
  \texttt{yiling@seas.harvard.edu} \\
  \And
  Amy Greenwald \\
  Department of Computer Science \\
    Brown University \\
  Providence, RI \\
   \texttt{ amy\_greenwald@brown.edu} \\
}

\begin{document}
\maketitle

\begin{abstract} 
\deni{We need to change to "infinite horizon exchange economy Markov pseudo-game" in the appendix or whatever the terminology is. Also ask Sadie for deanonymized code repos.}\sadie{Language is the appendix is fixed, and the code repo is deanonymized now.}
In this paper, we study a generalization of Markov games and pseudo-games that we call Markov pseudo-games, which like the former, captures time and uncertainty, and like the latter, allows for the players' actions to determine the set of actions available to the other players.
In the same vein as \citeauthor{arrow-debreu}, we intend for this model to be rich enough to encapsulate a broad mathematical framework for modeling economies.
We then prove the existence of a game-theoretic equilibrium in our model, which in turn implies the existence of a general equilibrium in the corresponding economies.
Finally, going beyond \citeauthor{arrow-debreu}, we introduce a solution method for Markov pseudo-games, and prove its polynomial-time convergence.


We then provide an application of Markov pseudo-games to infinite-horizon Markov exchange economies, a stochastic economic model that extends \citeauthor{radner1972existence}'s stochastic exchange economy 
and \citeauthor{magill1994infinite}'s infinite horizon incomplete markets model.
We show that under suitable assumptions, the solutions of any infinite horizon Markov exchange economy (i.e., recursive Radner equilibria---RRE) can be formulated as the solution to a concave Markov pseudo-game, thus establishing the existence of RRE, and providing first-order methods for approximating RRE. 
Finally, we demonstrate the effectiveness of our approach in practice by building the corresponding generative adversarial policy neural network, and using it to compute RRE in a variety of infinite-horizon Markov exchange economies.
\end{abstract}

\section{Introduction}

In \citeyear{walras}, L\'eon Walras formulated a mathematical model of markets as a system for resource allocation comprising supply and demand functions that map values for resources, called \mydef{prices}, to quantities of resources---\emph{ceteris paribus}, i.e., all else being equal.
Walras argued that any market would eventually settle into a steady state, which he called \mydef{competitive} (nowadays, also called \mydef{Walrasian}) \mydef{equilibrium}, as a collection of prices and associated supply and demand such that
the demand is \mydef{feasible}, i.e., the demand for each resource is less than or equal to its supply, and \mydef{Walras' law} holds, i.e., the value of the supply is equal to the value of the demand.
Unlike in Walras' model, real-world markets do not exist in isolation but are part of an \mydef{economy}.
Indeed, the supply and demand of resources in one market depend not only on prices in that market, but also on the supply and demand of resources in other markets.
If every market in an economy is simultaneously at a competitive equilibrium, Walras' law holds for the economy as a whole; this steady state, now a property of the economy, is called a \mydef{general equilibrium}.

Beyond Walras' early forays into competitive equilibrium analysis, foremost to the development of the theory of general equilibrium was the introduction of a broad mathematical framework for modeling economies, which is known today as the \mydef{Arrow-Debreu competitive economy}
\cite{arrow-debreu}.
In this same paper, \citeauthor{arrow-debreu} developed their seminal game-theoretic model, namely (quasi)concave pseudo-games, and proved the existence of generalized Nash equilibrium in this model.
Since this game-theoretic model is sufficiently rich to capture Arrow-Debreu economies, they obtained as a corollary the existence of general equilibrium in these economies.

%
In their model, \citeauthor{arrow-debreu} posit a set of resources, modeled as commodities, each of which is assigned a price; a set of consumers, each choosing a quantity of each commodity to consume in exchange for their endowment; and a set of firms, each choosing a quantity of each commodity to produce, with prices determining \mydef{(aggregate) demand}, i.e., the sum of the consumptions across all consumers, and \mydef{(aggregate) supply}, i.e., the sum of endowments and productions across all consumers and firms, respectively.
This model is \mydef{static}, as it comprises only a single period model, but it is nonetheless rich, as commodities can be state and time contingent, with each one representing a good or service which can be bought or sold in a single time period, but that encodes delivery opportunities at a finite number of distinct 
points in time. Following \citeauthor{arrow-debreu}'s seminal existence result, the literature slowly turned away from static economies, such as Arrow-Debreu competitive economies, which do not \emph{explicitly\/} involve time and uncertainty.

\citeauthor{arrow-debreu}'s model fails to provide a comprehensive account of the economic activity observed in the real world, especially that which is designed to account for \emph{time\/} and \emph{uncertainty}. 
Chief among these activities are 
\mydef{asset markets}, 
which allow consumers and firms to insure themselves against uncertainty about future states of the world. 
Indeed, while static economies with
state- and time-contingent commodities can \emph{implicitly\/} incorporate time and uncertainty, the assumption that a complete set of state- and time-contingent commodities are available at the time of trade is highly unrealistic. 
\citeauthor{arrow1964role} (\citeyear{arrow1964role}) thus proposed to 
enhance the Arrow-Debreu competitive economy 
with \mydef{assets} (or \mydef{securities} or \mydef{stocks}),%
\footnote{Some authors (e.g., \citet{geanakoplos1990introduction}) distinguish between assets, stocks, and securities, instead defining securities (resp. stocks) as those assets which are defined exogenously (resp.\@ endogenously), e.g., government bonds (resp.\@ company stocks). 
As this distinction makes no mathematical difference to our results, and is only relevant to stylized models, we make no such distinction.} 
i.e., contracts between two consumers, which 
promise the delivery of commodities by its seller to its buyer at a future date. 
In particular, \citeauthor{arrow1964role} introduced an asset type nowadays known as the \mydef{\textit{num\'eraire} Arrow security},
which transfers one unit of a designated commodity used as a unit of account---\mydef{the num\'eraire}---when a particular state of the world is observed, and nothing otherwise.
As the num\'eraire is often interpreted as money, assets which deliver only some amount of the num\'eraire, are called \mydef{financial assets}.


Formally, Arrow considered a \mydef{two-step stochastic exchange economy}.
In the initial state, consumers can buy or sell \textit{num\'eraire} Arrow securities in a \mydef{financial asset market}.
Following these trades, the economy stochastically transitions to one of finitely many other states in which consumers receive returns on their initial investment and participate in a \mydef{spot market}, i.e., a market for the immediate delivery of commodities, modeled as a static exchange economy---which, for our purposes, is better called an \mydef{exchange market}.%
\footnote{An (Arrow-Debreu) exchange economy is simply an (Arrow-Debreu) competitive economy without firms. Historically, for simplicity, 
it has become standard practice \emph{not\/} to model firms, as most, if not all, results extend directly to settings with firms. 
In line with this practice, we do not model firms, but note that our results and methods also extend directly to settings that include firms.}
A general equilibrium of this economy is then simply prices for financial assets \emph{and\/} 
commodities, which lead to a feasible allocation of all resources (i.e., financial assets and spot market commodities) that satisfies Walras' law. 

\citet{arrow1964role} demonstrated that the general equilibrium consumptions of an exchange economy with state- and time-contingent commodities can be implemented by the general equilibrium spot market consumptions of a two-step stochastic economy with a considerably smaller, yet \mydef{complete} set of \textit{num\'eraire} Arrow securities, i.e., a set of securities available for purchase in the first period that allow consumers to transfer wealth to \emph{all\/} possible states of the world that can be realized in the second period.
In conjunction with the welfare theorems \cite{debreu1951pareto, arrow1951extension}, this result implies that economies with \mydef{complete financial asset markets}, i.e., economies with such a complete set of securities, 
achieve a Pareto-optimal allocation of commodities by ensuring optimal risk-bearing via financial asset markets; and
conversely, any Pareto-optimal allocation of commodities in economies with time and uncertainty can be realized as a competitive equilibrium of a complete financial asset market.

Arrow's contributions led to the development of a new class of general equilibrium models, namely \mydef{stochastic economies} (or \mydef{dynamic stochastic general equilibrium---DSGE---models}) \cite{geanakoplos1990introduction}.%
\footnote{As these models incorporate both time and uncertainty, they are often referred to as dynamic stochastic general equilibrium models.
Nonetheless, we opt for the stochastic economy nomenclature, because, as we 
demonstrate in this paper, these economies can be seen as instances of (generalized) stochastic games.}
At a high-level, these models comprise a sequence of world states and spot markets,
which are linked across time by
asset markets, with each next state of the world (resp.\@ spot market) determined by a stochastic process that is independent of market interactions (resp.\@ dependent only on their asset purchase) in the current state. 
Mathematically, the key difference between a static and a stochastic economy is that consumers in a stochastic economy face a collection of budget constraints, one per time-step, rather than only one.
Indeed, \citet{arrow1964role}'s proof that general equilibrium consumptions in stochastic complete economies are equivalent to general equilibrium consumptions in static state- and time-contingent commodity economies relies on proving that the many budget constraints in a complete stochastic economy can be reduced to a single one.

Stochastic economies were introduced to model arbitrary finite time horizons \cite{radner1968competitive} and a variety of risky asset classes (e.g., stocks \cite{diamond1967incompletege}, risky assets \cite{lintner1975valuation}, derivatives \cite{black1973pricing}, capital assets \cite{mossin1966equilibrium}, debts \cite{modigliani1958cost} etc.), eventually leading to the emergence of \mydef{stochastic economies with incomplete asset markets} \cite{magill1991incomplete, magill2002theory, geanakoplos1990introduction}, or colloquially, \mydef{(incomplete) stochastic economies}.%
\footnote{While many authors have called these models incomplete economies \cite{geanakoplos1990introduction, magill2002theory, magill1991incomplete}, these models capture both incomplete and complete asset markets.
In contrast, we refer to stochastic economies with incomplete or complete asset markets as \mydef{stochastic economies}, adding the (in)complete epithet only when necessary to indicate that the asset market is (in)complete.}
Unlike in Arrow's stochastic economy, the asset market is not complete in such economies, so consumers cannot necessarily insure themselves against all future world states. 

The archetypal stochastic economy is the \mydef{Radner stochastic exchange economy}, deriving its name from Radner's proof of existence of a general equilibrium in his model \cite{radner1972existence}.
Radner's economy is a finite-horizon stochastic economy comprising a sequence of spot markets, linked across time by asset markets.
At each time period, a finite set of consumers observe a world state and trade in an asset market and a spot market, modeled as an exchange market. 
Each \mydef{asset market} comprises \mydef{assets}, modelled as time-contingent \mydef{generalized Arrow securities}, which specify quantities of the commodities the seller is obliged to transfer to its buyer, should the relevant state of the economy be realized at some specified future time.%
\footnote{Here, Arrow securities are ``generalized'' in the sense that they can deliver different quantities of \emph{many\/} commodities at different states of the world, rather than only one unit of a commodity at only one state of the world. Although \citet{arrow1964role} considered only \textit{num\'eraire} securities, 
his theory was subsequently generalized to 
generalized Arrow securities \cite{geanakoplos1990introduction}.} 
Consumers can buy and sell assets, thereby transferring their wealth across time, all the while insuring themselves against uncertainty about the future.
The canonical solution concept for stochastic economies, \mydef{Radner equilibrium}
(also called \mydef{sequential competitive equilibrium}%
\footnote{This terminology does not contradict the economy being at a competitive equilibrium, but rather indicates that at all times, the spot and asset markets are at a competitive equilibrium, hence implying the overall economy is at a general equilibrium.} 
\cite{mas-colell}, \mydef{rational expectations equilibrium} \cite{radner1979rational}, and \mydef{general equilibrium with incomplete markets} \cite{geanakoplos1990introduction}), is a collection of history-dependent prices for commodities and assets, as well as history-dependent consumptions of commodities and portfolios of assets, such that, for all histories, the aggregate demand for commodities and the \mydef{aggregate demand for assets} (i.e., the total number of assets bought) are feasible and satisfy Walras' law.

In spite of substantial interest in stochastic economies among microeconomists throughout the 1970s, the literature eventually trailed off, 
perhaps due to the difficulty of proving existence of a general equilibrium in simple economies with incomplete asset markets that allow assets to be sold short \cite{geanakoplos1990introduction}, or to the lack of a second welfare theorem \cite{dreze1974investment, hart1975optimality}.
Financial and macroeconomists stepped up, however, with financial economists seeking to further develop the theoretical aspects of stochastic economies (see, for instance, \citet{magill2002theory}), and macroeconomists seeking practical methods by which to solve stochastic economies in order to determine the impact of various policy choices (via simulation; see, for instance, \citet{sargent2000recursive}).

\mydef{Infinite horizon stochastic economies} are one of the new and interesting directions in this more recent work on stochastic economies.
Infinite horizon models come with one significant difficulty that has no counterpart in a finite horizon model, namely the possibility for agents to run a \mydef{Ponzi scheme} via asset markets, in which they borrow but then indefinitely postpone repaying their debts by refinancing them continually, from one period to the next.
From this perspective infinite horizon models represent very interesting objects of study, not only theoretically; it has also been argued that they are a better modeling paradigm for macroeconomists who employ simulations \cite{magill1994infinite}, because they facilitate the modeling of complex phenomena, such as asset bubbles \cite{huang2000asset}, which can be impacted by economic policy decisions.

\citet{magill1994infinite} introduced an extension of Radner's model to an infinite horizon setting, albeit with financial assets, 
and presented suitable assumptions under which a sequential competitive equilibrium is guaranteed to exist in this model.
Progress on the computational aspects of stochastic economies has been slow, however, and mostly confined to finite horizon settings
(see, \citet{sargent2000recursive} and Volume 2 of \citet{taylor1999handbook} for a standard survey, and \citet{FernandezVillaverde2023CompMethodsMacro}  for a more recent entry-level survey of computational methods used by macroeconomists).
Indeed, demands for novel computational methods for solving macroeconomic models, and theoretical frameworks in which to understand their computational complexity, have been repeatedly shared by macroeconomists \cite{taylor1999handbook}. 
This gap in the literature points to a novel research opportunity; however, it is challenging for non-macroeconomists 
to approach these problems with their computational tools. 

\paragraph{Contributions}


In \Cref{sec:gmg}, we introduce Markov pseudo-games, 
and we establish the existence of (pure) \mydef{generalized Markov perfect equilibria (\MPGNE)} in 
concave Markov pseudo-games (\Cref{thm:existence_of_mpgne}). 
This result can be seen as a stochastic generalization of \citet{arrow-debreu}'s existence result for (pure) generalized Nash equilibrium in concave pseudo-games \cite{facchinei2010generalized}.
To the best of our knowledge, it also implies the existence of pure (or deterministic) Markov perfect equilibria in a large class of continuous-action Markov games for which existence was heretofore known only in mixed (or randomized) policies \cite{fink1964equilibrium, takahashi1964equilibrium}.

Although the computation of \MPGNE{} is PPAD-hard in general, because \MPGNE{} generalize Nash equilibrium, we reduce this computational problem to generative adversarial learning between a generator, who produces a candidate equilibrium policy profile, and an adversary, who produces a policy profile of best responses to the candidate equilibrium \cite{goktas2023generative} (\Cref{obs:exploit_min_to_min_max}).
Assuming parameterized policies, and taking advantage of the recent progress on solving generative adversarial learning problems (e.g.,~\cite{lin2020gradient, daskalakis2020independent}), we show that a policy profile which is a stationary point of the exploitability (i.e., the players’ cumulative maximum regret) can be computed in polynomial time under mild assumptions (\Cref{thm:convergence_GNE}). 
This result implies that a policy profile that satisfies necessary first-order stationarity conditions for a \mpgne{} in Markov pseudo-games with a bounded best-response mismatch coefficient (\Cref{lemma:br_mismatch_coef})---i.e., those Markov pseudo-games in which states explored by any \mpgne{} are easily explored under the initial state distribution---can be computed in polynomial time, a result which is analogous known computational results for zero-sum Markov games \cite{daskalakis2020independent}.
As our theoretical computational guarantees apply to policies represented by neural networks, we obtain the first, to our knowledge, \amy{must check this again!!!} deep reinforcement learning algorithm with theoretical guarantees for general-sum games.




In \Cref{sec:infinite}, we introduce an extension of \citet{magill1994infinite}'s infinite horizon exchange economy, which we call the \mydef{infinite horizon Markov exchange economy}. 
On the one hand, our model generalizes \citeauthor{magill1994infinite}'s to a setting with arbitrary, not just financial assets; on the other hand, we restrict the transition model to be Markov.
The Markov restriction allows us to prove the existence of a \mydef{recursive Radner equilibrium (RRE)} \cite{mehra1977recursive} (\Cref{thm:existence_RRE}).
Our proof reformulates the set of RRE of any infinite horizon Markov exchange economy as the set of GMPE of an associated Markov pseudo-game (\Cref{thm:existence_RRE}).
To our knowledge, ours is the first result of its kind for such a general setting, as previous recursive competitive equilibrium existence proofs were restricted to economies with one consumer (also called the representative agent), one commodity, or one asset \cite{mehra1977recursive, prescott1980recursive}. 
The aforementioned results allow us to conclude that a stationary point of the exploitability of the Markov pseudo-game associated with any infinite horizon Markov exchange economy can be computed in polynomial time (\Cref{thm:compute_SRE}).

Finally, in \Cref{sec:expts} we implement our policy gradient method in the form of a generative adversarial policy network (GAPNet), and use it to search for RRE in three infinite horizon Markov exchange economies with three different types of utility functions.
Experimentally, we observe that GAPNet finds approximate equilibrium policies that are closer to \MPGNE{} than those produced by a standard baseline for solving stochastic economies.

\longversion{
\subsection{Outline}

This paper is organized as follows.
In Section~\ref{sec:prelim}, we define our notation and some requisite mathematical terminology.
Next, in Section~\ref{sec:gmg}, we develop our model of generalized Markov games (Section~\ref{sec:gmg_model}), for which we define appropriate solution concepts, i.e., equilibria, in Section~\ref{sec:gmg_exist}.
We then (Section~\ref{sec:gmg_conv}) present a gradient descent-ascent-based reinforcement learning algorithm (\Cref{alg:two_time_sgda}; TTSSGDA) which provably converges to these equilibria via a coupled min-max optmization formulation of the problem (Section~\ref{sec:gmg_minmax}).
Then, in Section~\ref{sec:dsge}, we apply our theory to infinite horizon stochastic exchange economies.
First, in Section~\ref{sec:static}, we develop static exchange economies, i.e., spot markets, as generalized (one-shot) games;
then, in Section~\ref{sec:infinite}, we develop infinite horizon---in general, incomplete---stochastic economies, by instantiating our model of generalized Markov games.
Finally, we invoke our main theorems for generalized Markov games to establish the existence of recursive competitive equilibria in incomplete stochastic economies, the first such result to our knowledge, as well as convergence of TTSSGDA to this equilibrium.
This latter result is notable because the equilibria that we learn are equilibria in Markov policies, as opposed to unwieldy non-stationary policies; and recursive, meaning they exhibit subgame-perfect-like behavior.}

\section{Preliminaries}
\label{sec:prelim}

\textbf{Notation.}
We use caligraphic uppercase letters to denote sets (e.g., $\calX$), bold uppercase letters to denote matrices (e.g., $\allocation$), bold lowercase letters to denote vectors (e.g., $\price$), lowercase letters to denote scalar quantities (e.g., $x$), and uppercase letters to denote random variables (e.g., $X$). 
We denote the $i$th row vector of a matrix (e.g., $\allocation$) by the corresponding bold lowercase letter with subscript $i$ (e.g., $\allocation[\buyer])$. 
Similarly, we denote the $j$th entry of a vector (e.g., $\price$ or $\allocation[\buyer]$) by the corresponding lowercase letter with subscript $j$ (e.g., $\price[\good]$ or $\allocation[\buyer][\good]$).
We denote functions by a letter determined by the value of the function, e.g., $f$ if the mapping is scalar valued, $\f$ if the mapping is vector valued, and $\calF$ if the mapping is set valued.

We denote the set $\left\{1, \hdots, n\right\}$ by $[n]$, the set $\left\{0, 1, \hdots, n\right\}$ by $[n^*]$, the set of natural numbers by $\N$, and the set of real numbers by $\R$. 
We denote the positive and strictly positive elements of a set using a $+$ or $++$ subscript, respectively, e.g., $\R_+$ and $\R_{++}$.

For any $n \in \N$, we denote the  $n$-dimensional vector of zeros and ones by $\zeros[n]$ and $\ones[n]$, respectively.
We let $\simplex[n] = \{\x \in \R_+^n \mid \sum_{i = 1}^n x_i = 1 \}$ denote the unit simplex in $\R^n$, and $\simplex(A)$ denote the set of all probability distributions over a given set $A$.
We also define the support of a probability density function $f \in \simplex(\calX)$ as $\supp(f) \doteq \left\{ \x \in \calX: f(\x) > 0 \right\}$.
Finally, we denote the orthogonal projection operator onto a set $C$ by $\project[C]$, i.e., $\project[C](\x) \doteq \argmin_{\y \in C} \left\|\x - \y \right\|^2$.

We define the subdifferential of a function $\obj: \calX \times \calY \to \R$ w.r.t.\@ variable $\x$ at a point $(\a, \b) \in \calX \times \calY$ by $\subdiff[\x] f (\bm{a}, \b) \doteq \{\h \mid f(\bm{x}, \b) \geq f (\bm{a}, \b) + \h^T (\bm{x} - \bm{a}) \}$, 
and we denote the derivative operator (resp.\@ partial derivative operator w.r.t.\@ $\x$) 
of a function $\g: \calX \times \calY \to \calZ$ by $\subgrad \g$ (resp.\@ $\subgrad[\x] \g$).


\textbf{Terminology.}
Fix any norm $\| \cdot \|$. Given $\calA \subset \R^d$, the function $\obj: \calA \to \R$ is said to be $\lipschitz[\obj]$-\mydef{Lipschitz-continuous} iff $\forall \x_1, \x_2 \in \calX, \left\| \obj(\x_1) - \obj(\x_2) \right\| \leq \lipschitz[\obj] \left\| \x_1 - \x_2 \right\|$.
If the gradient of $\obj$ is $\lipschitz[\grad \obj]$-Lipschitz-continuous, $\obj$ is called $\lipschitz[\grad \obj]$-\mydef{Lipschitz-smooth}. 

We require notions of stochastic convexity related to stochastic dominance of probability measures \cite{atakan2003valfunc}.
Given non-empty and convex parameter and outcome spaces $\calW$ and $\calO$ respectively, a conditional probability distribution $\w \mapsto \trans(\cdot \mid \w) \in \simplex (\calO)$ is said to be \mydef{stochastically convex} (resp.\ \mydef{stochastically concave}) in $\w \in \calW$ if for all continuous, bounded, and convex (resp.\ concave) functions $\statevalue: \calO \to \R$,  $\lambda \in (0,1)$, and $\w^\prime, \w^\dagger \in \calW$ s.t.\ $\mean[\w] = \lambda \w^\prime + (1-\lambda)  \w^\dagger$, it holds that $\Ex_{O \sim \trans(\cdot \mid \mean[\w])} \left[\statevalue(O)\right] \leq \text{ (resp.\@ $\geq$) } \lambda \Ex_{O \sim  \trans(\cdot \mid \w^\prime)} \left[\statevalue(O)\right] + (1-\lambda)\Ex_{O \sim \trans(\cdot \mid \w^\dagger)} \left[\statevalue(O)\right]$.

\deni{Define Gradient dominance?} \amy{if you use it, then yes, of course!}


\section{Markov Pseudo-Games}
\label{sec:gmg}

We begin by developing our formal game model.
The games we study are stochastic, in the sense of \citet{shapley1953stochastic}, \citet{fink1964equilibrium}, and \citet{takahashi1964equilibrium}.
Further, they are pseudo-games, in the sense of \citet{arrow-debreu}.
\citeauthor{arrow-debreu} introduced pseudo-games to establish the existence of competitive equilibrium in their seminal model of an exchange economy, where an auctioneer sets prices that determine the consumers' budget sets, and hence their feasible consumptions.
It is this dependency among the players' feasible actions that characterizes pseudo-games.
%
We model stochastic pseudo-games, and dub them \mydef{Markov pseudo-games}, as the games are Markov in that the stochastic transitions depend only on the most recent state and player actions.



\subsection{Model}
\label{sec:gmg_model}


An \mydef{(infinite horizon discounted) Markov pseudo-game}\longversion{%
\footnote{We use the term ``generalized game'' where others before us have used ``pseudo-game'' \cite{facchinei2010generalized}, ``abstract economy'' \cite{arrow-debreu}, and ``social system'' \cite{debreu1952social}. 
In generalized games, because the players' actions are constrained by one another's, a number of authors have argued that it can be hard to imagine a game where the players make their choices simultaneously and it so happens that all constraints are satisfied. 
Our choice of terminology is intended to reflect the fact that a generalized game can be seen as a game with a mediator who proposes an action profile from which the players can deviate, similar to the account given to justify correlated strategies in normal-form games \cite{Aumann1974welfare}.}} 
$\mgame \doteq (\numplayers, \numactions, \states, \actionspace, \actions, \rewards, \trans, \discount, \initstates)$ is an $\numplayers$-player dynamic
game played over an infinite discrete time horizon. 
The game starts at time $\numhorizon = 0$ in some initial state $\staterv[0] \sim \initstates \in \simplex (\states)$ drawn randomly from a set of states $\states \subseteq \R^{\numstates}$. 
At this and each subsequent time period $\numhorizon = 1, 2, \hdots$, the players encounter a state $\state[\numhorizon] \in \states$, in which each $\player \in \players$ simultaneously takes an \mydef{action} $\action[\player][][\numhorizon] \in \actions[\player] (\state[\numhorizon], \action[-\player][][\numhorizon])$ from a \mydef{set of feasible actions} $\actions[\player] (\state[\numhorizon], \action[-\player][][\numhorizon]) \subseteq \actionspace[\player] \subseteq \R^{\numactions}$, determined by a \mydef{feasible action correspondence} $\actions[\player]: \states\times \actionspace[-\player] \rightrightarrows \actionspace[\player]$, which takes as input the current state $\state[\numhorizon]$ and the other players' actions $\action[-\player][][\numhorizon] \in \actionspace[-\player]$, and outputs a subset of the $\player$th player's action space $\actionspace[\player]$. 
We define $\actions (\state, \action) \doteq \bigtimes_{\player \in \players} \actions[\player] (\state, \action[-\player])$. 
%

Once the players have taken their actions $\action[][][\numhorizon] \doteq (\action[1][][\numhorizon], \hdots, \action[\numplayers][][\numhorizon])$, each player $\player \in \players$ receives a \mydef{reward} $\reward[\player](\state[\numhorizon][], \action[][][\numhorizon])$ given by a \mydef{reward function} $\rewards: \states \times \actionspace \to \R^\numplayers$, after which the game either ends with probability $1-\discount$, where $\discount \in (0,1)$ is called the \mydef{discount factor},\longversion{%
\footnote{
Our results generalize to settings with per-player discount factors $\discount_\buyer \in (0, 1)$, where the discount rates express the players' intertemporal preferences over game outcomes at each time-step. \deni{Elaborate more!}.}}
or continues on to time period $\numhorizon + 1$, transitioning to a new state $\staterv[\numhorizon + 1] \sim \trans (\cdot \mid \state[\numhorizon], \action[][][\numhorizon])$, according to a \mydef{transition} probability function $\trans: \states \times \states \times \actionspace \to \R_+$, where $\trans (\state[\numhorizon + 1] \mid \state[\numhorizon], \action[][][\numhorizon]) \in [0,1]$ denotes the probability of transitioning to state $\state[\numhorizon + 1] \in \states$ from state $\state[\numhorizon] \in \states$ when action profile $\action[][][\numhorizon] \in \actionspace$ is played.

Our focus is on continuous-state and continuous-action Markov pseudo-games, where
the state and action spaces
are non-empty and compact, and the reward functions are continuous and bounded in each of $\state$ and $\action$, holding the other fixed.

A \mydef{history} $\hist[][][] \in \hists[\numhorizons] \doteq (\states \times \actionspace)^\numhorizons \times \states$ of length $\numhorizons \in \N$ is a sequence of states and action profiles $\hist[][][] = ((\state[\numhorizon], \action[][][\numhorizon])_{\numhorizon = 0}^{\numhorizons-1}, \state[\numhorizons])$
s.t.\@ a history of length $0$ corresponds only to the initial state of the game. %
For any history $\hist[][][] = ((\state[\numhorizon], \action[][][\numhorizon])_{\numhorizon = 0}^{\numhorizons-1}, \state[\numhorizons])$ of length $\numhorizons \in \N$, we denote by $\hist[: \othernumhorizons]$ the first $\othernumhorizons \in [\tau^*]$ steps of $\hist$, i.e., $\hist[: \othernumhorizons] = ((\state[\numhorizon], \action[][][\numhorizon])_{\numhorizon = 0}^{\othernumhorizons-1}, \state[\othernumhorizons])$.
Overloading notation, we define the \mydef{history space} $\hists \doteq \bigcup_{\numhorizons = 0}^\infty \hists[\numhorizons]$.
%
For any player $\player \in \players$,
a \mydef{policy} $\policy[\player]: \hists \to \actionspace[\player]$ is a mapping from histories of any length to $\player$'s space of (pure) actions.
We define the space of all (deterministic) policies as $\policies[\player] \doteq \{ \policy[\player]: \hists \to \actionspace[\player] \}$.\longversion{%
\footnote{
A mixed policy is simply a distribution over pure policies, i.e., an element of $\simplex (\policies[\player])$.
Moreover, any mixed policy can be equivalently represented as a mapping $\policy[\player][][\mathrm{mixed}]: \hists \to \simplex(\actionspace[\player])$ from histories to distributions over actions s.t. at any history $\hist \in \hists$, player $\player$ plays action $\action[\player] \sim \policy[\player] (\hist)$. 
An analogous definition extends directly to mixed Markov policies as well.}}
%
A \mydef{Markov policy} \cite{maskin2001markov} $\policy[\player]$ is a 
policy s.t.\@ $\policy[\player] (\state[\numhorizons]) = \policy[\player] (\hist[: \numhorizons])$, 
for all histories $\hist \in \hists[\numhorizons]$ of length $\numhorizons \in \N_+$, where $\state[\numhorizons]$ denotes the final state of history $\hist$. 
As Markov policies are only state-contingent, we can compactly represent the space of all Markov policies for player $\player \in \players$ as $\markovpolicies[\player] \doteq \{ \policy[\player]: \states \to \actionspace[\player] \}$.


Fixing player $\player \in \players$ and $\policy[-\player] \in \policies[-\player]$, we define the \mydef{feasible policy correspondence} $\fpolicies[\player] (\policy[-\player]) \doteq \{ \policy[\player] \in \policies[\player] \mid \forall \hist \in \hists, \policy[\player] (\hist) \in \actions[\player] (\state[\numhorizons], \policy[-\player] (\hist)) \}$, given history $\hist \in \hists[\numhorizons]$, 
and the \mydef{feasible subclass policy correspondence} $\fsubpolicies[\player] (\policy[-\player]) \doteq \{ \policy[\player] \in \subpolicies[\player] \mid \forall \state \in \states, \policy[\player] (\state) \in \actions[\player] (\state, \policy[-\player] (\state)) \}$, for any $\subpolicies \subseteq \markovpolicies$. 
Of particular interest is $\fmarkovpolicies[\player] (\policy[-\player])$ itself, obtained when $\subpolicies = \markovpolicies$.

Given a policy profile $\policy \in \policies$ and a history $\hist[][][] \in \hists[\numhorizons]$,
the \mydef{discounted history distribution that originates at state $\state$} is defined as $\histdistrib[\state][\policy, \numhorizons] (\hist[][][]) = \setindic[\state] (\state[0]) \prod_{\numhorizon = 0}^{\numhorizons-1} \discount^\numhorizon \trans (\state[\numhorizon  + 1] \mid \state[\numhorizon], \action[][][\numhorizon]) \setindic[{\{ \policy (\hist[:\numhorizon]) \}}] (\action[][][\numhorizon])$.
Furthermore, the \mydef{discounted history distribution given initial state distribution $\initstates$} is defined as 
$\histdistrib[\initstates][\policy, \numhorizons] (\hist[][][]) = \Ex_{\staterv \sim \initstates} \left[ \histdistrib[\staterv][\policy, \numhorizons] (\hist[][][]) \right]$. 
Next, we define the set of all realizable trajectories of length $\numhorizons$ under policy $\policy$ as $\hists[\policy][\numhorizons]_{\initstates} \doteq \supp (\histdistrib[\initstates][\policy][\numhorizons])$, i.e., the set of all histories that occur with non-zero probability given initial state distribution $\initstates$, and we let $\hists[\policy]_{\initstates} \doteq \hists[\policy][\infty]_{\initstates}$ and $\histdistrib[\initstates][\policy] \doteq \histdistrib[\initstates][\policy][\infty]$.
We can now write $\histrv[][] = \left( \staterv[0], (\actionrv[][][\numhorizon], \staterv[\numhorizon + 1] )_{\numhorizon = 0}^{\numhorizons - 1} \right)$ to denote a history $\hist \in \hists[\policy][\numhorizons]_{\initstates}$ sampled from $\histdistrib[\initstates][\policy][\numhorizons]$.

Now, given a policy profile $\policy \in \policies$, we define the \mydef{state-value function} $\vfunc[][\policy]: \states \to \R^\numplayers$ 
and the \mydef{action-value function} $\qfunc[][\policy]: \states \times \actionspace \to \R^\numplayers$, respectively, as
\begin{align}
    \vfunc[][\policy] (\state) 
    &\doteq \E_{\histrv \sim \histdistrib[\state][\policy]} \left[ \sum_{\numhorizon = 0}^{\infty} \rewards (\staterv[\numhorizon], \actionrv[][][\numhorizon]) \right] \\
    &= \E_{\histrv \sim \histdistrib[\initstates][\policy]} \left[ \sum_{\numhorizon = 0}^{\infty} \rewards (\staterv[\numhorizon], \actionrv[][][\numhorizon]) \mid \staterv[0] = \state \right] \\
    &= \int_{\histrv \in {\hists[\policy]_{\initstates}} : {\staterv[0]} = \state} \sum_{\numhorizon = 0}^{\infty} \rewards (\staterv[\numhorizon], \actionrv[][][\numhorizon]) \, \histdistrib[\initstates][\policy, \numhorizons] (\histrv[][][]) \, d \histrv
    \label{eq:state_value} \\
    \qfunc[][\policy] (\state, \action) 
    &\doteq \rewards (\state, \action) + \E_{{\staterv[][][\prime]} \sim \, \trans ({\staterv[][][\prime]} \mid \state, \action)} \left[ \vfunc[][\policy] (\staterv[][][\prime]) \right]
    \label{eq:action_value}
\end{align}

Finally, we define the \mydef{discounted state-visitation distribution} 
$\statedist[{\initstates}][\policy] (\state) \doteq \sum_{\numhorizons = 0}^\infty \int_{\hist \in \hists[\policy][\numhorizons]_{\initstates}: \state[\numhorizons] = \state} \histdistrib[\initstates][\policy][\numhorizons] (\hist[][][])$ and
the \mydef{(expected) 
payoff} of policy profile $\policy$ as $\payoff (\policy) \doteq \Ex_{\staterv \sim \initstates} \left[ \vfunc[][\policy] (\staterv) \right] = \Ex_{\staterv \sim {\statedist[{\initstates}][\policy]}} \left[ \rewards (\staterv, \policy (\staterv)) \right]$.

\subsection{Solution Concepts and Existence}
\label{sec:gmg_exist}

Having defined our game model, we now define two natural solution concepts, and establish their existence.
Our first solution concept is based on the usual notion of Nash equilibrium (\citeyear{nash1950existence}), yet applied to Markov pseudo-games.
Our second is based on the notion of subgame-perfect equilibrium in extensive-form games, a strengthening of Nash equilibrium with the additional requirement that an equilibrium be Nash not just at the start of the game, but at all states encountered during play.
In the context of stochastic games, such equilibria are called ``recursive,'' or ``Markov perfect.''
Following \citet{bellman1966dynamic} and \citet{arrow-debreu}, we identify natural assumptions that guarantee the existence of equilibrium in pure Markov
policies, meaning deterministic policies that depend only on the current state, not on the history.
When applied to incomplete stochastic economies, this theorem implies existence of recursive Radner (or competitive) equilibrium, to our knowledge the first result of its kind.


An $\varepsilon$-\mydef{generalized Markov perfect equilibrium ($\varepsilon$-\MPGNE)} $\policy[][][*] \in \fmarkovpolicies (\policy[][][*])$ is a Markov policy profile s.t.\@ for all states $\state \in \states$ and players $\player \in \players$, $\vfunc[\player][{\policy[][][*]}] (\state) \geq \max_{\policy[\player] \in \fpolicies[\player] (\policy[-\player][][*])} \vfunc[\player][{(\policy[\player], \policy[-\player][][*])}] (\state) - \varepsilon$.  
An $\varepsilon$-\mydef{generalized Nash equilibrium ($\varepsilon$-GNE)} $\policy[][][*] \in \fpolicies (\policy[][][*])$ is a policy profile s.t.\@ for all states $\state \in \states$ and players $\player \in \players$, $\payoff[\player] ({\policy[][][*]}) \geq \max_{\policy[\player] \in \fpolicies[\player] (\policy[-\player][][*])} \payoff[\player] (\policy[\player], \policy[-\player][][*]) - \varepsilon$. 
We call a $0$-\MPGNE{} ($0$-GNE) simply a \MPGNE{} (GNE).
As \MPGNE{} is a stronger notion than GNE, every $\varepsilon$-\MPGNE{} is an $\varepsilon$-GNE.

\begin{assumption}[Existence] \label{assum:existence_of_mpgne}
For all $\player \in \players$, assume
1.~$\actionspace[\player]$ is convex;
2.~$\actions[\player] (\state, \cdot)$ is upper- and lower-hemicontinuous, for all $\state \in \states$; 
3.~$\actions[\player] (\state, \action[-\player])$ is non-empty, convex, and compact, for all $\state \in \states$ and $\action[-\player] \in \actionspace[-\player]$; and
4.~for any policy $\policy \in \policies$, $\action[\player] \mapsto \qfunc[\player][\policy] (\state, \action[\player], \action[-\player])$ is continuous and concave over $\actions[\player] (\state, \action[-\player])$, for all $\state \in \states$ and $\action[-\player] \in \actionspace[-\player]$.
\end{assumption}

\begin{assumption}[Policy Class] \label{assum:policy_class_exist}
Given 
$\subpolicies \subseteq \markovpolicies$, assume
1.~$\subpolicies$ is non-empty, compact, and convex; and
2.~(Closure under policy improvement) for each $\policy \in \subpolicies$, there exists $\policy[][][+] \in \subpolicies$ s.t.\@ $\qfunc[\player][\policy] (\state, \policy[\player][][+] (\state), \policy[-\player] (\state)) = \max_{\policy[\player][][\prime] \in \fpolicies (\policy[-\player])} \qfunc[\player][\policy] (\state, \policy[\player][][\prime] (\state), \policy[-\player] (\state))$, for all $\player \in \players$ and $\state \in \states$. 


\end{assumption}


Assumption 2, introduced as Condition 1 in \citeauthor{bhandari2019global} (\citeyear{bhandari2019global}), ensures that the policy class under consideration (e.g., $\subpolicies \subseteq \markovpolicies$)
is expressive enough to include best responses.

\begin{restatable}{theorem}{thmexistmpgne}
\label{thm:existence_of_mpgne}
Let $\mgame$
be a Markov pseudo-game for which \Cref{assum:existence_of_mpgne} holds, 
and let $\subpolicies\subseteq \markovpolicies$ be a subspace of Markov policy profiles that satisfies \Cref{assum:policy_class_exist}. 
Then there exists a policy $\policy[][][*] \in \subpolicies$ such that $\policy[][][*]$ is an 
\MPGNE{} of $\mgame$.
\end{restatable}

\subsection{Equilibrium Computation}
\label{sec:gmg_minmax}

Our approach to computing a \MPGNE{} in a Markov pseudo-game $\mgame$ is to minimize a \mydef{merit function} associated with $\mgame$, i.e., a function whose minima coincide with the pseudo-game's \MPGNE.
Our choice of merit function, a common one in game theory, is \mydef{exploitability} $\gexploit: \policies \to \R_+$, defined as $\gexploit (\policy) \doteq \sum_{\player \in \players} \left[ \max_{\policy[\player][][\prime] \in \fmarkovpolicies[\player] (\policy[-\player])}
\payoff[\player] (\policy[\player][][\prime], \policy[-\player]) - \payoff[\player] (\policy) \right]$.
In words, exploitability is the sum of the players' maximal unilateral payoff deviations. 

Exploitability, however, is a merit function for GNE, \emph{not\/} \MPGNE; \mydef{state exploitability}, $\sexploit (\state, \policy) = \sum_{\player \in \players} [\max_{\policy[\player][][\prime] \in \fmarkovpolicies[\player] (\policy[-\player])} \vfunc[\player][{(\policy[\player][][\prime], \policy[-\player])}] (\state) - \vfunc[\player][\policy] (\state)]$ at all states $\state \in \states$, is a merit function for \MPGNE.
Nevertheless, as we show in the sequel, for a large class of Markov pseudo-games, namely those with a bounded best-response mismatch coefficient (see Section~\ref{sec:mismatch}), 
the set of Markov policies that minimize exploitability 
equals the set of \MPGNE, making our approach a sensible one.

We are not out of the woods yet, however, as exploitability is non-convex in general, even in one-shot finite games \cite{nash1950bargaining}.
Although Markov pseudo-games can afford a convex exploitability (see, for instance \cite{flam1994gne}), it is unlikely that all do, as GNE computation is PPAD-hard \cite{chen2009settling, daskalakis2009complexity}. 
Accordingly, we instead set our sights on computing a \mydef{stationary point} of the exploitability, i.e., a policy profile $\policy[][][*] \in \fmarkovpolicies(\policy[][][*])$ s.t. for any other policy $\policy[][][] \in \fmarkovpolicies(\policy[][][*])$, it holds that $\min_{\h\in \subdiff \gexploit(\policy[][][*])} \langle \h, \policy[][][*] - \policy[][][] \rangle \leq 0$.%
\footnote{While we provide a definition of a(n approximate) stationary point for expositional purposes at present, an observant reader might have noticed the exploitability 
is a mapping from a function space to the positive reals, and its Fr\^echet (sub)derivative is ill-specified without a clear definition of the normed vector space of policies on which exploitability is defined. Further, even when clearly specified, such a (sub)derivative might not exist. The precise meaning of a derivative of the exploitability and its stationary points will be introduced more rigorously after we suitably parameterize the policy spaces.}
Under suitable assumptions, such a point satisfies the necessary conditions of a \MPGNE. 

In this paper, we study Markov pseudo-games with possibly continuous state and action spaces. 
As such, we can only hope to compute an \emph{approximate\/} stationary point of exploitability in finite time. 
Defining a notion of approximate stationarity for exploitability is, however, a challenge.

Given an approximation parameter $\varepsilon \geq 0$, a natural definition of an $\varepsilon$-stationary point might be a policy profile $\policy[][][*] \in \fmarkovpolicies(\policy[][][*])$ s.t. for any other policy $\policy[][][] \in \fmarkovpolicies(\policy[][][*])$, it holds that $\min_{\h\in \subdiff \gexploit(\policy[][][*])} \langle \h, \policy[][][*] - \policy[][][] \rangle \leq \varepsilon$.
Exploitability is not necessarily Lipschitz-smooth, however, so in general it may not be possible 
to compute an $\varepsilon$-stationary point in $\poly(\nicefrac{1}{\varepsilon})$ evaluations of the (sub)gradient of the exploitability.%
\footnote{To see this, consider the convex minimization problem $\min_{x \in \R} f(x) = |x|$. 
The minimum of this optimization occurs at $x = 0$, which is a stationary point since a (sub)derivative of $f$ at $x= 0$ is $0$. 
However, for $x< 0$, we have $\frac{\partial f(x)}{\partial x} = -1$, and for $x> 0$, we have $\frac{\partial f(x)}{\partial x} = 1$. 
Hence, any $x \in \R \setminus \{ 0 \}$ can at best be a $1$-stationary point, i.e., $\left| \frac{\partial f(x)}{\partial x} \right| = 1$. 
Hence, for this optimization problem, it is not even possible to guarantee the existence of an $\varepsilon$-stationary point distinct from $x= 0$, assuming $\varepsilon \in (0, 1)$, let alone the computation of an $\varepsilon$-stationary point $x^*$ s.t. $\left| \frac{\partial f(x^*)}{\partial x} \right| \leq \varepsilon$.}
 
To address, this challenge, a common approach in the optimization literature (see, for instance Appendix H, Definition 19 of \citet{liu_first-order_2021}) is to consider an alternative definition known as $(\varepsilon, \delta)$-stationarity. 
Given approximation parameters $\varepsilon, \delta \geq 0$, an $(\varepsilon, \delta)$-stationary point of exploitability is a policy profile  $\policy[][][*] \in \fmarkovpolicies(\policy[][][*])$ for which there exists a $\delta$-close policy $\policy[][][\dagger] \in \policies$ with $\| \policy[][][\dagger] - \policy[][][*]\| \leq \delta$ s.t.\@ for any other policy $\policy[][][] \in \fmarkovpolicies(\policy[][][\dagger])$, it holds that $\min_{\h\in \subdiff \gexploit(\policy[][][\dagger])} \langle \h, \policy[][][\dagger] - \policy[][][] \rangle \leq \varepsilon$. 
The exploitability minimization method we introduce can indeed compute such an approximate stationary point in polynomial time: i.e., a point in the neighborhood of an approximate stationary point of exploitability.
Furthermore, asymptotically, our method is guaranteed to converge to an exact stationary point of exploitability.

More precisely, following \citet{goktas2022exploit}, who minimize exploitability to solve for variational equilibria in (one-shot) pseudo-games, we first formulate our problem as the quasi-optimization problem of minimizing exploitability,%
\footnote{Here, ``quasi'' refers to the fact that a solution to this problem is both a minimizer of exploitability and a fixed point of an operator, such as $\fpolicies$ or $\fmarkovpolicies$.}
and then transform this problem into a coupled min-max optimization (i.e., a two-player zero-sum game) whose objective is cumulative regret, rather than the potentially ill-behaved exploitability.
Under suitable parametrization, such problems are amenable to polynomial-time solutions via simultaneous gradient descent ascent~\cite{arrow1958studies}, assuming the objective is Lipschitz smooth in both players' decision variables and gradient dominated in the inner player's.
We thus formulate the requisite assumptions to ensure these properties hold of cumulative regret in our game, which in turn allows us to show that two time scale simultaneous stochastic gradient descent ascent (TTSSGDA)
converges to an $(\varepsilon, O(\varepsilon))$-stationary point of exploitability in $\poly(\nicefrac{1}{\varepsilon})$ gradient steps.

\subsubsection{Exploitability Minimization}

Given a Markov pseudo-game $\mgame$ and two policy profiles $\policy, \policy[][][\prime] \in \policies$, 
we define the \mydef{state cumulative regret} at state $\state \in \states$ as $\scumulreg (\state, \policy, \policy[][][\prime]) = \sum_{\player \in \players} \left[\vfunc[\player][{(\policy[\player][][\prime], \policy[-\player])}] (\state) - \vfunc[\player][\policy] (\state) \right]$;
the \mydef{expected cumulative regret} as $\scumulreg (\diststates, \policy, \policy[][][\prime]) = \Ex_{\staterv \sim \diststates} \left[ \scumulreg (\staterv, \policy, \policy[][][\prime]) \right]$, 
for an arbitrary state distribution $\diststates \in \simplex(\states)$, 
and the \mydef{cumulative regret} as $\gcumulreg (\policy, \policy[][][\prime]) = \scumulreg (\initstates, \policy, \policy[][][\prime])$.
Additionally, we define the \mydef{state exploitability} of a policy profile $\policy$ at state $\state \in \states$ as $\sexploit (\state, \policy) = \sum_{\player \in \players} \max_{\policy[\player][][\prime] \in \fmarkovpolicies[\player] (\policy[-\player])} \vfunc[\player][{(\policy[\player][][\prime], \policy[-\player])}] (\state) - \vfunc[\player][\policy] (\state)$;
the \mydef{expected state exploitability} of a policy profile $\policy$ as $\sexploit (\diststates, \policy) = \Ex_{\staterv \sim \diststates} \left[ \sexploit (\staterv, \policy) \right]$, for an arbitrary state distribution $\diststates \in \simplex(\states)$, 
and the (global) \mydef{exploitability} as $\gexploit(\policy)=\sum_{\player\in \players} \max_{\policy[\player][][\prime] \in \fmarkovpolicies[\player] (\policy[-\player])} \payoff[\player](\policy[\player][][\prime], \policy[-\player])$. 

In what follows, we restrict our attention to the subclass $\markovpolicies \subseteq \policies$ of (pure) Markov policies.
This restriction is without loss of generality, because finding an optimal policy that maximizes a state-value or payoff function, while the other players' policies remain fixed, reduces to solving a Markov decision process (MDP), and an optimal (possibly history-dependent) policy in an MDP is guaranteed to exist in the space of (pure) Markov policies
under very mild continuity and compactness assumptions \cite{puterman2014markov}. 
Indeed, the next lemma justifies this restriction.

\begin{restatable}{lemma}{lemmaexploitGNE}\label{lemma:exploit_GNE}
    Given a Markov pseudo-game $\mgame$, a Markov policy profile $\policy[][][*] \in \fmarkovpolicies (\policy[][][*])$ is a \MPGNE{} if and only if $\sexploit (\state, \policy[][][*]) = 0$, for all states $\state \in \states$.
    Similarly, a policy profile $\policy[][][*] \in \fpolicies (\policy[][][*])$ is an GNE if and only if $\gexploit (\policy[][][*]) =0$.
\end{restatable}

This lemma tells us that we can reformulate the problem of computing a \MPGNE{} as the quasi-minimization problem of minimizing state exploitability, i.e., $\min_{\policy \in \fmarkovpolicies (\policy)} \sexploit (\state, \policy)$, at all states $\state \in \states$ simultaneously.
The same is true of computing a GNE and exploitability.

This straightforward reformulation of \MPGNE{} (resp.\@ GNE) in terms of state exploitability (resp.\@ exploitability) does not immediately lend itself to computation, as exploitability minimization is non-trivial, because
exploitability is neither convex nor differentiable in general.
Following \citet{goktas2022exploit},
we can reformulate these problems yet again, this time as coupled quasi-min-max optimization problems \cite{wald1945statistical}.
We proceed to do so now; however, we restrict our attention to exploitability, and hence GNE, knowing that we will later show that minimizing exploitability suffices to minimize state exploitability, and thereby find \MPGNE.

\begin{restatable}{observation}{obsminmax}
\label{obs:exploit_min_to_min_max}
Given a Markov pseudo-game $\mgame$,
    \begin{align}
        \min_{\policy \in \fpolicies (\policy)} \gexploit (\policy)
        &= \min_{\policy \in \fpolicies (\policy)} \max_{\policy[][][\prime] \in \fmarkovpolicies (\policy)} \gcumulreg (\policy, \policy[][][\prime])
        \label{eq:min_max_cumul_regret_formulation}
    \enspace .
    \end{align}   
\end{restatable}

While the above observation makes progress towards our goal of reformulating 
exploitability minimization in a tractable manner, the problem remains challenging to solve for two reasons: first, a fixed point computation is required to solve the outer player's minimization problem; second, the inner player's policy space depends on the choice of outer policy.
We overcome these difficulties by choosing suitable policy parameterizations.

\subsubsection{Policy Parameterization}
\label{sec:policy_parameterization}

In a coupled min-max optimization problem, any solution to the inner player's maximization problem is implicitly parameterized by the outer player's decision. 
We restructure the jointly feasible Markov policy class to represent this dependence explicitly.

Define the class of \mydef{dependent policies} $\depolicies \doteq \{ \depolicy: \states \times \actionspace \to \actionspace \mid \forall (\state, \action) \in \states \times \actionspace, \; \depolicy (\state, \action) \in \actions (\state, \action) \}= \bigtimes_{\player \in \players} \{ \depolicy[\player]: \states \times \actionspace[\player] \to \actionspace[-\player] \mid \forall(\state, \action[-\player]) \in \states \times \actionspace[-\player], \; \depolicy[\player] (\state, \action[-\player]) \in \actions[\player] (\state, \action[-\player]) \}$. 
With this definition in hand we arrive at an \emph{uncoupled\/} quasi-min-max optimization problem:

\begin{restatable}{lemma}{lemmauncoupledminmax}
\label{lemma:independent_min_max}
    Given a Markov pseudo-game $\mgame$,
    \begin{align}
        \min_{\policy \in \fpolicies (\policy)} \max_{\policy[][][\prime] \in \fmarkovpolicies (\policy)} \gcumulreg (\policy, \policy[][][\prime])
        =
        \min_{\policy \in \fpolicies (\policy)}
        \max_{\depolicy \in \depolicies} \gcumulreg (\policy, \depolicy (\cdot, \policy (\cdot)))
        \label{eq:uncoupled_min_max}
    \enspace .
    \end{align}
\end{restatable}

It can be expensive to represent the aforementioned dependence in policies explicitly.
This situation can be naturally rectified, however, by a suitable policy parameterization. 
A suitable policy parameterization can also allow us to represent the set of fixed points s.t. $\policy \in \fmarkovpolicies (\policy)$ more efficiently in practice \cite{goktas2023generative}. 

Define a \mydef{parameterization scheme} $(\policy, \depolicy, \params, \deparams)$ as comprising a unconstrained parameter space $\params$ and parametric policy profile function $\policy: \states \times \params \to \actionspace$ for the outer player, and an unconstrained parameter space $\deparams$ and parametric policy profile function $\depolicy: \states\times \actionspace \times \deparams \to \actionspace$ for the inner player.
Given such a scheme, we restrict the players' policies to be parameterized: i.e., the outer player's space of policies $\policies[][\params] = \{ 
\policy: \states \times \params \to \actionspace \mid \param \in \params \} \subseteq \markovpolicies$, while the inner player's space of policies $\depolicies[][\deparams] = \{ \depolicy: \states\times \actionspace \times \deparams \to \actionspace
\mid \deparam \in \deparams \}$.
Using these parameterizations, we redefine $\vfunc[][\param] \doteq \vfunc[][{\policy (\cdot; \param)}]$, $\qfunc[][\param] \doteq \qfunc[][{\policy (\cdot; \param)}]$, $\payoff (\param) = \payoff ({\policy (\cdot; \param)})$, and $\histdistrib[\initstates][\param] = \histdistrib[\initstates][{\policy (\cdot; \param)}]$; and
$\vfunc[][\deparam(\param)] \doteq \vfunc[][{\depolicy (\cdot, \policy (\cdot; \param); \deparam)}]$; 
$\qfunc[][\deparam(\param)] \doteq \qfunc[][{\depolicy (\cdot, \policy (\cdot; \param); \deparam)}]$; $\payoff (\deparam(\param)) = \payoff ({\depolicy (\cdot, \policy (\cdot; \param); \deparam)})$; $\histdistrib[\initstates][\deparam(\param)] = \histdistrib[\initstates][{\depolicy (\cdot, \policy (\cdot; \param); \deparam)}]$; and so on.


\begin{assumption}[Parameterization for Min-Max Optimization] 
\label{assum:param_min_max}
Given a Markov pseudo-game $\mgame$ and a parameterization scheme $(\policy, \depolicy, \params, \deparams)$, assume 
1.~for all $\param \in \params$, 
$\policy (\state; \param) \in \actions (\state, \policy (\state; \param))$, for all $\state \in \states$; and
2.~for all $\deparam \in \deparams$, 
$\depolicy (\state, \action; \deparam) \in \actions (\state, \action)$, for all $(\state, \action) \in \states\times \actionspace$. 
\end{assumption}


Assuming a policy parameterization scheme that satisfies \Cref{assum:param_min_max}, we restate our goal, state exploitability minimization, one last time as the following min-max optimization problem:
\begin{align}
\label{eq:new_min_max_opt}
    \min_{\param \in \params}
        \max_{\deparam \in \deparams} 
        \gcumulreg (\param, \deparam)
        \doteq \gcumulreg (\policy (\cdot; \param), \depolicy (\cdot, \policy (\cdot; \param); \deparam)) \enspace .
\end{align}

In summary, by choosing a suitable parameterization scheme, namely one that satisfies \Cref{assum:param_min_max}, we resolve the two challenges highlighted in \Cref{obs:exploit_min_to_min_max}.
First, the unconstrained parameter space facilitates an efficient representation of the outer player's policy space, i.e., the set of fixed points $\{\policy \in \markovpolicies \mid \policy \in \fmarkovpolicies(\policy)\}$. 
Second, the parameterization provides an explicit representation of the class of dependent policies, thereby eliminating the dependence of the inner player's policy space on the outer player's policy.


Now, given an unconstrained parameter space, we are able to simplify our definition of \mydef{$(\varepsilon, \delta)$-stationary point of exploitability}, namely, a policy parameter $\param[][][*]\in \params$ for which there exists a $\delta$-close policy parameter $\param[][][\dagger]\in \params$ with $\|\param[][][*]-\param[][][\dagger]\|\leq \delta$ s.t. $\min_{\h \in \subdiff \exploit (\param[][][\dagger])} \|\h\| \leq \varepsilon$.

\subsubsection{State Exploitability Minimization}
\label{sec:mismatch}

Returning to our objective, namely \emph{state\/} exploitability minimization, we now turn our attention to obtaining a tractable characterization of this goal.
Specifically, we argue that it suffices to minimize exploitability, rather than state exploitability, as any policy profile that is a stationary point of exploitability is also a stationary point of state exploitability across all states simultaneously, under suitable assumptions.

Our first lemma states that a stationary point of exploitability is almost surely also a stationary point of state exploitability at all states.
Moreover, if the initial state distribution has full support, then any $(\varepsilon, \delta)$-stationary point of exploitability can be converted into an $(\nicefrac{\varepsilon}{\alpha}, \delta)$-stationary point of \emph{state\/} exploitability, with probability at least $1-\alpha$.



\begin{restatable}{lemma}{lemmafullsupport}
\label{lem:full_support}
Given a Markov pseudo-game $\mgame$ and a parameterization scheme $(\policy, \depolicy, \params, \deparams)$ with $\sexploit(\state, \cdot)$ 
differentiable at $\param \in \params$, for all $\state \in \states$.
If $\| \grad[\param] \gexploit (\param) \| \, = 0$, then $\| \grad[\param] \sexploit (\state, \param) \| \, = 0$  $\initstates$-almost surely, for all states $\state \in \states$, i.e., $\initstates (\{ \state \in \states \mid \|\grad[\param] \sexploit (\state, \param) \| \, = 0 \}) = 1$.
Moreover, for any $\varepsilon > 0$ and $\delta \in [0, 1]$, if $\supp (\initstates) = \states$ and $\| \grad[\param] \gexploit (\param) \| \, \leq \varepsilon$, then with probability at least $1-\delta$, $\| \grad[\param] \sexploit (\state, \param) \| \, \leq \nicefrac{\varepsilon}{\delta}$, for all states $\state \in \states$. 
\end{restatable}

In fact, we can strengthen this probabilistic equivalence to a deterministic one by restricting our attention to Markov pseudo-games with bounded best-response mismatch coefficients.
Our best-response mismatch coefficient generalizes the minimax mismatch coefficient in two-player settings \cite{daskalakis2020independent} and the distribution mismatch coefficient in single-agent settings \cite{agarwal2020optimality}.

Given $\mgame$ with initial state distribution $\initstates$ and alternative state distribution $\diststates \in \simplex (\states)$, 
and letting $\brmap[\player] (\policy[-\player]) \doteq \argmax_{\policy[\player][][\prime] \in \fmarkovpolicies[\player](\policy[-\player])} \payoff[\player] (\policy[\player][][\prime], \policy[-\player])$ denote the set of best response policies for player $\player$ when the other players play policy profile $\policy[-\player]$,
we define the \mydef{best-response mismatch coefficient} for policy profile $\policy$ as $\brmismatch (\policy, \initstates, \diststates) \doteq \max_{\player\in \players} \max_{\policyp[\player] \in \brmap[\player](\policy[-\player])} \left( \nicefrac{1}{1-\discount} \right)^2 \Vert \nicefrac{\statedist[\diststates][{(\policy[\player][][\prime], \policy[-\player])}]}{\initstates} \Vert_{\infty} \Vert \nicefrac{\statedist[\diststates][\policy]}{\initstates} \Vert_\infty$.

\begin{restatable}{lemma}{lemmabrmismatch}
\label{lemma:br_mismatch_coef}
Let $\mgame$ be a Markov pseudo-game with initial state distribution $\initstates$. 
Given policy parameter $\param \in \params$, for any state distribution $\diststates \in \simplex (\states)$, if both $\sexploit (\diststates, \cdot)$ and $\gexploit (\cdot)$ 
are differentiable at $\param$, then $\| \grad \sexploit (\diststates, \param) \| \leq \brmismatch (\policy (\cdot; \param[][][]), \initstates, \diststates) \| \grad \gexploit(\param) \|$.
In particular, for any $\state \in \states$, if $\diststates_{\state}$ is the Dirac distribution on $\states$ centered at $\state$, then $\| \grad \sexploit (\state, \param) \| \leq \brmismatch (\policy (\cdot; \param[][][]), \initstates, \diststates_{\state})
\| \grad \gexploit(\param) \|$.
\label{lem:arb_dist}
\end{restatable}

Once again, \Cref{lem:full_support} states that any stationary point of
exploitability is also a stationary point of state exploitability with high probability, while \Cref{lemma:br_mismatch_coef} upper bounds 
state exploitability in terms of exploitability, when the best-response mismatch coefficient is bounded.
In other words, a policy profile that minimizes exploitability likewise minimizes state exploitability at all states simultaneously, and thus satisfies the necessary conditions of a GMPE.

\subsubsection{Algorithmic Assumptions}

We are nearly ready to describe our reinforcement learning algorithm for computing a stationary point of \Cref{eq:new_min_max_opt}, and thereby finding a policy profile that satisfies the necessary conditions of a \MPGNE.
As \Cref{eq:new_min_max_opt} is a two-player zero-sum game, our method is a variant of simultaneous gradient descent ascent (GDA) \cite{arrow1958studies}, meaning it adjusts its parameters based on first-order information until it reaches a (first-order) stationary point.
Polynomial-time convergence of GDA typically requires that the objective be Lipschitz smooth in both decision variables, and gradient dominated in the inner one, which in our application, translates to the cumulative regret $\gcumulreg (\param, \deparam)$ being Lipschitz smooth in $(\param, \deparam)$ and gradient dominated in $\deparam$. 
These conditions are ensured, under the following assumptions on the Markov pseudo-game.


\begin{assumption}[Lipschitz Smooth Payoffs]
\label{assum:param_lipschitz}
Given a Markov pseudo-game $\mgame$ and a parameterization scheme $(\policy, \depolicy, \params, \deparams)$, assume
1.~$\params$ and $\deparams$ are non-empty, compact, and convex,
2.~$\param \mapsto \policy (\state; \param)$ is twice continuously differentiable, for all $\state \in \states$, and $\deparam \mapsto \depolicy (\state, \action; \deparam)$ is twice continuously differentiable, for all $(\state, \action) \in \states \times \actionspace$;
3.~$\action \mapsto \reward (\state, \action)$ is twice continuously differentiable, for all $\state \in \states$;
4.~$\action \mapsto \trans
(\state[][][\prime] \mid \state, 
\action)$ 
is twice continuously differentiable, for all $\state, \state[][][\prime] \in \states$.
\end{assumption}

\begin{assumption}[Gradient Dominance Conditions]
\label{assum:param_gradient_dominance}
Given a Markov pseudo-game $\mgame$ together with a parameterization scheme $(\policy, \depolicy, \params, \deparams)$, assume
1.~(Closure under policy improvement) 
For each $\param\in \params$, there exists $\deparam\in \deparams$ s.t. $\qfunc[\player][\param](\state, \depolicy[\player](\state, \policy(\state;\param);\deparam), \policy[-\player](\state;\param)) 
= \max_{\policy[\player][][\prime]\in \fpolicies[\player](\policy(\cdot;\param))} 
\qfunc[\player][\param](\state, \policy[\player][][\prime](\state), \policy[-\player](\state;\param)) 
$ for all $\player\in \players$, $\state\in \states$.
2.~(Concavity of action-value) $\deparam \mapsto \qfunc[\player][{\param[][][\prime]}] (\state, 
    \depolicy[\player] (\state, \policy[-\player] (\state; \param); \deparam), 
    \policy[-\player] (\state; \param) 
    )$ is concave, for all $\state \in \states$ and $\param, \param[][][\prime] \in \params$. 
\end{assumption}

\subsubsection{Algorithm and Convergence}

Finally, we present our algorithm for finding an approximate stationary point of exploitability, and thus state exploitability.
The algorithm we use is two time-scale stochastic simultaneous gradient descent-ascent (TTSSGDA), first analyzed by
\citet{lin2020gradient, daskalakis2020independent}, for which we prove best-iterate convergence to an $(\varepsilon, O(\varepsilon))$-stationary point of exploitability after taking $\poly(\nicefrac{1}{\varepsilon})$ gradient steps 
under Assumptions \ref{assum:param_lipschitz} and \ref{assum:param_gradient_dominance}.

\begin{wrapfigure}{L}{0.55\textwidth}
    \begin{minipage}{0.55\textwidth}
    \begin{algorithm}[H]
    \caption{Two time-scale 
    simultaneous 
    SGDA (TTSSGDA)}
    \textbf{Inputs:} $\mgame, (\policy, \depolicy, \params, \deparams), \learnrate[\param][ ], \learnrate[\deparam][ ],  \param[][][(0)], \deparam[][][(0)], \numiters$ \\
    \textbf{Outputs:} $(\param[][][(\numhorizon)], \deparam[][][(\numhorizon)])_{\numhorizon = 0}^\numiters$
    \label{alg:two_time_sgda}
    \begin{algorithmic}[1]
    
    \State Build gradient estimator $\estG$ associated with $\mgame$ 
    
    \For{$\numhorizon = 0, \hdots, \numiters - 1$}
            
            \State 
            $\hist \sim \histdistrib[][\param]$, $\hist[][\prime] \sim \bigtimes_{\player \in \players} \histdistrib[][{(\deparam[\player] (\param[-\player]), \param[-\player])}]$ 
            
            \State  $\param[][][(\numhorizon + 1)] \gets 
            \param[][][(\numhorizon)] - \learnrate[\param][ ]  \estG[\param] (\param[][][(\numhorizon)], \deparam[][][(\numhorizon)]; \hist, \hist[][\prime]) $
        
            \State  $\deparam[][][(\numhorizon + 1)]  \gets  
            \deparam[][][(\numhorizon)] + \learnrate[\deparam][ ] \estG[\deparam] (\param[][][(\numhorizon)], \deparam[][][(\numhorizon)]; \hist, \hist[][\prime]) $
    
            
    \EndFor
    \State \Return $(\param[][][(\numhorizon)], \deparam[][][(\numhorizon])_{\numhorizon = 0}^\numiters$
    \end{algorithmic}
    \end{algorithm}
    \end{minipage}
\end{wrapfigure}

Recall that \Cref{assum:param_lipschitz} guarantees Lipschitz smoothness w.r.t.\@ to both $\param$ and $\deparam$, while \Cref{assum:param_gradient_dominance} guarantees gradient dominance w.r.t\@ $\deparam$.
As the gradient of cumulative regret involves an expectation over histories, we assume that we can simulate trajectories of play $\hist \sim \histdistrib[\initstates][\policy]$ according to the history distribution $\histdistrib[\initstates][\policy]$, for any policy profile $\policy$, and that doing so provides both value and gradient information for the rewards and transition probabilities along simulated trajectories.
That is, we rely on a differentiable game simulator (see, for instance, \citet{suh2022differentiable}), meaning a stochastic first-order oracle that returns the gradients of the rewards and transition probabilities, which we query to estimate deviation payoffs, and ultimately cumulative regrets.


Under this assumption, we estimate these values using realized trajectories from the 
history distribution $\hist[][] \sim \histdistrib[\initstates][\param]$ induced by the outer player's policy, and the
deviation history distribution $\hist[][\deparam] 
\sim \bigtimes_{\player \in \players} \histdistrib[\initstates][{(\deparam[\player] (\param[-\player]), \param[-\player])}]$ induced by the inner player's policy.
More specifically, for all policies $\policy \in \markovpolicies$ and histories $\hist \in \hists[\numhorizons]$,
the \mydef{payoff estimator} 
for player $\player \in \players$ is given by
$\estpayoff[\player] (\policy; \hist) \doteq \sum_{\numhorizon = 0}^{\numhorizons - 1} \initstates (\state[0]) \reward[\player] (\state[t], \policy[][][\prime] (\state[t]))
\prod_{k = 0}^{\numhorizon-1} \discount^k \trans (\state[k  + 1] \mid \state[k], (\state[k]))$.
%
Furthermore, for all $\param \in \params$, $\deparam \in \deparams$, $\hist[][][] \sim \histdistrib[\initstates][\param]$, and $\hist[][\deparam] = (\hist[1][\deparam], \cdots, \hist[\numplayers][\deparam]) \sim \bigtimes_{\player \in \players} \histdistrib[\initstates][{(\deparam[\player] (\param[-\player]), \param[\player])}]$, the \mydef{cumulative regret estimator} is given by
$\estgcumulreg (\param, \deparam; \hist, \hist[][\prime]) \doteq 
    \sum_{\player \in \players}
    \estpayoff[\player] (\depolicy[\player] ({} \cdot{}, \policy[-\player] (\cdot; \param); \deparam),
    \policy[-\player] ({} \cdot{}, \param); \hist[\player][\deparam])
    - \estpayoff[\player] (\policy (\cdot; \param); \hist[][])$, 
while the \mydef{cumulative regret gradient estimator} is given by $\estG (\param, \deparam; \hist, \hist[][\deparam]) \doteq 
(\grad[\param] \estgcumulreg (\param, \deparam; \hist, \hist[][\prime]), \grad[\deparam] \estgcumulreg (\param, \deparam; \hist, \hist[][\deparam]))$.

Our main theorem requires one final definition, namely the \mydef{equilibrium distribution mismatch coefficient} $\|\nicefrac{\partial\statedist[\initstates][{\policy[][][*]}]}{\partial \initstates} \|_\infty$, defined as the Radon-Nikodym derivative of the state-visitation distribution of the GNE $\policy[][][*]$ w.r.t.\@ the initial state distribution $\initstates$.
This coefficient, which measures the inherent difficulty of visiting states under the equilibrium policy $\policy[][][*]$---without knowing $\policy[][][*]$---is closely related to other distribution mismatch coefficients used in the analysis of policy gradient methods \citep{agarwal2020optimality}. 

We now state our main theorem, namely that, under the assumptions outlined above, \Cref{alg:two_time_sgda} computes values for the policy parameters that nearly satisfy the necessary conditions for an MGPNE 
in polynomially many gradient steps, or equivalently, calls to the differentiable simulator.


\begin{restatable}{theorem}{thmconvergence} \label{thm:convergence_GNE} 
Given a Markov pseudo-game $\mgame$ and a parameterization scheme $(\policy, \depolicy, \params, \deparams)$, assume Assumptions~\ref{assum:existence_of_mpgne}, \ref{assum:param_lipschitz}, and \ref{assum:param_gradient_dominance} hold. 
For any $\delta > 0$, set $\varepsilon = \delta \|\brmismatch (\cdot, \initstates, \cdot)\|_\infty^{-1}$.
If \Cref{alg:two_time_sgda} is run with inputs that satisfy
$\learnrate[\param][ ], \learnrate[\deparam][ ] \asymp  \poly(\varepsilon, \| \nicefrac{\partial \statedist[\initstates][{\policy[][][*]}]}{\partial \initstates} \|_\infty, \frac{1}{1-\discount}, \lipschitz[\grad \gcumulreg]^{-1}, \lipschitz[\gcumulreg]^{-1})$, then there exists $\numiters \in \poly \left( \varepsilon^{-1}, (1-\discount)^{-1}, \| \nicefrac{\partial \statedist[\initstates][{\policy[][][*]}]}{\partial \initstates} \|_\infty, \lipschitz[{\grad \gcumulreg}], \lipschitz[\gcumulreg], \diam (\params \times \deparams), \learnrate[\param][ ]^{-1}\right)$ and $k \le T$ s.t.\ $\bestiter[{\param}][\numiters] = \param[][][(k)]$ is an $(\varepsilon, \nicefrac{\varepsilon}{2 \lipschitz[\gcumulreg]})$-stationary point of exploitability, i.e., there exists $\param[][][*]\in \params$ s.t. $\| \bestiter[\param][T] - \param[][][*] \| \leq \nicefrac{\varepsilon}{2 \lipschitz[\gcumulreg]}$ and $\min_{\h \in \subdiff \gexploit(\param[][][*])} \| \h \| \leq \varepsilon$.
Moreover, for any distribution $\diststates \in \simplex(\states)$, if $\sexploit(\diststates, \cdot)$ is differentiable at $\param[][][*]$, then $\| \grad[\param] \gexploit (\diststates, \param[][][*]) \| \leq \delta$, i.e., $\bestiter[{\param}][\numiters]$ is an $(\varepsilon, \delta)$-stationary point of expected state exploitability $\sexploit(\diststates, \cdot)$.
\end{restatable}


In other words, by running \Cref{alg:two_time_sgda} on $\mgame$, we compute a policy profile $\bestiter[{\param}][\numiters]$ in the neighborhood of $\param[][][*]$, an approximate stationary point of exploitability. 
By \Cref{lemma:br_mismatch_coef}, 
$\param[][][*]$ is also an approximate stationary point of state exploitability at all states in $\mgame$ simultaneously, and therefore approximately satisfies the necessary conditions of a \MPGNE{}.
Therefore, \Cref{alg:two_time_sgda}  converges to a point $\bestiter[{\param}][\numiters]$ in the neighborhood of a point $\param[][][*]$ that approximately satisfies the necessary conditions of an \MPGNE{}.
While arguably a relatively weak theoretical conclusion in finite time, in the limit, \Cref{alg:two_time_sgda}  converges to a point that exactly satisfies the necessary conditions of an \MPGNE{}.
Moreover, our experiments (Section~\ref{sec:expts}) demonstrate that in practice our method succeeds at approximating \MPGNE{} in exchange economy Markov pseudo-Games.

\section{Incomplete Markov Economies}
\label{sec:infinite}

Having developed a mathematical formalism for Markov pseudo-games, along with a proof of existence of \MPGNE{} as well as an algorithm that computes them, we now move on to our main agenda, namely modeling incomplete stochastic economies in this formalism.
We establish the first proof, to our knowledge, of the existence of recursive competitive equilibria in standard incomplete stochastic economies, and we provide a polynomial-time algorithm for approximating them.

\subsection{Static Exchange Economies} 

A \mydef{static exchange economy} (or \mydef{market%
\footnote{Although a static exchange ``market'' is an economy, we prefer the term ``market'' for the static components of an infinite horizon Markov exchange economy, a dynamic exchange economy in which each time-period comprises one static market among many.}})
$(\numbuyers, \numcommods, \numtypes, \consumptions, \consendowspace, \typespace, \util, \consendow, \type)$, abbreviated by $(\consendow, \type)$ when clear from context, comprises a finite set of $\numbuyers \in \N_+$ \mydef{consumers} and  $\numcommods \in \N_+$ \mydef{commodities}. 
Each consumer $\buyer \in \buyers$ arrives at the market with an \mydef{endowment} of commodities represented as vector $\consendow[\buyer] = \left(\consendow[\buyer][1], \dots, \consendow[\buyer][\numcommods] \right) \in \consendowspace[\buyer]$, where $\consendowspace[\buyer] \subset \R^\numcommods$ is called the \mydef{endowment space}.%
\footnote{Commodities are assumed to include labor services. 
Further, for any consumer $\buyer$ and endowment $\consendow[\buyer] \in \consendowspace[\buyer]$, $\consendow[\buyer][\commod] \geq 0$ denotes the quantity of commodity $\commod$ in consumer $\buyer$'s possession, while $\consendow[\buyer][\commod] < 0$ denotes consumer $\buyer$'s debt, in terms of commodity $\commod$.}
Any consumer $\buyer$ can sell its endowment $\consendow[\buyer] \in \consendowspace[\buyer]$ at \mydef{prices} $\price \in \simplex[\numcommods]$, where $\price[\commod] \geq 0$ represents the value (resp.\@ cost) of selling (resp.\@ buying) a unit of commodity $\commod \in \commods$, to purchase a consumption $\consumption[\buyer] \in \consumptions[\buyer]$ of commodities in its \mydef{consumption space} $\consumptions[\buyer] \subseteq \R^{\numcommods}$.%
\footnote{We note that, for any labor service $\commod$, consumer $\buyer$'s consumption $\consumption[\buyer][\commod]$ is negative and restricted by its consumption space to be lower bounded by the negative of $\buyer$'s endowment, i.e., $\consumption[\buyer][\commod] \in [- \consendow[\buyer][\commod], 0]$. 
This modeling choice allows us to model a consumer's preferences over the labor services she can provide. 
More generally, the consumption space models the constraints imposed on consumption by the ``physical properties'' of the world. 
That is, it rules out impossible combinations of commodities, such as strictly positive quantities of a commodity that is not available in the region where a consumer resides, or a supply of labor that amounts to more than 24 labor hours in a given day.}
Every consumer is constrained to buy a consumption with a cost weakly less than the value of its endowment, i.e., consumer $\buyer$'s \mydef{budget set}---the set of consumptions $\buyer$ can afford with its endowment $\consendow[\buyer] \in \consendowspace[\buyer]$ at prices $\price \in \simplex[\numcommods]$---is determined by its \mydef{budget correspondence} $\budgetset[\buyer] (\consendow[\buyer], \price) \doteq \{\consumption[\buyer] \in \consumptions[\buyer] \mid  \consumption[\buyer] \cdot \price \leq \consendow[\buyer] \cdot \price \}$.

Each consumer's consumption preferences are determined by its type-dependent preference relation $\succeq_{\buyer, \type[\buyer]}$ on $\consumptions[\buyer]$, represented by a type-dependent \mydef{utility function} $\consumption[\buyer] \mapsto \util[\buyer] (\consumption[\buyer]; \type[\buyer])$, for \mydef{type} $\type[\buyer] \in \typespace[\buyer]$ that characterizes consumer $\buyer$'s preferences within the \mydef{type space} $\typespace[\buyer] \subset \R^{\numtypes}$ of possible preferences.%
\footnote{In the sequel, we will be assuming, for any consumer $\buyer$ with any type $\type[\buyer] \in \typespace[\buyer]$, the type-dependent utility function $\consumption[\buyer] \mapsto \util[\buyer] (\consumption[\buyer]; \type[\buyer])$ is continuous, which implies that it can represent any type-dependent preference relation $\succeq_{\buyer, \type[\buyer]}$ on $\R^\numcommods$ that is complete, transitive, and continuous \cite{debreu1954representation}.}
The goal of each consumer $\buyer$ is thus to buy a consumption $\consumption[\buyer] \in \budgetset[\buyer] (\consendow[\buyer], \price)$ that maximizes its utility function $\consumption[\buyer] \mapsto \util[\buyer] (\consumption[\buyer]; \type[\buyer])$ over its budget set $\budgetset[\buyer] (\consendow[\buyer], \price)$.



We denote any \mydef{endowment profile} (resp.\@ \mydef{type profile} and \mydef{consumption profile}) as $\consendow \doteq \left(\consendow[1], \hdots, \consendow[\numbuyers] \right)^T \in \consendowspace$ (resp.\@ $\type \doteq (\type[1], \hdots, \type[\numbuyers])^T \in \typespace$ and $\consumption[][][][] \doteq (\consumption[1][][][], \hdots, \consumption[\numbuyers][][][])^T \in \consumptions$).
The \mydef{aggregate demand} (resp.\@ \mydef{aggregate supply}) of a \mydef{consumption profile} $\consumption \in \consumptions$ (resp.\@ \mydef{an endowment profile} $\consendow \in \consendowspace$) is defined as the sum of consumptions (resp.\@ endowments) across all consumers, i.e., $\sum_{\buyer \in \buyers} \consumption[\buyer]$ (resp.\@ $\sum_{\buyer \in \buyers} \consendow[\buyer])$.

\subsection{Infinite Horizon Markov Exchange Economies}
\label{sec:dsge}

An \mydef{infinite horizon Markov exchange economy} $\economy \doteq (\numbuyers, \numcommods, \numassets, \numtypes, \states, \consumptions, \portfoliospace, \consendowspace, \typespace,  \util, \discount, \trans, \returnset, \initstates)$, comprises $\numbuyers \in \N$ consumers who, over an infinite discrete time horizon $\numhorizon = 0, 1, 2, \hdots$, repeatedly encounter the opportunity to buy a consumption of $\numcommods \in \N$ commodities and a portfolio of $\numassets \in \N$ assets, with their collective decisions leading them through a \mydef{state space} $\states \doteq \worldstates \times (\consendowspace \times \typespace)$.
This state space comprises a \mydef{world state space} $\worldstates$ and 
a \mydef{spot market space} $\consendowspace \times \typespace$. 
The spot market space is a collection of \mydef{spot markets}, each one a static exchange market $(\consendow, \type) \in \consendowspace \times \typespace \subseteq \R^\numcommods \times \R^\numtypes$. 

Each \mydef{asset} $\asset \in \assets$ is a \mydef{generalized Arrow security}, i.e., a divisible contract that transfers to its owner a quantity 
of the $\commod$th commodity
at any world state $\worldstate \in \worldstates$ determined by a matrix of asset returns $\returns[\worldstate] \doteq \left(\returns[\worldstate][1], \hdots, \returns[\worldstate][\numassets] \right)^T \in \R^{\numassets \times \numcommods}$
s.t.\@ $\returns[\worldstate][\asset][\commod] \in \R$ denotes the quantity of commodity $\commod$ transferred at world state $\worldstate$ for one unit of asset $\asset$.
The collection of asset returns across all world states is given by $\returnset \doteq \{\returns[\worldstate] \}_{\worldstate \in \worldstates}$. 
At any time step $\numhorizon = 0, 1, 2, \hdots$, a consumer $\buyer \in \buyers$ can invest in an \mydef{asset portfolio} $\portfolio[\buyer] \in \portfoliospace[\buyer]$ from a \mydef{space of asset portfolios (or investments)} $\portfoliospace[\buyer] \subset \R^\numassets$ that define the \mydef{asset market}, where $\portfolio[\buyer][\asset] \geq 0$ denotes the units of asset $\asset$ bought (long) by consumer $\buyer$, while $\portfolio[\buyer][\asset] < 0$ denotes units that are sold (short).  
Assets are assumed to be \mydef{short-lived} \cite{magill1994infinite}, meaning that any asset purchased at time $\numhorizon$ pays its dividends in the subsequent time period $\numhorizon + 1$, and then expires.%
\footnote{While for ease of exposition we assume that assets are short-lived, our results generalize to infinitely-lived generalized Arrow securities \cite{huang2004implementing} (i.e., securities that never expire, so yield returns and can be resold in every subsequent time period following their purchase) with appropriate modifications to the definitions of the budget constraints and Walras' law. 
In contrast, our results do \emph{not\/} immediately generalize to $k$-\mydef{period-living generalized Arrow securities} (i.e., securities that yield returns and can be resold in the $k$ subsequent time periods following their purchase, until their expiration), as such securities introduce non-stationarities into the economy. 
To accommodate such securities would require that we generalize our Markov game model and methods to accommodate policies that depend on histories of length $k$.}
 %
Assets allow consumers to insure themselves against future realizations of 
the spot market (i.e., types and endowments), by allowing it to transfer wealth across world states.


The economy starts at time period $\numhorizon = 0$ in an \mydef{initial state} $\staterv[0] \sim \initstates$ determined by an initial state distribution $\initstates \in \simplex(\states)$.
%
%
At each time step $\numhorizon = 0, 1, 2, \hdots$, the state of the economy is $\state[\numhorizon] \doteq (\worldstate[\numhorizon], \consendow[][][\numhorizon], \type[][][\numhorizon]) \in \states$. 
Each consumer $\buyer \in \buyers$, observes the world state $\worldstate[\numhorizon] \in \worldstates$, and participates in a spot market $(\consendow[][][\numhorizon], \type[][][\numhorizon])$, where it purchases \mydef{a consumption} $\consumption[\buyer][][\numhorizon] \in \consumptions[\buyer]$ at \mydef{commodity prices} $\price[][\numhorizon] \in \simplex[\numgoods]$, and an \mydef{asset market} where it invests in an \mydef{asset portfolio} $\portfolio[\buyer][][\numhorizon] \in \portfoliospace[\buyer]$ at \mydef{assets prices} $\assetprice[][\numhorizon] \in \R^\numassets$.\longversion{%
\footnote{In general, asset prices can be negative. 
This modeling assumption is in line with the real world: e.g., it is common for energy futures to see negative prices because of costs associated with overproduction and limited storage capacity \cite{sheppard2020us}.}} 
Every consumer is constrained to buy a consumption $\consumption[\buyer][][\numhorizon] \in \consumptions[\buyer]$ and invest in an asset portfolio $\portfolio[\buyer][][\numhorizon] \in \portfoliospace[\buyer]$ with a total cost weakly less than the value of its current endowment $\consendow[\buyer][][\numhorizon] \in \consendowspace[\buyer]$.
Formally, the set of consumptions and investment portfolios that a consumer $\buyer$ can afford with its current endowment $\consendow[\buyer][][\numhorizon] \in \consendowspace[\buyer]$ at current commodity prices $\price[][\numhorizon] \in \simplex[\numcommods]$ and current asset prices $\assetprice[][\numhorizon] \in \R^\numassets$, i.e., its \mydef{budget set} $\budgetset[\buyer] (\consendow[\buyer][][\numhorizon], \price[][\numhorizon], \assetprice[][\numhorizon])$, is determined by its \mydef{budget correspondence} $\budgetset[\buyer] (\consendow[\buyer], \price, \assetprice) \doteq \{(\consumption[\buyer], \portfolio[\buyer]) \in \consumptions[\buyer] \times \portfoliospace[\buyer] \mid  \consumption[\buyer] \cdot \price + \portfolio[\buyer] \cdot \assetprice  \leq \consendow[\buyer] \cdot \price \}$. 

After the consumers make their consumption and investment decisions, they each receive \mydef{reward} $\util[\buyer] (\consumption[\buyer][][t]; \type[\buyer][][t])$ as a function of their consumption and type, and then
the economy either collapses with probability $1 - \discount$, or survives with probability $\discount$, where $\discount \in (0, 1)$ is called the \mydef{discount rate}.
\footnote{While for ease of exposition we assume a single discount factor for all consumers, our results extend to a setting in which each consumer $\buyer \in \buyers$ has a potentially unique discount factor $\discount_\buyer \in (0, 1)$ by incorporating the discount rates into the consumers' payoffs in the Markov pseudo-game defined in \Cref{sec_app:stochastc_exchange_economy}, rather than the history distribution.} 
If the economy survives to see another day, then a new state is realized, namely 
$(\worldstaterv[][][\prime], \consendowrv[][][][\prime], \typerv[][][][\prime]) \sim \trans(\cdot \mid \state[\numhorizon], \portfolio[][][\numhorizon])$, according to a \mydef{transition probability function} 
$\trans: \states \times \states \times \portfoliospace \to [0, 1]$ that depends on the consumers' 
investment portfolio profile $\portfolio[][][\numhorizon] \doteq (\portfolio[1][][\numhorizon], \hdots, \portfolio[\numbuyers][][\numhorizon])^T \in \portfoliospace$, after which the economy transitions to a new state $\staterv[\numhorizon + 1] \doteq (\worldstaterv[][][\prime], \consendowrv[][][][\prime] + \portfolio[][][\numhorizon] \returns[{\worldstaterv[][][\prime]}], \typerv[][][][\prime])$, where the consumers' new endowments depends on their returns $\portfolio[][][\numhorizon] \returns[{\worldstaterv[][][\prime]}] \in \R^{\numbuyers \times \numcommods}$ on their investments.



\begin{remark}
If only one commodity is delivered in exchange for assets,
i.e., for all world states $\worldstate \in \worldstates$, $\rank(\returns[\worldstate]) \leq 1$, then the generalized Arrow securities are \mydef{num\'eraire generalized Arrow securities}, and the assets are called \mydef{financial assets}.%
\footnote{Recall that the num\'eraire is a fixed commodity that is used to standardize the value of other commodities, while a num\'eraire generalized Arrow security is a generalized Arrow security that delivers its returns in terms of the num\'erarire. 
If the assets deliver exactly one commodity, i.e., $\rank (\returns[\worldstate]) = 1$ at all world states $\worldstate$, we take that commodity to be the num\'eraire for the corresponding spot markets.
On the other hand, if the assets deliver no commodity, i.e., $\rank (\returns[\worldstate]) = 0$ at world state $\worldstate$, then we can take any arbitrary commodity to be the num\'eraire, in which case, the assets vacuously ``deliver'' zero units of the num\'erarire, and no units of any other commodities either.}
An infinite horizon Markov exchange economy is \mydef{world-state-contingent} iff the cardinality of the world state space is weakly greater than that of the spot market space, i.e., $|\worldstates| \, \geq |\consendowspace \times \typespace|$. 
Intuitively, when this condition holds,
there exists a surjection from world states to spot market states, which implies that spot market states are implicit in
world states, so that the spot market states can be dropped from the state space, i.e., $\states \doteq \worldstates$.
An infinite horizon Markov exchange economy has \mydef{complete asset markets} if it is world-state-contingent, and assets can deliver some commodity at all world states, i.e., for all world states $\worldstate \in \worldstates$, $\rank(\returns[\worldstate]) \geq 1$.
Otherwise, it has \mydef{incomplete asset markets}.
Colloquially, we call an infinite horizon exchange economy with (in)complete asset markets an \mydef{(in)complete exchange economy}. 
Intuitively, in complete exchange economies, consumers can insure themselves against all future realizations of the spot market---uncertainty regarding their endowments and types---since a complete exchange economy is world-state contingent.
Further, when there is only a single commodity, s.t. $\numcommods = 1$, and only one financial asset which is a risk-free bond s.t. $\numassets = 1$, and the return matrix for all world states $\worldstate \in \worldstates$ (now a scalar since there is only one commodity and one financial asset) is given by $\returns[\worldstate][ ][ ] \doteq \alpha$, for some $\alpha \in \R$, we obtain the standard incomplete market model \amy{INCREDIBLE!!! NO WONDER YOU ARE PROUD!} \cite{blackwell1965discounted, lucas1971investment}. 
\end{remark}

A \mydef{history} $\hist[][][] \in \hists[\numhorizons] \doteq (\states \times \consumptions \times \portfoliospace \times \simplex[\numcommods]\times \R^\numassets)^\numhorizons \times \states$ 
is a sequence  $\hist[][][] = ((\state[\numhorizon], \consumption[][][\numhorizon], \portfolio[][][\numhorizon], \price[][\numhorizon], \assetprice[][\numhorizon])_{\numhorizon = 0}^{\numhorizons-1}, \state[\numhorizons])$  of tuples comprising states, consumption profiles, investment profiles, commodity price, and asset prices s.t.\@ a history of length $0$ corresponds only to the initial state of the economy. 
For any history $\hist[][][] \in \hists[\numhorizons]$, 
we denote by $\hist[:p]$ the first $p \in [\numhorizons^{*}]$ steps of $\hist$, i.e., $\hist[:p]\doteq ((\state[\numhorizon], \consumption[][][\numhorizon], \portfolio[][][\numhorizon], \price[][\numhorizon], \assetprice[][\numhorizon])_{\numhorizon = 0}^{p-1}, \state[p])$.
Overloading notation, we define the \mydef{history space} $\hists \doteq \bigcup_{\numhorizons = 0}^\infty \hists[\numhorizons]$, and
then \mydef{consumption}, \mydef{investment}, \mydef{commodity price} and \mydef{asset price policies} as mappings  $\consumption[\buyer][][][]: \hists \to \consumptions[\buyer]$, $\portfolio[\buyer][][]: \hists \to \portfoliospace[\buyer]$, $\price: \hists \to \simplex[\numgoods]$, and $\assetprice: \hists \to \R^\numassets$ from histories to consumptions, investments, commodity prices, and asset prices, respectively, s.t.\@ $(\consumption[\buyer][][][], \portfolio[\buyer][][])(\hist)$ is the consumption-investment decision of consumer $\buyer \in \buyers$, and $(\price, \assetprice)(\hist)$ are commodity and asset prices, both at history $\hist \in \hists$. 
A \mydef{consumption policy profile} (resp.\@ \mydef{investment policy profile}) $\consumption(\hist) \doteq (\consumption[1], \hdots, \consumption[\numbuyers])(\hist)^T$ (resp.\@ $\portfolio(\hist) \doteq (\portfolio[1], \hdots, \portfolio[\numbuyers])(\hist)^T$) is a collection of consumption (resp.\@ investment) policies for all consumers.
A consumption policy $\consumption[\buyer]: \states \to \consumptions[\buyer]$ is  \mydef{Markov} if it depends only on the last state of the history, i.e., $\consumption[\buyer][][][] (\hist) = \consumption[\buyer][][][] (\state[\numhorizons])$, for all histories $\hist \in \hists[\numhorizons]$ of all lengths $\numhorizons \in \N$.
An analogous definition extends to investment, commodity price, and asset price policies.


Given $\policy \doteq (\consumption, \portfolio, \price, \assetprice)$ 
and a history $\hist[][][] \in \hists[\numhorizons]$, we define the \mydef{discounted history distribution} assuming initial state distribution $\initstates$ as
%
    $
    \histdistrib[\initstates][\policy][\numhorizons] (\hist[][][]) 
    = \initstates (\state[0]) \prod_{\numhorizon = 0}^{\numhorizons-1} \discount^\numhorizon \trans (\worldstate[\numhorizon +1], \consendow[][][\numhorizon + 1] + \portfolio[][][\numhorizon] \returns[{\worldstate[\numhorizon + 1][][]}], \type[][][\numhorizon + 1] \mid \state[\numhorizon], \portfolio[][][\numhorizon]) \setindic[{\{\portfolio[][][](\hist[:\numhorizon]) \}}]  (\portfolio[][][\numhorizon])
    $.
Overloading notation, we define the set of all realizable trajectories $\hists[\policy][\numhorizons]$ of length $\numhorizons$ under policy profile $\policy$ as $\hists[\policy][\numhorizons] \doteq \supp (\histdistrib[\initstates][\policy][\numhorizons])$, i.e., the set of all histories that occur with non-zero probability, 
and we let $\histrv[][] = \left((\staterv[\numhorizon], \actionrv[][][\numhorizon])_{\numhorizon = 0}^{\numhorizons - 1},
\staterv[\numhorizons]  \right)$ be any randomly sampled history from $\histdistrib[\initstates][\policy][\numhorizons]$.
Finally, we abbreviate $\histdistrib[\initstates][\policy] \doteq \histdistrib[\initstates][\policy][\infty]$.


\subsection{Solution Concepts and Existence}
\label{sec:inf_eqa}

An \mydef{outcome} $(\consumption[][][][], \portfolio[][][][], \price, \assetprice): \hists \to \consumptions \times \portfoliospace \times \simplex[\numgoods] \times \R^\numassets$ of an infinite horizon Markov exchange economy is a tuple%
\footnote{Instead of expressing this tuple as $\consumptions^{\hists} \times \portfoliospace^{\hists} \times \simplex[\numgoods]^{\hists} \times {\R^\numassets}^{\hists}$, we sometimes write $(\consumption[][][][], \portfolio[][][][], \price, \assetprice): \hists \to \consumptions \times \portfoliospace \times \simplex[\numgoods] \times \R^\numassets$.}
consisting of a commodity prices policy, an asset prices policy, a consumption policy profile, and an investment policy profile. 
An outcome is \mydef{Markov} if all its constituent policies are Markov: i.e., if it depends only on the last state of the history, i.e., $(\consumption[][][][], \portfolio[][][][], \price, \assetprice)(\hist) = (\consumption[][][][], \portfolio[][][][], \price, \assetprice)(\state[\numhorizons])$, for all histories $\hist \in \hists[\numhorizons]$ of all lengths $\numhorizons \in \N$. 


We now introduce a number of properties of infinite horizon Markov exchange economies outcomes, which we use to define our solution concepts. 
While these properties are defined broadly for (in general, history-dependent) outcomes, they also apply {in the special case of Markov outcomes. 



Given a consumption and investment profile $(\allocation, \portfolio)$, 
the \mydef{consumption state-value function} $\vfunc[\buyer][{(\allocation, \portfolio, \price, \assetprice)}]: \states \to \R$ for consumer $\buyer$ is defined as:

\begin{align}
\vfunc[\buyer][{(\allocation, \portfolio, \price, \assetprice)}] (\state) \doteq \Ex_{\histrv \sim \histdistrib[\state][{( \allocation, \portfolio, \price, \assetprice)}]} \left[ \sum_{\numhorizon = 0}^\infty \discount^\numhorizon \util[\buyer] \left( \allocation[\buyer] (\histrv[:\numhorizon][][]); \typerv[][][\numhorizon] \right) \right]
\enspace .
\end{align}

\begin{align}
    \vfunc[\buyer][{(\allocation, \portfolio, \price, \assetprice)}] (\state) \doteq 
    \Ex_{(\worldstaterv[][][\prime], \consendowrv[][][][\prime], \typerv[][][][\prime]) \sim \trans(\cdot \mid \staterv[\numhorizon], \portfolio(\histrv[:\numhorizon][][]))} \left[
    \sum_{\numhorizon = 0}^\infty 
\discount^\numhorizon \util[\buyer] \left( \allocation[\buyer] (\histrv[:\numhorizon][][]); \typerv[][][\numhorizon] \right) \bigg| \right. \nonumber \\
\left.
\staterv[0] = \state, 
\histrv[:\numhorizon]=((\staterv[k], \actionrv[][][k])_{k=0}^{\numhorizon-1}, \staterv[\numhorizon]),
 \actionrv[][][\numhorizon] = (\allocation, \portfolio, \price, \assetprice)(\histrv[:\numhorizon]),
 \staterv[\numhorizon+1]=(\worldstaterv[][][\prime], \consendowrv[][][][\prime] + \portfolio[][][\numhorizon] \returns[{\worldstaterv[][][\prime]}], \typerv[][][][\prime]))
 \right]
\end{align}

%
%
An outcome $(\consumption[][][][*], \portfolio[][][][*], \price[][][*], \assetprice[][][*])$ is \mydef{optimal} for $\buyer$ 
if $\buyer$'s \mydef{expected cumulative utility}   $\cumulutil[\buyer] (\allocation, \portfolio, \price, \assetprice) \doteq \Ex_{\state \sim \initstates} \left[ \vfunc[\buyer][{(\allocation, \portfolio, \price, \assetprice)}] (\state) \right]$ is maximized over all affordable consumption and investment policies, i.e.,
\begin{align}
    (\allocation[\buyer][][][*], \portfolio[\buyer][][][*]) \in \argmax_{\substack{(\allocation[\buyer][][][], \portfolio[\buyer][][]): \hists \to \consumptions[\buyer] \times \portfoliospace[\buyer], \forall \numhorizon \in \N, \hist[][][] \in \hists[\numhorizon] \\(\allocation[\buyer], \portfolio[\buyer])(\hist[:\numhorizon]) \in \budgetset[\buyer] (\consendow[\buyer][][\numhorizon], \price[][][*] (\hist[:\numhorizon]), \assetprice[][][*] (\hist[:\numhorizon]))
    }}\cumulutil[\buyer] (\allocation[\buyer][][][], \allocation[-\buyer][][][*], 
    \portfolio[\buyer][][], \portfolio[-\buyer][][][*], \price[][][*], \assetprice[][][*]) \enspace .
\label{eq:opt_consum}
\end{align}

\noindent
A Markov outcome $(\consumption[][][][*], \portfolio[][][][*], \price[][][*], \assetprice[][][*])$ is \mydef{Markov perfect} for $\buyer$ 
if $\buyer$ maximizes its consumption state-value function over all affordable consumption and investment policies, i.e.,
\begin{align}
    (\allocation[\buyer][][][*], \portfolio[\buyer][][][*]) \in \argmax_{\substack{(\allocation[\buyer], \portfolio[\buyer]): \states \to \consumptions[\buyer] \times \portfoliospace[\buyer]: \forall \state \in \states, \\(\allocation[\buyer], \portfolio[\buyer])(\state) \in \budgetset[\buyer] (\consendow[\buyer], \price[][][*] (\state), \assetprice[][][*] (\state))
    }} \left\{ \vfunc[\buyer][{(\allocation[\buyer][][][], \allocation[-\buyer][][][*], 
    \portfolio[\buyer][][], \portfolio[-\buyer][][][*], \price[][][*], \assetprice[][][*])}] (\state)  \right\} \enspace .
\label{eq:Markov_consum}
\end{align}

A consumption policy $\consumption$ is said to be \mydef{feasible} iff for all time horizons $\numhorizons \in \N$ and histories $\hist \in \hists[\numhorizons]$ of length $\numhorizons$,
$\sum_{\buyer \in \buyers} \consumption[\buyer][][][] (\hist) - \sum_{\buyer \in \buyers} \consendow[\buyer][][\numhorizons] \leq \zeros[\numcommods]$,
where $\consendow[\buyer][][\numhorizons] \in \consendowspace[\buyer]$ is consumer $\buyer$'s endowment at the end of history $\hist$, i.e., at state $\state[\numhorizons]$. 
Similarly, an investment policy is \mydef{feasible} if $\sum_{\buyer \in \buyers} \portfolio[\buyer][][] (\hist) \leq \zeros[\numassets]$.
If all the consumption and investment policies associated with an outcome are feasible, we will colloquially refer to the outcome as \mydef{feasible} as well.

An outcome $(\consumption[][][][], \portfolio[][][][], \price, \assetprice)$ is said to satisfy \mydef{Walras' law} iff for all time horizons $\numhorizons \in \N$ and histories $\hist \in \hists[\numhorizons]$ of length $\numhorizons$,
        $\price(\hist) \cdot \left( \sum_{\buyer \in \buyers} \consumption[\buyer][][][] (\hist) - \sum_{\buyer \in \buyers} \consendow[\buyer][][\numhorizons] \right)  +  \assetprice[][] (\hist) \cdot \left(\sum_{\buyer \in \buyers} \portfolio[\buyer][][] (\hist) \right) 
        = 0$,
where, as above, $\consendow[\buyer][][\numhorizons] \in \consendowspace[\buyer]$ is consumer $\buyer$'s endowment at state $\state[\numhorizons]$. 

The canonical solution concept for stochastic economies is the Radner equilibrium.

\begin{definition}[Radner Equilibrium]
    A \mydef{Radner (or sequential competitive) equilibrium (RE)} \cite{radner1972existence} of an infinite horizon Markov exchange economy $\economy$ is an outcome $(\consumption[][][][*], \portfolio[][][][*], \price[][][*], \assetprice[][][*])$
    that is
    1.~optimal for all consumers, i.e., \Cref{eq:opt_consum} is satisfied, for all consumers $\buyer \in \buyers$; 
    2.~feasible; and 
    3.~satisfies Walras' law.
\end{definition}

As a Radner equilibrium is in general infinite dimensional, we are often interested in a recursive Radner equilibrium which is a \emph{Markov\/} outcome, i.e., one that depends only on the last state of the history rather than the entire history.

\begin{definition}[Recursive Radner Equilibrium]
\label{def:rre}
    A \mydef{recursive Radner (or Walrasian or competitive) equilibrium (RRE)} \cite{mehra1977recursive, prescott1980recursive} of an infinite horizon Markov exchange economy $\economy$ is a Markov outcome $(\consumption[][][][*], \portfolio[][][][*], \price[][][*], \assetprice[][][*])$
    that is 
    1.~Markov perfect for all consumers, i.e., \Cref{eq:Markov_consum} is satisfied, for all consumers $\buyer \in \buyers$; 
    2.~feasible; and 
    3.~satisfies Walras' law.
\end{definition}

The following assumptions are standard in the literature (see, for instance, \citet{geanakoplos1990introduction}).
We prove the existence of a recursive Radner equilibrium under these assumptions.


\begin{assumption}
\label{assum:existence_RRE}
Given an infinite horizon Markov exchange economy $\economy$,
assume for all 
$\buyer \in \buyers$,
\begin{enumerate}
    \item $\consumptions$,  $\portfoliospace$,
    $\consendowspace$, are non-empty, closed, convex, with $\consendowspace$ additionally bounded;
    
    \item $(\type[\buyer], \consumption[\buyer]) \mapsto\util[\buyer] (\consumption[\buyer]; \type[\buyer])$ is continuous and 
    concave, and $(\state, \portfolio[\buyer]) \mapsto\trans(\state[][][\prime] \mid \state, \portfolio[\buyer], \portfolio[-\buyer])$ is continuous and stochastically concave, for all $\state[][][\prime] \in \states$ and $\portfolio[-\buyer] \in \portfoliospace[-\buyer]$;
    
    \item for all $\consendow[\buyer] \in \consendowspace[\buyer]$, the correspondence $$(\price, \assetprice) \rightrightarrows \budgetset[\buyer] (\consendow[\buyer], \price, \assetprice) \cap \left\{ (\consumption[\buyer], \portfolio[\buyer]) \, \Bigg| \sum_{\buyer \in \buyers} \consumption[\buyer] \leq \sum_{\buyer \in \buyers} \consendow[\buyer], \sum_{\buyer \in \buyers} \portfolio[\buyer] \leq \zeros[\numcommods], (\consumption, \portfolio) \in \consumptions \times \portfoliospace \right\}$$ is continuous%
    \footnote{\ssadie{}{One way to ensure this condition holds is to assume that for all $\state=(\worldstate, \consendow, \type)\in \states$, returns from assets are positive, i.e., $\returns[\worldstate] \geq \zeros[\numcommods\numassets]$, and for all consumers $\buyer \in \buyers$, there exists $(\consumption[\buyer], \portfolio[\buyer]) \in \consumptions[\buyer] \times \portfoliospace[\buyer]$, s.t. $\consumption[\buyer] < \consendow[\buyer]$ and $\portfolio[\buyer] < 0$.}}
    and non-empty, convex, and compact, for all $\price\in \simplex[\numgoods]$ and $\assetprice\in \R^\numassets$;%
    \footnote{\ssadie{}{One way to ensure this condition holds is to assume that for all $\state=(\worldstate, \consendow, \type)\in \states$, returns from assets are positive, i.e., $\returns[\worldstate] \geq \zeros[\numcommods\numassets]$, and $\consumptions, \portfoliospace$ are bounded from below.}}
    
    \item (no saturation) there exists an $\consumption[\buyer][][][+] \in \consumptions[\buyer]$ s.t. $\util[\buyer] (\consumption[\buyer][][][+]; \type[\buyer]) > \util[\buyer] (\consumption[\buyer]; \type[\buyer])$, for all $\consumption[\buyer] \in \consumptions[\buyer]$ and $\type[\buyer] \in \typespace[\buyer]$.
\end{enumerate}
\end{assumption}

Next we associate an \mydef{exchange economy Markov pseudo-game} $\mgame$ with a given infinite horizon Markov exchange economy $\economy$.

\begin{definition}[Exchange Economy Markov Pseudo-Game]
Let $\economy$
be an infinite horizon Markov exchange economy. 
The corresponding \mydef{exchange economy Markov pseudo-game} $\mgame = (\numbuyers+1, 
\numcommods+\numassets, 
\states, 
\bigtimes_{\player\in \players}(\consumptions[\player] \times \portfoliospace[\player]) \times (\pricespace \times \assetpricespace), 
\newbudgetset, 
\newreward, 
\newtrans, 
\discount', 
\initstates')$ is defined as
\begin{itemize}
\setlength{\itemindent}{-5mm}
    \item The $\numbuyers+1$ players comprise $\numbuyers$ consumers, players $1, \ldots, \numbuyers$, and one auctioneer, player $\numbuyers+1$.
    
    \item The set of states $\states = \worldstates \times \consendowspace \times \typespace$.
    At each state $\state = (\worldstate, \consendow, \type) \in \states$, 
    \begin{itemize}
    \setlength{\itemindent}{-5mm}
        \item each consumer $\buyer \in \buyers$ chooses an action $\action[\buyer] = (\consumption[\buyer], \portfolio[\buyer]) \in \newbudgetset[\buyer] \left( \state, \action[-\buyer] \right) \subseteq \consumptions[\buyer] \times \portfoliospace[\buyer]$ from a set of feasible actions $\newbudgetset[\buyer] (\state, \action[-\buyer]) = \budgetset[\buyer] (\consendow[\buyer], \action[\numbuyers+1]) \cap \{(\consumption[\buyer], \portfolio[\buyer]) \mid \sum_{\buyer \in \buyers} \consumption[\buyer] \leq \sum_{\buyer \in \buyers} \consendow[\buyer], \sum_{\buyer \in \buyers} \portfolio[\buyer] \leq \zeros[\numcommods], (\consumption, \portfolio) \in \consumptions \times \portfoliospace\}$ and receives reward $\newreward[\buyer] (\state, \action) \doteq \util[\buyer] (\consumption[\buyer]; \type[\buyer])$; and
        
        \item  the auctioneer $\numbuyers+1$ chooses an action $\action[\numbuyers+1] = (\price, \assetprice) \in \newbudgetset[\numbuyers+1] \left( \state, \action[-(\numbuyers+1)] \right) \doteq \pricespace \times \assetpricespace$ 
        where $\pricespace
        \doteq \simplex[\numcommods]$ and  $\assetpricespace \subseteq [0, \max_{\consendow \in \consendowspace} \sum_{\buyer \in \buyers} \sum_{\commod \in \commods} \consendow[\buyer][\commod]]^\numassets$, and receives reward $\newreward[\numbuyers+1] (\state, \action) \doteq \price\cdot \left(\sum_{\buyer\in \buyers}\consumption[\buyer]-\sum_{\buyer\in \buyers}\consendow[\buyer] \right) + \assetprice\cdot \left(\sum_{\buyer\in \buyers}\portfolio[\buyer] \right)$. 
    \end{itemize}
    
    \item The transition probability function 
    is defined as $\newtrans(\state[][][\prime] \mid \state, \action) \doteq \trans(\state[][][\prime] \mid \state, \portfolio)$.

    \item The discount rate $\discount' = \discount$ and the initial state distribution $\initstates' = \initstates$.
\end{itemize}
\end{definition}

Our existence proof reformulates the set of recursive Radner equilibria of any infinite horizon Markov exchange economy as the set of \MPGNE{} of the exchange economy Markov pseudo-game.

\begin{restatable}{theorem}{thmexistRRE}
\label{thm:existence_RRE}
    Consider an infinite horizon Markov exchange economy $\economy$. 
    Under \Cref{assum:existence_RRE}, the set of recursive Radner equilibria of $\economy$ is equal to the set of \MPGNE{} of the associated exchange economy Markov pseudo-game $\mgame$.
\end{restatable}

We can define the exploitability (resp. expected state exploitability) of any infinite horizon Markov exchange economy $\economy$ satisfying \Cref{assum:existence_RRE} as the exploitability (resp. expected state exploitability) of its associated exchange economy Markov pseudo-game $\mgame$. By \Cref{thm:existence_RRE}, an outcome that minimizes the exploitability of the economy $\economy$ is a Radner equilibrium (RE), while an outcome that minimizes state exploitability at all states simultaneously is a recursive Radner equilibrium (RRE).
The following corollary now follows from Theorem~\ref{thm:existence_of_mpgne}.

\begin{restatable}{corollary}{corexistRRE}
\label{cor:existence_RRE}
    Under \Cref{assum:existence_RRE}, the set of recursive Radner equilibria of an infinite horizon Markov exchange economy is non-empty.
\end{restatable}

\subsection{Equilibrium Computation}
\label{sec:computation}




Since a recursive Radner equilibrium 
is infinite-dimensional when the state space is continuous, its computation is FNP-hard \cite{murty1987some}. 
As such, it is generally believed that the best we can hope to find in polynomial time is a solution that approximately satisfies the necessary conditions of a recursive Radner equilibrium.
Since the set of recursive Radner equilibria of any infinite horizon Markov exchange economy $\economy$ is equal to the set of \MPGNE{} of the associated exchange economy Markov pseudo-game $\mgame$ (\Cref{thm:existence_RRE}), we can apply \Cref{thm:convergence_GNE} to compute a policy profile with the following computational complexity guarantees for \Cref{alg:two_time_sgda}, when run on the exchange economy Markov pseudo-game associated with an infinite horizon Markov exchange economy.

\begin{restatable}{theorem}{thmcomputeSRE}
\label{thm:compute_SRE}
    Given an infinite horizon Markov exchange economy $\economy$ for which Assumption~\ref{assum:existence_RRE} holds, and the associated exchange economy Markov pseudo-game $\mgame$. 
    If $(\policy, \depolicy, \params, \deparams)$ is a parametrization scheme for $\mgame$ such that Assumptions~\ref{assum:param_lipschitz} and \ref{assum:param_gradient_dominance} hold,%
\footnote{\ssadie{}{While for generality and ease of exposition we state Assumptions~\ref{assum:param_lipschitz} and \ref{assum:param_gradient_dominance} for $\mgame$, we note that when $\economy$ satisfies \Cref{assum:existence_RRE}, to ensure that $\mgame$ satisfies \Cref{assum:param_lipschitz} and \ref{assum:param_gradient_dominance}, 
it suffices to assume that the parametric policy functions $(\policy, \depolicy)$ are affine; 
the policy parameter spaces $(\params, \deparams)$ are non-empty, compact, and convex; 
for all players $\player \in \players$ and types $\type[\buyer] \in \typespace[\buyer]$, the utility function $\consumption[\buyer] \mapsto\util[\buyer] (\consumption[\buyer]; \type[\buyer])$ is twice-continuously differentiable;
and for all $\state, \state[][][\prime] \in \states$, the transition function $\portfolio \mapsto\trans(\state[][][\prime] \mid \state, \portfolio)$ is twice-continuously differentiable.}}    
then the convergence results in \Cref{thm:convergence_GNE} hold, meaning \Cref{alg:two_time_sgda} converges to a point in the neighborhood of a point that approximately satisfies the necessary conditions of an \MPGNE{} in $\mgame$, which is likewise a point that approximately satisfies the necessary conditions of an recursive Radner equilibrium of $\economy$.
Moreover, beyond its finite-time guarantees, in the limit, \Cref{alg:two_time_sgda} converges to a point that satisfies these conditions exactly.
\end{restatable}

\section{Experiments}
\label{sec:expts}

Given an infinite horizon Markov exchange economy $\economy$, we associate with it an exchange economy Markov pseudo-game $\mgame$, and we then construct a deep learning network to solve $\mgame$.
To do so, we assume a parametrization scheme $(\policy, \depolicy, \params, \deparams)$, where the parametric policy profiles $(\policy, \depolicy)$ are represented by neural networks with $(\params, \deparams)$ as the corresponding network weights.
Computing an RRE via \Cref{alg:two_time_sgda} can then be seen as the result of training a generative adversarial neural network \cite{goodfellow2014gan}, where $\policy$ (resp.\@ $\depolicy$) is the output of the generator (resp.\@ adversarial) network.
We call such a neural representation a \mydef{generative adversarial policy network (GAPNet)}.

Following this approach, we built GAPNet to approximate the RRE in two types of infinite-horizon Markov exchange economies: one with a deterministic transition probability function and another with a stochastic transition probability function. 
Within each type, we experimented with three randomly sampled economies, each with 10 consumers, 10 commodities, 1 asset, 5 world states, and characterized by a distinct class of reward functions, which impart different smoothness properties onto the state-value function: 
\mydef{linear}: $\util[\buyer](\allocation[\buyer]; \type[\buyer]) = \sum_{\good \in \goods} \type[\buyer][\good] \allocation[\buyer][\good]$; 
\mydef{Cobb-Douglas}:  $\util[\buyer](\allocation[\buyer]; \type[\buyer]) = \prod_{\good \in \goods} {\allocation[\buyer][\good]}^{\type[\buyer][\good]}$; 
and \mydef{Leontief}:  $\util[\buyer](\allocation[\buyer]; \type[\buyer]) = \min_{\good \in \goods} \left\{ \frac{\allocation[\buyer][\good]}{\type[\buyer][\good]}\right\}$. 
\footnote{Full details of our experimental setup appear in \Cref{sec_app:experiments}, including hyperparameter search values, final experimental configurations, and visualization code.
See also our code repository: \texttt{\coderepo}.} 

We compare our results with a classic neural projection method (also known as deep equilibrium nets \cite{azinovic2022deep}), which macroeconomists and others use to solve stochastic economies.
In this latter method, one seeks a policy profile that minimizes the norm of the system of first-order necessary and sufficient conditions that characterize RRE
(see for instance, \cite{FernandezVillaverde2023CompMethodsMacro}).%
\footnote{We describe the neural projection method in \Cref{sec_app:npm}.}
We use the same network architecture for both methods, and select hyperparameters through grid search.
In all experiments, we evaluate the performance of the computed policy profiles using three metrics: total first-order violations, average Bellman errors,%
\footnote{The definitions of these two metrics can be found in \Cref{sec_app:npm}.} 
and exploitability. 

\amy{move to appendix:}
\ssadie{For each metric, we randomly sample 50 policy profiles, record their corresponding values, and normalize the results by dividing by the average.}{}

\begin{figure}
    \begin{subfigure}{\textwidth}
        \centering
        \includegraphics[width=0.75\textwidth]{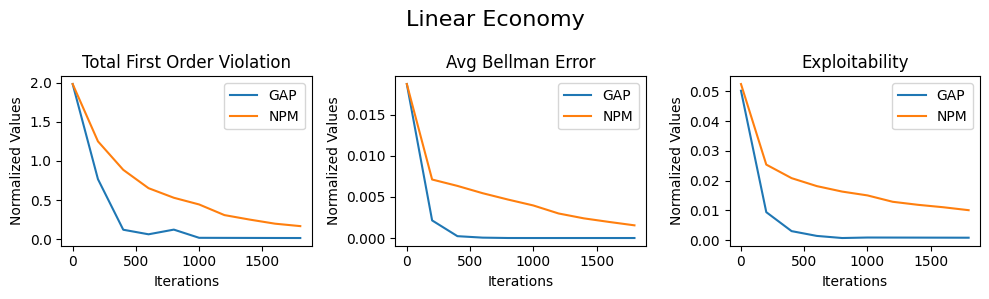}
        \label{fig:linear_stochastic}
    \end{subfigure}
    
    
    \begin{subfigure}{\textwidth}
        \centering
        \includegraphics[width=0.75\textwidth]{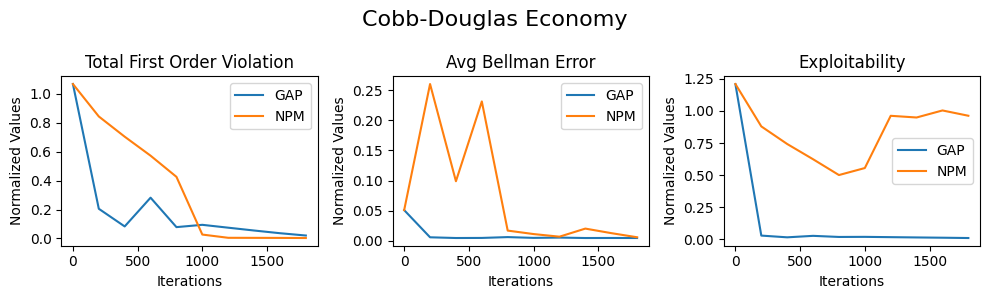}
        \label{fig:cd_stochastic}
    \end{subfigure}
    
    
    \begin{subfigure}{\textwidth}
        \centering
        \includegraphics[width=0.75\textwidth]{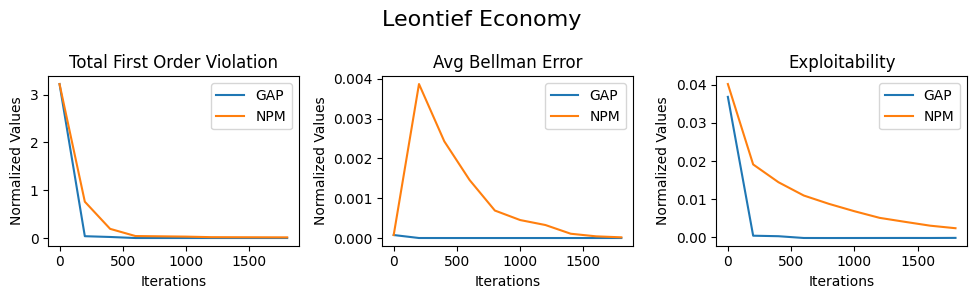}
        \label{fig:leontief_stochastic}
    \end{subfigure}

    \caption{Normalized Metrics for Economies with Stochastic Transition Functions}
    \label{fig:stochastic}
\end{figure}

\Cref{fig:nonstochastic} (\Cref{sec_app:experiments}) depicts our results for economies with deterministic transition functions.
Perhaps unsurprisingly, while GAPNet demonstrates a clear advantage in minimizing exploitability,
the neural projection method (NPM) slightly outperforms GAPNet in minimizing total first-order violations and average Bellman error, the metrics they are specifically designed to minimize.
Furthermore, in all three economies, GAPNet's exploitability is near 0, highlighting its ability to at least approximate the necessary conditions of an RRE.
\Cref{fig:stochastic} presents our results for economies with stochastic transition functions. 
These results indicate that stochasticity hinders NPM's ability to minimize the three metrics, even though the method is explictly designed to minimize two of them.
In contrast, GAPNet successfully minimizes all three metrics across all economies.

\yc{Wow! GAP performed so well.}

\section{Conclusion}
\label{sec:conc}

In this paper, we tackled the problem of computing general equilibrium in dynamic stochastic economies. 
We showed that the computation of a recursive Radner equilibrium in an infinite horizon Markov exchange economy can be reduced to the computation of a generalized Markov perfect equilibrium in an associated Markov pseudo-game.
This reduction allowed us to develop a polynomial-time algorithm to approximate recursive Radner equilibria. 
Perhaps more importantly, our work connects recent developments in deep reinforcement learning to macroeconomics, thereby uncovering myriad potential new research directions.

\bibliographystyle{plainnat}  
\bibliography{references}  

\begin{thebibliography}{124}
\providecommand{\natexlab}[1]{#1}
\providecommand{\url}[1]{\texttt{#1}}
\expandafter\ifx\csname urlstyle\endcsname\relax
  \providecommand{\doi}[1]{doi: #1}\else
  \providecommand{\doi}{doi: \begingroup \urlstyle{rm}\Url}\fi

\bibitem[Acemoglu(2008)]{acemoglu2008introduction}
Daron Acemoglu.
\newblock \emph{Introduction to modern economic growth}.
\newblock Princeton university press, 2008.

\bibitem[Achdou et~al.(2022)Achdou, Han, Lasry, Lions, and Moll]{achdou2022income}
Yves Achdou, Jiequn Han, Jean-Michel Lasry, Pierre-Louis Lions, and Benjamin Moll.
\newblock Income and wealth distribution in macroeconomics: A continuous-time approach.
\newblock \emph{The review of economic studies}, 89\penalty0 (1):\penalty0 45--86, 2022.

\bibitem[Agarwal et~al.(2020)Agarwal, Kakade, Lee, and Mahajan]{agarwal2020optimality}
Alekh Agarwal, Sham~M Kakade, Jason~D Lee, and Gaurav Mahajan.
\newblock Optimality and approximation with policy gradient methods in markov decision processes.
\newblock In \emph{Conference on Learning Theory}, pages 64--66. PMLR, 2020.

\bibitem[Aiyagari(1994)]{aiyagari1994uninsured}
S~Rao Aiyagari.
\newblock Uninsured idiosyncratic risk and aggregate saving.
\newblock \emph{The Quarterly Journal of Economics}, 109\penalty0 (3):\penalty0 659--684, 1994.

\bibitem[Arrow and Debreu(1954)]{arrow-debreu}
Kenneth Arrow and Gerard Debreu.
\newblock Existence of an equilibrium for a competitive economy.
\newblock \emph{Econometrica: Journal of the Econometric Society}, pages 265--290, 1954.

\bibitem[Arrow(1951)]{arrow1951extension}
Kenneth~J Arrow.
\newblock An extension of the basic theorems of classical welfare economics.
\newblock In \emph{Proceedings of the second Berkeley symposium on mathematical statistics and probability}, volume~2, pages 507--533. University of California Press, 1951.

\bibitem[Arrow(1964)]{arrow1964role}
Kenneth~J Arrow.
\newblock Le role des valeurs boursieres pour la repartition la meilleure des risques, econometrie, 41-47, english translation as the role of securities in the optimal allocation of risk-bearing.
\newblock \emph{Review of Economic Studies}, 31:\penalty0 91--96, 1964.

\bibitem[Arrow et~al.(1958)Arrow, Hurwicz, and Uzawa]{arrow1958studies}
Kenneth~Joseph Arrow, Leonid Hurwicz, and Hirofumi Uzawa.
\newblock Studies in linear and non-linear programming.
\newblock 1958.

\bibitem[Atakan(2003{\natexlab{a}})]{atakan2003stochastic}
Alp~E Atakan.
\newblock Stochastic convexity in dynamic programming.
\newblock \emph{Economic Theory}, 22:\penalty0 447--455, 2003{\natexlab{a}}.

\bibitem[Atakan(2003{\natexlab{b}})]{atakan2003valfunc}
Alp~E. Atakan.
\newblock Stochastic convexity in dynamic programming.
\newblock \emph{Economic Theory}, 22\penalty0 (2):\penalty0 447--455, 2003{\natexlab{b}}.
\newblock ISSN 09382259, 14320479.
\newblock URL \url{http://www.jstor.org/stable/25055693}.

\bibitem[Auclert et~al.(2021)Auclert, Bard{\'o}czy, Rognlie, and Straub]{auclert2021using}
Adrien Auclert, Bence Bard{\'o}czy, Matthew Rognlie, and Ludwig Straub.
\newblock Using the sequence-space jacobian to solve and estimate heterogeneous-agent models.
\newblock \emph{Econometrica}, 89\penalty0 (5):\penalty0 2375--2408, 2021.

\bibitem[Azinovic et~al.(2022)Azinovic, Gaegauf, and Scheidegger]{azinovic2022deep}
Marlon Azinovic, Luca Gaegauf, and Simon Scheidegger.
\newblock Deep equilibrium nets.
\newblock \emph{International Economic Review}, 63\penalty0 (4):\penalty0 1471--1525, 2022.

\bibitem[Bei et~al.(2015)Bei, Garg, and Hoefer]{bei2015tatonnement}
Xiaohui Bei, Jugal Garg, and Martin Hoefer.
\newblock Tatonnement for linear and gross substitutes markets.
\newblock \emph{CoRR abs/1507.04925}, 2015.

\bibitem[Bellman(1966)]{bellman1966dynamic}
Richard Bellman.
\newblock Dynamic programming.
\newblock \emph{Science}, 153\penalty0 (3731):\penalty0 34--37, 1966.

\bibitem[Bewley(1983)]{bewley1983difficulty}
Truman Bewley.
\newblock A difficulty with the optimum quantity of money.
\newblock \emph{Econometrica: Journal of the Econometric Society}, pages 1485--1504, 1983.

\bibitem[Bhandari and Russo(2019)]{bhandari2019global}
Jalaj Bhandari and Daniel Russo.
\newblock Global optimality guarantees for policy gradient methods.
\newblock \emph{arXiv preprint arXiv:1906.01786}, 2019.

\bibitem[Black and Scholes(1973)]{black1973pricing}
Fischer Black and Myron Scholes.
\newblock The pricing of options and corporate liabilities.
\newblock \emph{Journal of political economy}, 81\penalty0 (3):\penalty0 637--654, 1973.

\bibitem[Blackwell(1965)]{blackwell1965discounted}
David Blackwell.
\newblock Discounted dynamic programming.
\newblock \emph{The Annals of Mathematical Statistics}, 36\penalty0 (1):\penalty0 226--235, 1965.

\bibitem[Blanchard and Kahn(1980)]{blanchard1980solution}
Olivier~Jean Blanchard and Charles~M Kahn.
\newblock The solution of linear difference models under rational expectations.
\newblock \emph{Econometrica: Journal of the Econometric Society}, pages 1305--1311, 1980.

\bibitem[Blondel et~al.(2021)Blondel, Berthet, Cuturi, Frostig, Hoyer, Llinares-L{\'o}pez, Pedregosa, and Vert]{jaxopt_implicit_diff}
Mathieu Blondel, Quentin Berthet, Marco Cuturi, Roy Frostig, Stephan Hoyer, Felipe Llinares-L{\'o}pez, Fabian Pedregosa, and Jean-Philippe Vert.
\newblock Efficient and modular implicit differentiation.
\newblock \emph{arXiv preprint arXiv:2105.15183}, 2021.

\bibitem[Bradbury et~al.(2018)Bradbury, Frostig, Hawkins, Johnson, Leary, Maclaurin, Necula, Paszke, Vander{P}las, Wanderman-{M}ilne, and Zhang]{jax2018github}
James Bradbury, Roy Frostig, Peter Hawkins, Matthew~James Johnson, Chris Leary, Dougal Maclaurin, George Necula, Adam Paszke, Jake Vander{P}las, Skye Wanderman-{M}ilne, and Qiao Zhang.
\newblock {JAX}: composable transformations of {P}ython+{N}um{P}y programs, 2018.
\newblock URL \url{http://github.com/google/jax}.

\bibitem[Br{\^a}nzei et~al.(2021)Br{\^a}nzei, Devanur, and Rabani]{branzei2021proportional}
Simina Br{\^a}nzei, Nikhil Devanur, and Yuval Rabani.
\newblock Proportional dynamics in exchange economies.
\newblock In \emph{Proceedings of the 22nd ACM Conference on Economics and Computation}, pages 180--201, 2021.

\bibitem[Cass(1984)]{cass1984competitive}
David Cass.
\newblock \emph{Competitive equilibrium with incomplete financial markets}.
\newblock University of Pennsylvania, Center for Analytic Research in Economics and~…, 1984.

\bibitem[Cass(1985)]{cass1985number}
David Cass.
\newblock \emph{On the" number" of equilibrium allocations with incomplete financial markets}.
\newblock University of Pennsylvania, Center for Analytic Research in Economics and~…, 1985.

\bibitem[Chen and Deng(2006)]{chen2006settling}
Xi~Chen and Xiaotie Deng.
\newblock Settling the complexity of two-player nash equilibrium.
\newblock In \emph{2006 47th Annual IEEE Symposium on Foundations of Computer Science (FOCS'06)}, pages 261--272. IEEE, 2006.

\bibitem[Chen et~al.(2009)Chen, Deng, and Teng]{chen2009settling}
Xi~Chen, Xiaotie Deng, and Shang-Hua Teng.
\newblock Settling the complexity of computing two-player nash equilibria.
\newblock \emph{Journal of the ACM (JACM)}, 56\penalty0 (3):\penalty0 1--57, 2009.

\bibitem[Childers et~al.(2022)Childers, Fern{\'a}ndez-Villaverde, Perla, Rackauckas, and Wu]{childers2022differentiable}
David Childers, Jes{\'u}s Fern{\'a}ndez-Villaverde, Jesse Perla, Christopher Rackauckas, and Peifan Wu.
\newblock Differentiable state-space models and hamiltonian monte carlo estimation.
\newblock Technical report, National Bureau of Economic Research, 2022.

\bibitem[Christiano et~al.(2018)Christiano, Eichenbaum, and Trabandt]{christiano2018dsge}
Lawrence~J Christiano, Martin~S Eichenbaum, and Mathias Trabandt.
\newblock On dsge models.
\newblock \emph{Journal of Economic Perspectives}, 32\penalty0 (3):\penalty0 113--140, 2018.

\bibitem[Clarida et~al.(2000)Clarida, Gali, and Gertler]{clarida2000monetary}
Richard Clarida, Jordi Gali, and Mark Gertler.
\newblock Monetary policy rules and macroeconomic stability: evidence and some theory.
\newblock \emph{The Quarterly journal of economics}, 115\penalty0 (1):\penalty0 147--180, 2000.

\bibitem[Cloyne et~al.(2020)Cloyne, Ferreira, and Surico]{cloyne2020monetary}
James Cloyne, Clodomiro Ferreira, and Paolo Surico.
\newblock Monetary policy when households have debt: new evidence on the transmission mechanism.
\newblock \emph{The Review of Economic Studies}, 87\penalty0 (1):\penalty0 102--129, 2020.

\bibitem[Codenotti et~al.(2005)Codenotti, McCune, and Varadarajan]{codenotti2005market}
Bruno Codenotti, Benton McCune, and Kasturi Varadarajan.
\newblock Market equilibrium via the excess demand function.
\newblock In \emph{Proceedings of the thirty-seventh annual ACM symposium on Theory of computing}, pages 74--83, 2005.

\bibitem[Codenotti et~al.(2006)Codenotti, Saberi, Varadarajan, and Ye]{codenotti2006leontief}
Bruno Codenotti, Amin Saberi, Kasturi Varadarajan, and Yinyu Ye.
\newblock Leontief economies encode nonzero sum two-player games.
\newblock In \emph{SODA}, volume~6, pages 659--667, 2006.

\bibitem[Cox and Ross(1976)]{cox1976valuation}
John~C Cox and Stephen~A Ross.
\newblock The valuation of options for alternative stochastic processes.
\newblock \emph{Journal of financial economics}, 3\penalty0 (1-2):\penalty0 145--166, 1976.

\bibitem[Cox et~al.(1979)Cox, Ross, and Rubinstein]{cox1979option}
John~C Cox, Stephen~A Ross, and Mark Rubinstein.
\newblock Option pricing: A simplified approach.
\newblock \emph{Journal of financial Economics}, 7\penalty0 (3):\penalty0 229--263, 1979.

\bibitem[Cox et~al.(1985)Cox, Ingersoll~Jr, and Ross]{cox1985intertemporal}
John~C Cox, Jonathan~E Ingersoll~Jr, and Stephen~A Ross.
\newblock An intertemporal general equilibrium model of asset prices.
\newblock \emph{Econometrica: Journal of the Econometric Society}, pages 363--384, 1985.

\bibitem[Curry et~al.()Curry, Trott, Phade, Bai, and Zheng]{curry2023learning}
Michael Curry, Alexander Trott, Soham Phade, Yu~Bai, and Stephan Zheng.
\newblock Learning solutions in large economic networks using deep multi-agent reinforcement learning.

\bibitem[Daskalakis et~al.(2009)Daskalakis, Goldberg, and Papadimitriou]{daskalakis2009complexity}
Constantinos Daskalakis, Paul~W Goldberg, and Christos~H Papadimitriou.
\newblock The complexity of computing a nash equilibrium.
\newblock \emph{SIAM Journal on Computing}, 39\penalty0 (1):\penalty0 195--259, 2009.

\bibitem[Daskalakis et~al.(2020)Daskalakis, Foster, and Golowich]{daskalakis2020independent}
Constantinos Daskalakis, Dylan~J Foster, and Noah Golowich.
\newblock Independent policy gradient methods for competitive reinforcement learning.
\newblock \emph{Advances in neural information processing systems}, 33:\penalty0 5527--5540, 2020.

\bibitem[Davis et~al.(2018)Davis, Drusvyatskiy, MacPhee, and Paquette]{davis2018subgradient}
Damek Davis, Dmitriy Drusvyatskiy, Kellie~J MacPhee, and Courtney Paquette.
\newblock Subgradient methods for sharp weakly convex functions.
\newblock \emph{Journal of Optimization Theory and Applications}, 179:\penalty0 962--982, 2018.

\bibitem[Debreu(1951)]{debreu1951pareto}
Gerard Debreu.
\newblock The coefficient of resource utilization.
\newblock \emph{Econometrica}, 19\penalty0 (3):\penalty0 273--292, 1951.
\newblock ISSN 00129682, 14680262.
\newblock URL \url{http://www.jstor.org/stable/1906814}.

\bibitem[Debreu et~al.(1954)]{debreu1954representation}
Gerard Debreu et~al.
\newblock Representation of a preference ordering by a numerical function.
\newblock \emph{Decision processes}, 3:\penalty0 159--165, 1954.

\bibitem[Deng and Du(2008)]{deng2008computation}
Xiaotie Deng and Ye~Du.
\newblock The computation of approximate competitive equilibrium is ppad-hard.
\newblock \emph{Information Processing Letters}, 108\penalty0 (6):\penalty0 369--373, 2008.

\bibitem[{Devanur} et~al.(2002){Devanur}, {Papadimitriou}, {Saberi}, and {Vazirani}]{devanur2002market}
N.~R. {Devanur}, C.~H. {Papadimitriou}, A.~{Saberi}, and V.~V. {Vazirani}.
\newblock Market equilibrium via a primal-dual-type algorithm.
\newblock In \emph{The 43rd Annual IEEE Symposium on Foundations of Computer Science, 2002. Proceedings.}, pages 389--395, 2002.
\newblock \doi{10.1109/SFCS.2002.1181963}.

\bibitem[Diamond(1980)]{diamond1980income}
Peter Diamond.
\newblock Income taxation with fixed hours of work.
\newblock \emph{Journal of Public Economics}, 13\penalty0 (1):\penalty0 101--110, 1980.

\bibitem[Diamond(1967)]{diamond1967incompletege}
Peter~A. Diamond.
\newblock The role of a stock market in a general equilibrium model with technological uncertainty.
\newblock \emph{The American Economic Review}, 57\penalty0 (4):\penalty0 759--776, 1967.
\newblock ISSN 00028282.
\newblock URL \url{http://www.jstor.org/stable/1815367}.

\bibitem[Diamond and Boyd(2016)]{diamond2016cvxpy}
Steven Diamond and Stephen Boyd.
\newblock {CVXPY}: {A} {P}ython-embedded modeling language for convex optimization.
\newblock \emph{Journal of Machine Learning Research}, 17\penalty0 (83):\penalty0 1--5, 2016.

\bibitem[Dreze(1974)]{dreze1974investment}
Jacques~H Dreze.
\newblock Investment under private ownership: optimality, equilibrium and stability.
\newblock In \emph{Allocation under Uncertainty: Equilibrium and Optimality: Proceedings from a Workshop sponsored by the International Economic Association}, pages 129--166. Springer, 1974.

\bibitem[Duffie(1987)]{duffie1987stochastic}
Darrell Duffie.
\newblock Stochastic equilibria with incomplete financial markets.
\newblock \emph{Journal of Economic Theory}, 41\penalty0 (2):\penalty0 405--416, 1987.

\bibitem[Duffie and Shafer(1985)]{duffie1985equilibrium}
Darrell Duffie and Wayne Shafer.
\newblock Equilibrium in incomplete markets: I: A basic model of generic existence.
\newblock \emph{Journal of Mathematical Economics}, 14\penalty0 (3):\penalty0 285--300, 1985.

\bibitem[Duffie and Shafer(1986)]{duffie1986equilibrium}
Darrell Duffie and Wayne Shafer.
\newblock Equilibrium in incomplete markets: Ii: Generic existence in stochastic economies.
\newblock \emph{Journal of Mathematical Economics}, 15\penalty0 (3):\penalty0 199--216, 1986.

\bibitem[Durbin and Koopman(2012)]{durbin2012time}
James Durbin and Siem~Jan Koopman.
\newblock \emph{Time series analysis by state space methods}, volume~38.
\newblock OUP Oxford, 2012.

\bibitem[Eskelinen(2021)]{eskelinen2021monetary}
Maria Eskelinen.
\newblock Monetary policy, agent heterogeneity and inequality: Insights from a three-agent new keynesian model.
\newblock 2021.

\bibitem[Facchinei and Kanzow(2010)]{facchinei2010generalized}
Francisco Facchinei and Christian Kanzow.
\newblock Generalized nash equilibrium problems.
\newblock \emph{Annals of Operations Research}, 175\penalty0 (1):\penalty0 177--211, 2010.

\bibitem[Fan(1952)]{Fan1952FixedPoint}
Ky~Fan.
\newblock Fixed-point and minimax theorems in locally convex topological linear spaces.
\newblock \emph{Proceedings of the National Academy of Sciences of the United States of America}, 38\penalty0 (2):\penalty0 121–126, 1952.
\newblock ISSN 0027-8424.

\bibitem[Fernández-Villaverde et~al.(2016)Fernández-Villaverde, Rubio-Ramírez, and Schorfheide]{fernandez1016solve}
J.~Fernández-Villaverde, J.F. Rubio-Ramírez, and F.~Schorfheide.
\newblock Chapter 9 - solution and estimation methods for dsge models.
\newblock volume~2 of \emph{Handbook of Macroeconomics}, pages 527--724. Elsevier, 2016.
\newblock \doi{https://doi.org/10.1016/bs.hesmac.2016.03.006}.
\newblock URL \url{https://www.sciencedirect.com/science/article/pii/S1574004816000070}.

\bibitem[Fernández-Villaverde(2023)]{FernandezVillaverde2023CompMethodsMacro}
Jesús Fernández-Villaverde.
\newblock \emph{Computational Methods for Macroeconomics}.
\newblock University of Pennsylvania, 2023.
\newblock URL \url{https://www.sas.upenn.edu/~jesusfv/teaching.html}.
\newblock Lecture notes for one-year course on computational methods for economists.

\bibitem[Fiacco and Kyparisis(1986)]{fiacco1986convexity}
Anthony~V Fiacco and Jerzy Kyparisis.
\newblock Convexity and concavity properties of the optimal value function in parametric nonlinear programming.
\newblock \emph{Journal of optimization theory and applications}, 48\penalty0 (1):\penalty0 95--126, 1986.

\bibitem[Fink(1964)]{fink1964equilibrium}
Arlington~M Fink.
\newblock Equilibrium in a stochastic $ n $-person game.
\newblock \emph{Journal of science of the hiroshima university, series ai (mathematics)}, 28\penalty0 (1):\penalty0 89--93, 1964.

\bibitem[Flam and Ruszczynski(1994)]{flam1994gne}
Sjur Flam and Andrzej Ruszczynski.
\newblock Noncooperative convex games: Computing equilibrium by partial regularization.
\newblock Working papers, International Institute for Applied Systems Analysis, 1994.
\newblock URL \url{https://EconPapers.repec.org/RePEc:wop:iasawp:wp94042}.

\bibitem[Gao and Kroer(2020)]{gao2020polygm}
Yuan Gao and Christian Kroer.
\newblock First-order methods for large-scale market equilibrium computation.
\newblock In Hugo Larochelle, Marc'Aurelio Ranzato, Raia Hadsell, Maria{-}Florina Balcan, and Hsuan{-}Tien Lin, editors, \emph{Advances in Neural Information Processing Systems 33: Annual Conference on Neural Information Processing Systems 2020, NeurIPS 2020, December 6-12, 2020, virtual}, 2020.
\newblock URL \url{https://proceedings.neurips.cc/paper/2020/hash/f75526659f31040afeb61cb7133e4e6d-Abstract.html}.

\bibitem[Garg and Kapoor(2004)]{garg2004auction}
Rahul Garg and Sanjiv Kapoor.
\newblock Auction algorithms for market equilibrium.
\newblock In \emph{Proceedings of the thirty-sixth annual ACM symposium on Theory of computing}, pages 511--518, 2004.

\bibitem[Geanakoplos(1990)]{geanakoplos1990introduction}
John Geanakoplos.
\newblock An introduction to general equilibrium with incomplete asset markets.
\newblock \emph{Journal of mathematical economics}, 19\penalty0 (1-2):\penalty0 1--38, 1990.

\bibitem[Geanakoplos and Polemarchakis(1986)]{geanakoplos1986walrasian}
John~D Geanakoplos and Herakles~M Polemarchakis.
\newblock Walrasian indeterminacy and keynesian macroeconomics.
\newblock \emph{The Review of Economic Studies}, 53\penalty0 (5):\penalty0 755--779, 1986.

\bibitem[Goktas and Greenwald(2022)]{goktas2022exploit}
Denizalp Goktas and Amy Greenwald.
\newblock Exploitability minimization in games and beyond.
\newblock In \emph{Advances in Neural Information Processing Systems}, 2022.

\bibitem[Goktas et~al.(2023{\natexlab{a}})Goktas, Parkes, Gemp, Marris, Piliouras, Elie, Lever, and Tacchetti]{goktas2023generative}
Denizalp Goktas, David~C Parkes, Ian Gemp, Luke Marris, Georgios Piliouras, Romuald Elie, Guy Lever, and Andrea Tacchetti.
\newblock Generative adversarial equilibrium solvers.
\newblock \emph{arXiv preprint arXiv:2302.06607}, 2023{\natexlab{a}}.

\bibitem[Goktas et~al.(2023{\natexlab{b}})Goktas, Zhao, and Greenwald]{goktas2023tatonn}
Denizalp Goktas, Jiayi Zhao, and Amy Greenwald.
\newblock T\^atonnement in homothetic {F}isher markets.
\newblock \emph{arXiv preprint arXiv:2306.04890}, 2023{\natexlab{b}}.

\bibitem[Goodfellow et~al.(2014)Goodfellow, Pouget-Abadie, Mirza, Xu, Warde-Farley, Ozair, Courville, and Bengio]{goodfellow2014gan}
Ian Goodfellow, Jean Pouget-Abadie, Mehdi Mirza, Bing Xu, David Warde-Farley, Sherjil Ozair, Aaron Courville, and Yoshua Bengio.
\newblock Generative adversarial nets.
\newblock In Z.~Ghahramani, M.~Welling, C.~Cortes, N.~Lawrence, and K.Q. Weinberger, editors, \emph{Advances in Neural Information Processing Systems}, volume~27. Curran Associates, Inc., 2014.
\newblock URL \url{https://proceedings.neurips.cc/paper/2014/file/5ca3e9b122f61f8f06494c97b1afccf3-Paper.pdf}.

\bibitem[Greenwald and Stiglitz(1986)]{greenwald1986externalities}
Bruce~C Greenwald and Joseph~E Stiglitz.
\newblock Externalities in economies with imperfect information and incomplete markets.
\newblock \emph{The quarterly journal of economics}, 101\penalty0 (2):\penalty0 229--264, 1986.

\bibitem[Han et~al.(2021)Han, Yang, et~al.]{han2021deepham}
Jiequn Han, Yucheng Yang, et~al.
\newblock Deepham: A global solution method for heterogeneous agent models with aggregate shocks.
\newblock \emph{arXiv preprint arXiv:2112.14377}, 2021.

\bibitem[Harris et~al.(2020)Harris, Millman, van~der Walt, Gommers, Virtanen, Cournapeau, Wieser, Taylor, Berg, Smith, Kern, Picus, Hoyer, van Kerkwijk, Brett, Haldane, Fernandez~del Rio, Wiebe, Peterson, Gérard-Marchant, Sheppard, Reddy, Weckesser, Abbasi, Gohlke, and Oliphant]{numpy}
Charles~R. Harris, K.~Jarrod Millman, Stéfan~J van~der Walt, Ralf Gommers, Pauli Virtanen, David Cournapeau, Eric Wieser, Julian Taylor, Sebastian Berg, Nathaniel~J. Smith, Robert Kern, Matti Picus, Stephan Hoyer, Marten~H. van Kerkwijk, Matthew Brett, Allan Haldane, Jaime Fernandez~del Rio, Mark Wiebe, Pearu Peterson, Pierre Gérard-Marchant, Kevin Sheppard, Tyler Reddy, Warren Weckesser, Hameer Abbasi, Christoph Gohlke, and Travis~E. Oliphant.
\newblock Array programming with {NumPy}.
\newblock \emph{Nature}, 585:\penalty0 357–362, 2020.
\newblock \doi{10.1038/s41586-020-2649-2}.

\bibitem[Hart(1975)]{hart1975optimality}
Oliver~D Hart.
\newblock On the optimality of equilibrium when the market structure is incomplete.
\newblock \emph{Journal of economic theory}, 11\penalty0 (3):\penalty0 418--443, 1975.

\bibitem[Hennigan et~al.(2020)Hennigan, Cai, Norman, Martens, and Babuschkin]{haiku2020github}
Tom Hennigan, Trevor Cai, Tamara Norman, Lena Martens, and Igor Babuschkin.
\newblock {H}aiku: {S}onnet for {JAX}, 2020.
\newblock URL \url{http://github.com/deepmind/dm-haiku}.

\bibitem[Huang and Werner(2000)]{huang2000asset}
Kevin~XD Huang and Jan Werner.
\newblock Asset price bubbles in arrow-debreu and sequential equilibrium.
\newblock \emph{Economic Theory}, 15:\penalty0 253--278, 2000.

\bibitem[Huang and Werner(2004)]{huang2004implementing}
Kevin~XD Huang and Jan Werner.
\newblock Implementing arrow-debreu equilibria by trading infinitely-lived securities.
\newblock \emph{Economic Theory}, 24:\penalty0 603--622, 2004.

\bibitem[Huberman(1982)]{huberman1982simple}
Gur Huberman.
\newblock A simple approach to arbitrage pricing theory.
\newblock \emph{Journal of Economic Theory}, 28\penalty0 (1):\penalty0 183--191, 1982.

\bibitem[Huggett(1993)]{huggett1993risk}
Mark Huggett.
\newblock The risk-free rate in heterogeneous-agent incomplete-insurance economies.
\newblock \emph{Journal of economic Dynamics and Control}, 17\penalty0 (5-6):\penalty0 953--969, 1993.

\bibitem[Hunter(2007)]{matplotlib}
J.~D. Hunter.
\newblock Matplotlib: A 2d graphics environment.
\newblock \emph{Computing in Science and Engineering}, 9\penalty0 (3):\penalty0 90--95, 2007.
\newblock \doi{10.1109/MCSE.2007.55}.

\bibitem[Jain et~al.(2005)Jain, Vazirani, and Ye]{jain2005market}
Kamal Jain, Vijay~V Vazirani, and Yinyu Ye.
\newblock Market equilibria for homothetic, quasi-concave utilities and economies of scale in production.
\newblock In \emph{SODA}, volume~5, pages 63--71, 2005.

\bibitem[Judd(1992)]{judd_projection_1992}
Kenneth~L Judd.
\newblock Projection methods for solving aggregate growth models.
\newblock \emph{Journal of Economic Theory}, 58\penalty0 (2):\penalty0 410--452, December 1992.
\newblock ISSN 0022-0531.
\newblock \doi{10.1016/0022-0531(92)90061-L}.
\newblock URL \url{https://www.sciencedirect.com/science/article/pii/002205319290061L}.

\bibitem[Kim and Kim(2003)]{kim2003spurious}
Jinill Kim and Sunghyun~Henry Kim.
\newblock Spurious welfare reversals in international business cycle models.
\newblock \emph{journal of International Economics}, 60\penalty0 (2):\penalty0 471--500, 2003.

\bibitem[King et~al.(2002)King, Plosser, and Rebelo]{king2002production}
Robert~G King, Charles~I Plosser, and Sergio~T Rebelo.
\newblock Production, growth and business cycles: Technical appendix.
\newblock \emph{Computational Economics}, 20:\penalty0 87--116, 2002.

\bibitem[Kydland and Prescott(1982)]{kydland1982time}
Finn~E Kydland and Edward~C Prescott.
\newblock Time to build and aggregate fluctuations.
\newblock \emph{Econometrica: Journal of the Econometric Society}, pages 1345--1370, 1982.

\bibitem[Lin et~al.(2020)Lin, Jin, and Jordan]{lin2020gradient}
Tianyi Lin, Chi Jin, and Michael Jordan.
\newblock On gradient descent ascent for nonconvex-concave minimax problems.
\newblock In \emph{International Conference on Machine Learning}, pages 6083--6093. PMLR, 2020.

\bibitem[Lintner(1975)]{lintner1975valuation}
John Lintner.
\newblock The valuation of risk assets and the selection of risky investments in stock portfolios and capital budgets.
\newblock In \emph{Stochastic optimization models in finance}, pages 131--155. Elsevier, 1975.

\bibitem[Liu et~al.(2021)Liu, Rafique, Lin, and Yang]{liu_first-order_2021}
Mingrui Liu, Hassan Rafique, Qihang Lin, and Tianbao Yang.
\newblock First-order {Convergence} {Theory} for {Weakly}-{Convex}-{Weakly}-{Concave} {Min}-max {Problems}, July 2021.
\newblock URL \url{http://arxiv.org/abs/1810.10207}.
\newblock arXiv:1810.10207.

\bibitem[Long~Jr and Plosser(1983)]{long1983real}
John~B Long~Jr and Charles~I Plosser.
\newblock Real business cycles.
\newblock \emph{Journal of political Economy}, 91\penalty0 (1):\penalty0 39--69, 1983.

\bibitem[Loong and Zeckhauser(1982)]{loong1982pecuniary}
Lee~Hsien Loong and Richard Zeckhauser.
\newblock Pecuniary externalities do matter when contingent claims markets are incomplete.
\newblock \emph{The Quarterly Journal of Economics}, 97\penalty0 (1):\penalty0 171--179, 1982.

\bibitem[Lucas~Jr(1978)]{lucas1978asset}
Robert~E Lucas~Jr.
\newblock Asset prices in an exchange economy.
\newblock \emph{Econometrica: journal of the Econometric Society}, pages 1429--1445, 1978.

\bibitem[Lucas~Jr and Prescott(1971)]{lucas1971investment}
Robert~E Lucas~Jr and Edward~C Prescott.
\newblock Investment under uncertainty.
\newblock \emph{Econometrica: Journal of the Econometric Society}, pages 659--681, 1971.

\bibitem[Magill and Quinzii(1994)]{magill1994infinite}
Michael Magill and Martine Quinzii.
\newblock Infinite horizon incomplete markets.
\newblock \emph{Econometrica: Journal of the Econometric Society}, pages 853--880, 1994.

\bibitem[Magill and Quinzii(2002)]{magill2002theory}
Michael Magill and Martine Quinzii.
\newblock \emph{Theory of incomplete markets}, volume~1.
\newblock Mit press, 2002.

\bibitem[Magill and Shafer(1991)]{magill1991incomplete}
Michael Magill and Wayne Shafer.
\newblock Incomplete markets.
\newblock \emph{Handbook of mathematical economics}, 4:\penalty0 1523--1614, 1991.

\bibitem[Maliar et~al.(2015)Maliar, Maliar, Taylor, and Tsener]{maliar2015tractable}
Lilia Maliar, Serguei Maliar, John Taylor, and Inna Tsener.
\newblock A tractable framework for analyzing a class of nonstationary markov models.
\newblock Technical report, National Bureau of Economic Research, 2015.

\bibitem[Maliar et~al.(2021)Maliar, Maliar, and Winant]{maliar_deep_2021}
Lilia Maliar, Serguei Maliar, and Pablo Winant.
\newblock Deep learning for solving dynamic economic models.
\newblock \emph{Journal of Monetary Economics}, 122:\penalty0 76--101, September 2021.
\newblock ISSN 0304-3932.
\newblock \doi{10.1016/j.jmoneco.2021.07.004}.
\newblock URL \url{https://www.sciencedirect.com/science/article/pii/S0304393221000799}.

\bibitem[Mas-Colell et~al.(1995)Mas-Colell, Whinston, and Green]{mas-colell}
Andreu Mas-Colell, Michael~D. Whinston, and Jerry~R. Green.
\newblock \emph{{Microeconomic Theory}}.
\newblock Number 9780195102680 in OUP Catalogue. Oxford University Press, 1995.
\newblock ISBN ARRAY(0x4cf9c5c0).
\newblock URL \url{https://ideas.repec.org/b/oxp/obooks/9780195102680.html}.

\bibitem[Maskin and Tirole(2001)]{maskin2001markov}
Eric Maskin and Jean Tirole.
\newblock Markov perfect equilibrium: I. observable actions.
\newblock \emph{Journal of Economic Theory}, 100\penalty0 (2):\penalty0 191--219, 2001.

\bibitem[Mehra and Prescott(1977)]{mehra1977recursive}
R~Mehra and EC~Prescott.
\newblock Recursive competitive equilibria and capital asset pricing.
\newblock \emph{Essays in Financial Economics}, 1977.

\bibitem[Modigliani and Miller(1958)]{modigliani1958cost}
Franco Modigliani and Merton~H Miller.
\newblock The cost of capital, corporation finance and the theory of investment.
\newblock \emph{The American economic review}, 48\penalty0 (3):\penalty0 261--297, 1958.

\bibitem[Mossin(1966)]{mossin1966equilibrium}
Jan Mossin.
\newblock Equilibrium in a capital asset market.
\newblock \emph{Econometrica: Journal of the econometric society}, pages 768--783, 1966.

\bibitem[Murty and Kabadi(1987)]{murty1987some}
Katta~G Murty and Santosh~N Kabadi.
\newblock Some np-complete problems in quadratic and nonlinear programming.
\newblock \emph{Mathematical Programming}, 39:\penalty0 117--129, 1987.

\bibitem[Nash(1950{\natexlab{a}})]{nash1950bargaining}
John~F. Nash.
\newblock The bargaining problem.
\newblock \emph{Econometrica}, 18\penalty0 (2):\penalty0 155--162, 1950{\natexlab{a}}.
\newblock ISSN 00129682, 14680262.
\newblock URL \url{http://www.jstor.org/stable/1907266}.

\bibitem[Nash(1950{\natexlab{b}})]{nash1950existence}
John~F. Nash.
\newblock Equilibrium points in <i>n</i>-person games.
\newblock \emph{Proceedings of the National Academy of Sciences}, 36\penalty0 (1):\penalty0 48--49, 1950{\natexlab{b}}.
\newblock \doi{10.1073/pnas.36.1.48}.
\newblock URL \url{https://www.pnas.org/doi/abs/10.1073/pnas.36.1.48}.

\bibitem[Newbery and Stiglitz(1982)]{newbery1982theory}
David~MG Newbery and Joseph~E Stiglitz.
\newblock \emph{The theory of commodity price stabilization. A study in the economics of risk.}
\newblock 1982.

\bibitem[Parra-Alvarez(2015)]{parra2015solution}
Juan~Carlos Parra-Alvarez.
\newblock \emph{Solution Methods and Inference in Continuous-time Dynamic Equilibrium Economies:(with Applications in Asset Pricing and Income Fluctuation Models): a PhD Thesis Submitted to School of Business and Social Sciences, Aarhus University, in Partial Fulfilment of the Requirements of the PhD Degree in Economics and Business}.
\newblock Department of Economics and Business, Aarhus University, 2015.

\bibitem[Prescott and Mehra(1980)]{prescott1980recursive}
Edward~C. Prescott and Rajnish Mehra.
\newblock Recursive competitive equilibrium: The case of homogeneous households.
\newblock \emph{Econometrica}, 48\penalty0 (6):\penalty0 1365--1379, 1980.
\newblock ISSN 00129682, 14680262.
\newblock URL \url{http://www.jstor.org/stable/1912812}.

\bibitem[Puterman(2014)]{puterman2014markov}
Martin~L Puterman.
\newblock \emph{Markov decision processes: discrete stochastic dynamic programming}.
\newblock John Wiley \& Sons, 2014.

\bibitem[Radner(1968)]{radner1968competitive}
Roy Radner.
\newblock Competitive equilibrium under uncertainty.
\newblock \emph{Econometrica: Journal of the Econometric Society}, pages 31--58, 1968.

\bibitem[Radner(1972)]{radner1972existence}
Roy Radner.
\newblock Existence of equilibrium of plans, prices, and price expectations in a sequence of markets.
\newblock \emph{Econometrica: Journal of the Econometric Society}, pages 289--303, 1972.

\bibitem[Radner(1979)]{radner1979rational}
Roy Radner.
\newblock Rational expectations equilibrium: Generic existence and the information revealed by prices.
\newblock \emph{Econometrica: Journal of the Econometric Society}, pages 655--678, 1979.

\bibitem[Ross(1976)]{ross1976options}
Stephen~A Ross.
\newblock Options and efficiency.
\newblock \emph{The Quarterly Journal of Economics}, 90\penalty0 (1):\penalty0 75--89, 1976.

\bibitem[Sargent and Ljungqvist(2000)]{sargent2000recursive}
Thomas~J Sargent and Lars Ljungqvist.
\newblock Recursive macroeconomic theory.
\newblock \emph{Massachusetss Institute of Technology}, 2000.

\bibitem[Sauzet(2021)]{sauzet_projection_2021}
Maxime Sauzet.
\newblock Projection {Methods} via {Neural} {Networks} for {Continuous}-{Time} {Models}, December 2021.
\newblock URL \url{https://papers.ssrn.com/abstract=3981838}.

\bibitem[Selby(1990)]{duffie1990stochmodels}
M.~J.~P. Selby.
\newblock \emph{Economica}, 57\penalty0 (227):\penalty0 413--415, 1990.
\newblock ISSN 00130427, 14680335.
\newblock URL \url{http://www.jstor.org/stable/2554945}.

\bibitem[Shapley(1953)]{shapley1953stochastic}
Lloyd~S Shapley.
\newblock Stochastic games.
\newblock \emph{Proceedings of the national academy of sciences}, 39\penalty0 (10):\penalty0 1095--1100, 1953.

\bibitem[Silver et~al.(2014)Silver, Lever, Heess, Degris, Wierstra, and Riedmiller]{silver2014deterministic}
David Silver, Guy Lever, Nicolas Heess, Thomas Degris, Daan Wierstra, and Martin Riedmiller.
\newblock Deterministic policy gradient algorithms.
\newblock In \emph{International conference on machine learning}, pages 387--395. Pmlr, 2014.

\bibitem[Stiglitz(1982)]{stiglitz1982self}
Joseph~E Stiglitz.
\newblock Self-selection and pareto efficient taxation.
\newblock \emph{Journal of public economics}, 17\penalty0 (2):\penalty0 213--240, 1982.

\bibitem[Stokey(1989)]{stokey1989recursive}
Nancy~L Stokey.
\newblock \emph{Recursive methods in economic dynamics}.
\newblock Harvard University Press, 1989.

\bibitem[Suh et~al.(2022)Suh, Simchowitz, Zhang, and Tedrake]{suh2022differentiable}
Hyung~Ju Suh, Max Simchowitz, Kaiqing Zhang, and Russ Tedrake.
\newblock Do differentiable simulators give better policy gradients?
\newblock In \emph{International Conference on Machine Learning}, pages 20668--20696. PMLR, 2022.

\bibitem[Takahashi(1964)]{takahashi1964equilibrium}
Masayuki Takahashi.
\newblock Equilibrium points of stochastic non-cooperative $ n $-person games.
\newblock \emph{Journal of Science of the Hiroshima University, Series AI (Mathematics)}, 28\penalty0 (1):\penalty0 95--99, 1964.

\bibitem[Taylor and Woodford(1999)]{taylor1999handbook}
John~B Taylor and Michael Woodford.
\newblock \emph{Handbook of macroeconomics}, volume~1.
\newblock Elsevier, 1999.

\bibitem[Van~Rossum and Drake~Jr(1995)]{van1995python}
Guido Van~Rossum and Fred~L Drake~Jr.
\newblock \emph{Python tutorial}.
\newblock Centrum voor Wiskunde en Informatica Amsterdam, The Netherlands, 1995.

\bibitem[Wald(1945)]{wald1945statistical}
Abraham Wald.
\newblock Statistical decision functions which minimize the maximum risk.
\newblock \emph{Annals of Mathematics}, pages 265--280, 1945.

\bibitem[Walras(1896)]{walras}
Leon Walras.
\newblock \emph{Elements de l'economie politique pure, ou, Theorie de la richesse sociale}.
\newblock F. Rouge, 1896.

\bibitem[Werner(1985)]{werner1985equilibrium}
Jan Werner.
\newblock Equilibrium in economies with incomplete financial markets.
\newblock \emph{Journal of Economic Theory}, 36\penalty0 (1):\penalty0 110--119, 1985.

\end{thebibliography}

\appendix
\section{Related Work}
\label{sec:related}

Beyond the works mentioned earlier, our paper is close to two literature on stochastic economies, one in financial economics which theoretical and focuses on understanding mathematical properties of general equilibrium competitive equilibrium in incomplete markets \cite{duffie1987stochastic, duffie1990stochmodels, duffie1985equilibrium,duffie1986equilibrium}, and another one in macroeconomics which focuses on the computation of sequential or recursive competitive equilibrium in incomplete stochastic economies to simulate various macroeconomic issues; see, for instance, \citeauthor{kydland1982time} (\citeyear{kydland1982time}) and \citeauthor{lucas1971investment} (\citeyear{lucas1971investment}).%
\footnote{Since the 90s, a sizable body of work in financial economics (see for instance \cite{fernandez1016solve, auclert2021using} has considered computational approaches to solving general equilibrium models of financial markets; however, much of this work can be seen as extension of the macroeconomics literature.} 

\paragraph{Financial economics}
Regarding the literature in financial economics, we refer the reader to the survey work of \citet{magill2002theory}, and mention here only a few of some the influential models for the development of stochastic economies. Following the initial interest of the early 1970, the literature on stochastic economies in financial economics mostly focused on stochastic economies with two time-periods up to the end of the 1980s. In the early 1980s, there was an explosion of option pricing studies and arbitrage pricing in the early 1980s (See, for
example, \citet{cox1976valuation, ross1976options}  \citet{cox1979option}, \citet{cox1985intertemporal}, and \citet{huberman1982simple}.] 
By the mid-1980s, the theory of stochastic economies made great strides, with two influential papers, \citeauthor{cass1984competitive} (\citeyear{cass1984competitive, cass1985number}) showing that existence of a general equilibrium could be guaranteed if all the assets promise delivery
in fiat money, and he showed that with such financial assets there could be a multiplicity of equilibrium. In contrast, our existence result does not assume the existence of fiat money. Almost simultaneously \citet{werner1985equilibrium} also gave a proof of existence of equilibrium with financial assets, and \citet{geanakoplos1986walrasian} showed the same for economies with real assets that promise delivery in the same consumption good.\citet{duffie1987stochastic} then extended the existence results for purely financial assets to arbitrary finite horizon stochastic economies. As stochastic economies with incomplete asset markets have been shown to not satisfy a first welfare theorem of economics, following preliminary insight from \citet{diamond1967incompletege} the literature turned its attention definine notions of constrained efficiency. Successive refinements of the definition were given by \citet{diamond1980income}, \citet{loong1982pecuniary}, \citet{newbery1982theory}, \citet{stiglitz1982self}, and \citet{greenwald1986externalities} with a mostly accepted definition of constrained efficiency becoming becoming clear by the late 1980, with \citet{geanakoplos1990introduction} eventually proving that sequential competitive equilibrium are is constrained efficient inefficient.

\paragraph{Macroeconomics}

The literature on stochastic economies in macroeconomics is known under the name of \mydef{dynamic stochastic general equilibrium (DSGE) models}. Stochastic economies have received interest in macroeconomics after \citeauthor{lucas1978asset}'s (\citeyear{lucas1978asset}) seminal work, in which he derived a recursive competitive equilibrium in closed form in a stochastic economy with one commodity and and one consumer allowing him to analyze asset prices in his model. Unfortunately, beyond Lucas' simpler model, it became apparent that analyzing the solutions of stochastic economies required the use computation. 
One of the earliest popular stochastic economy models in economics which was solved via computational methods is the \mydef{Real Business Cycle (RBC)} model \cite{kydland1982time, long1983real}. The RBC model is a parameterized stochastic economy whose parameters are calibrated to accurately model the US economy. RBC models are characterized by demand generated by a representative infinitely-lived agent, with supply generated exogeneously by a standard (or Solow) growth model \cite{acemoglu2008introduction}, or by a representative firm.
These models have fallen out of favor, because some of their assumptions were invalidated by data (see, for example, Section 2 of \citet{christiano2018dsge}).
They were replaced by a class of DSGE models known as \mydef{Representative Agent New Keynesian (RANK)} models (see, for instance, \citet{clarida2000monetary}). As RBC and RANK models derive their modelling assumptions from two different schools of macroeconomic thought (i.e., the New Keynesian and New Classical schools, respectively),%
from a mathematical and computational perspective they can be seen as the same, as both are characterized by a representative consumer and an exogenous growth model, or a representative firm.

Following the financial crisis of 2008, these representative-agent models, too, fell out of favor, and the literature turned to modeling heterogeneity, because of its importance in understanding inequality, in particular across consumers.
\mydef{Heterogeneous agent new Keynesian models (HANK)} are stochastic economies which 
are built on top \mydef{Bewley-Huggett-Aiyagari} models \cite{bewley1983difficulty, huggett1993risk,aiyagari1994uninsured}, and are characterized by demand and supply generated by an infinite population of agents with differing characteristics. These models are mathematically and computationally much more different than the RBC and RANK models and have been shown to be possible to model as single population mean-field games \cite{achdou2022income}, i.e., games with an infinite population of players. 
More recently, a new class of stochastic economies called \mydef{Many Agent New Keynesian (MANK)} has emerged. This class of models bridges the gap between the infinite population regime of heterogeneous agent models and the single agent regime of representative agent models. These models are characterized by a demand and a supply generated by multiple consumers and firms, but are arguably more interpretable \cite{eskelinen2021monetary} (see, for instance, \citet{cloyne2020monetary, eskelinen2021monetary}) and have shown to approximate the solutions of heterogeneous agent models effectively when the number of agents in the economy is large enough \cite{han2021deepham}.
That is MANK models are sufficiently expressive to capture a range of models, corresponding to RANK at one extreme, and to HANK at the other.
Ignoring the stylized details of the aforementioned stochastic economies, all of them feature static markets, linked over time, often although not always, by incomplete financial asset markets, and differ in the number and heterogeneity of the agents, firms and good in the economy, as well as the types of transitions they employ, whether it be transition functions which model \mydef{aggregate shocks} (i.e., transitions functions which change the state of each consumer and firm in the economy in the same way)  or \mydef{idiosyncratic shock} (i.e. transition function which model transition the state of each consumer in the economy in distinct way).
\emph{The infinite horizon Markov exchange economy that we develop in this paper corresponds to a many agent stochastic economy model, and can be coupled with either the New Keynesian or New Classical paradigm to capture most of the models proposed in the literature.}

\deni{Revisit this paragraph to add references.}

\paragraph{Computation of competitive equilibrium}
The study of the computational complexity of competitive equilibria was initiated by \citet{devanur2002market}, who provided a polynomial-time method for computing competitive equilibrium in a special case of the Arrow-Debreu (exchange) market model, namely Fisher markets, when buyers utilities are linear.
\citet{jain2005market} subsequently showed that a large class of Fisher markets with homogeneous utility functions could be solved in polynomial-time using interior point methods.
\citet{gao2020polygm} studied an alternative family of first-order methods for solving Fisher markets, assuming linear, quasilinear, and Leontief utilities, as such methods can be more efficient when markets are large. 
More recently, \citet{goktas2023tatonn} showed that t\^atonnement converges to competitive equilibrium in homothetic Fisher markets, assuming bounded elasticity of Hicksian demand.

Devising algorithms for the computation of competitive equilibrium in general Arrow-Debreu markets is still an active area of research.
While the computation of competitive equilibrium is PPAD-hard in general \cite{chen2006settling, daskalakis2009complexity}, the computation of competitive equilibrium in Arrow-Debreu markets with Leontief buyers is equivalent to the computation of Nash equilibrium in bimatrix games \cite{codenotti2006leontief, deng2008computation}, and hence PPAD-hard as well, there exist polynomial-time algorithms to compute competitive equilibrium in special cases of Arrow-Debreu markets, including markets whose excess demand satisfies the weak gross substitutes condition \cite{codenotti2005market, bei2015tatonnement} and Arrow-Debreu markets with buyers whose utilities are linear \cite{garg2004auction, branzei2021proportional} or satisfy constant elasticity of substitution, which gives rise to weak gross substitute demands \cite{branzei2021proportional}.


\paragraph{Solution methods in macroeconomics.}

As stochastic economies can be analytically intractable to solve without restrictive assumptions, such as homogeneous consumers (e.g., representative agent new Keynesian models  models, for a survey, see \citet{sargent2000recursive}), researchers have attempted to solve them via dynamic programming.
These methods often discretize the continuous state and action spaces, and then apply variants of value and policy iteration \cite{stokey1989recursive, sargent2000recursive, auclert2021using}. 
Unfortunately, this approach is unwieldy when applied to incomplete markets with multiple commodities and/or heterogeneous consumers \cite{fernandez1016solve}.
As a result, many of these methods lack optimality guarantees, and thus might not produce correct solutions, which may lead to drastically different policy recommendations, as inaccurate solutions to stochastic economies have been known to cause spurious welfare reversal \cite{kim2003spurious}.
Perhaps even more importantly, while static markets afford efficient, i.e., polynomial-time, algorithms for computing competitive equilibrium under suitable assumption (see, for instance \citet{jain2005market} or \citet{goktas2023tatonn} for a more recent survey), to the best of our knowledge, there is no known class of stochastic economies (excluding the special case of static economies) for which the computation of a sequential or recursive competitive equilibrium is polynomial-time. 
Yet the macroeconomics literature speaks to the need for efficient methods to solve these models, or at least better understand the trade-offs between the speed and accuracy of proposed solution techniques \cite{fernandez1016solve}.

We describe only a few of the most influential computational approaches to solving stochastic economies in macroeconomics, and refer the reader to \citet{fernandez1016solve} for a detailed survey.
\citeauthor{durbin2012time} \citeyear{durbin2012time} developed an extended path algorithm.
The idea was to solve, for a terminal date sufficiently far into the future, the path of endogenous variables using a shooting algorithm. 
Recently, \citeauthor{maliar2015tractable} (\citeyear{maliar2015tractable}) extended this idea, developing the extended function path (EFP) algorithm, applicable to models that do not admit stationary Markov equilibria.
\citeauthor{kydland1982time} (\citeyear{kydland1982time}) exploit the fact that their model admits a Pareto-optimal recursive equilibrium, and thus they solve the social planner's problem, instead of solving for an equilibrium.
To do so, they rely on a linear quadratic approximation, and exploit the fast algorithms known to solve that class of optimization problems. 
\citeauthor{king2002production} (\citeyear{king2002production}) (in the widely disseminated technical appendix, not published until 2002), building on \citeauthor{blanchard1980solution} (\citeyear{blanchard1980solution})'s approach, linearized the equilibrium conditions of their model (optimality conditions, market clearing conditions, etc.), and solved the resulting system of stochastic linear difference equations.
More recently, a growing literature has been applying deep learning methods in attempt to stochastic economies (see, for instance, \citet{curry2023learning, han2021deepham, childers2022differentiable}).
There also exists a large literature in macroeconomics on solution methods in continuous rather than discrete time, which is out of the scope of this paper.
We refer the interested reader to \citet{parra2015solution}.

\section{Omitted Results and Proofs}

\subsection{Omitted Results and Proofs from \Cref{sec:gmg}}

\thmexistmpgne*
\begin{proof}

    First, by Part 3 of \Cref{assum:existence_of_mpgne}, we know that for any $\player\in \players$, $\fpolicies[\player][{\mathrm{sub}}](\policy[-\player])$ is non-empty, convex, and compact, for all $\policy[-\player]\in \policies[-\player]$. Moreover, 2 of \Cref{assum:existence_of_mpgne}, $\fpolicies[][{\mathrm{sub}}]$ is upper-hemicontinuous.
    Therefore, by the Fan's fixed-point theorem \cite{Fan1952FixedPoint}, the set $\fpolicies[][{\mathrm{sub}}]\doteq \{\policy\in \policies[][{\mathrm{sub}}]\mid \policy\in \fpolicies[][{\mathrm{sub}}](\policy)\}$ is non-empty.
    
    For any player $\player \in \players$ and state $\state \in \states$, we define the \mydef{individual state best-response correspondence} $\brmap[\player][\state]: \subpolicies \rightrightarrows \actionspace[\player]$ by 
    \begin{align}
        \brmap[\player][\state] (\policy)
        &\doteq 
        \argmax_{\action[\player] \in \actions[\player] (\state, \policy[-\player] (\state))}  
        \reward[\player] (\state, \action[\player], \policy[-\player] (\state)) +\Ex_{\staterv[][][\prime] \sim \trans (\cdot\mid \state, \action[\player], \policy[-\player] (\state))}[ \discount \vfunc[\player][\policy] (\staterv[][][\prime]) ] \\
        &= \argmax_{\action[\player] \in \actions[\player] (\state, \policy[-\player] (\state))}
        \qfunc[\player][\policy] (\state, \action[\player], \policy[-\player] (\state))
    \end{align}
    
    Then, for any player $\player \in \players$, we define the \mydef{restricted individual best-response correspondence} $\brmap[\player]: \subpolicies \rightrightarrows \subpolicies[\player]$ as the Cartesian product of individual state best-response correspondences across the states restricted to $\subpolicies$: 
    \begin{align}
        \brmap[\player] (\policy)
        &= \left(\bigtimes_{\state \in \states} \brmap[\player][\state] (\policy) \right) \bigcap \subpolicies[\player]
        \\
        &= \{ \policy[\player] \in \subpolicies[\player] \mid \policy[\player] (\state) \in \brmap[\player][\state] (\policy), \forall\; \state \in \states \}
    \end{align}

    Finally, we can define the \mydef{multi-player best-response correspondence} $\brmap: \subpolicies \rightrightarrows \subpolicies$ as the Cartesian product of the individual correspondences, i.e., $\brmap(\policy) \doteq \bigtimes_{\player \in \players} \brmap[\player] (\policy)$.

    To show the existence of \MPGNE{}, we first want to show that there exists a fixed point $\policy[][][*] \in \subpolicies$ such that $\policy[][][*] \in \brmap(\policy[][][*])$. 
    To this end, we need to show that 1.~for any $\policy\in \subpolicies$, $\brmap(\policy)$ is non-empty, compact, and convex; 2.~$\brmap$ is upper hemicontinuous.
    
    Take any $\policy\in \subpolicies$. 
    Fix $\player \in \players, \state \in \states$, 
     we know that $\action[\player] \mapsto \qfunc[\player][{\policy}] (\state, \action[\player], \policy[-\player] (\state))$ is concave over $\actions[\player] (\state, \policy[-\player] (\state))$, and $\actions[\player] (\state, \policy[-\player] (\state))$ is non-empty, convex, and compact by \Cref{assum:existence_of_mpgne}, then by Proposition 4.1 of \citet{fiacco1986convexity}, $\brmap[\player][\state] (\policy)$ is non-empty, compact, and convex. 
    
    Now, notice $\bigtimes_{\state \in \states} \brmap[\player][\state] (\policy)$ is compact and convex as it is a Cartesian product of compact, convex sets. Thus, as $\subpolicies$ is also compact and convex by \Cref{assum:policy_class_exist}, we know that $\brmap[\player] (\policy) = \left(\bigtimes_{\state \in \states} \brmap[\player][\state] (\policy) \right) \bigcap \subpolicies[\player]$ is compact and convex. 
    By the assumption of \emph{closure under policy improvement} under \Cref{assum:policy_class_exist}, we know that since $\policy \in \subpolicies$, there exists $\policy[][][+] \in \subpolicies$ such that 
    $$\policy[\player][][+] \in \argmax_{\policy[\player][][\prime] \in \fmarkovpolicies[\player] (\policy[-\player])} \qfunc[\player][\policy] (\state, \policy[\player][][\prime] (\state), \policy[-\player] (\state))$$ for all $\state \in \states$, and that means $\policy[\player][][+] (\state) \in \brmap[\player][\state] (\policy)$ for all $\state \in \states$. Thus, $\brmap[\player] (\policy)$ is also non-empty. 
     Since Cartesian product preserves non-emptiness, compactness, and convexity, we can conclude that 
    $\brmap(\policy) = \bigtimes_{\player \in \players} \brmap[\player] (\policy)$ is non-empty, compact, and convex. 
    
     Similarly, fix $\player \in \players, \state \in \states$, for any $\policy\in \subpolicies$, since $\actions[\player] (\state, \cdot)$ is continuous (i.e. both upper and lower hemicontinuous), by the Maximum theorem, $\brmap[\player][\state]$ is upper hemicontinuous. 
    $\policy\mapsto \bigtimes_{\state \in \states} \brmap[\player][\state] (\policy)$ is upper hemicontinuous as it is a Cartesian product of upper hemicontinuous correspondence, and consequently, $\policy\mapsto \left(\bigtimes_{\state \in \states} \brmap[\player][\state] (\policy) \right) \bigcap \subpolicies$ is also upper hemicontinuous. 
    Therefore, $\brmap$ is also upper hemicontinuous.

    Since $\brmap(\policy)$ is non-empty, compact, and convex for any $\policy\in \subpolicies$ and $\brmap$ is upper hemicontinuous, by Fan's fixed-point theorem \cite{Fan1952FixedPoint}, $\brmap$ admits a fixed point. 

    Finally, say $\policy[][][*] \in \subpolicies$ is a fixed point of $\brmap$, and we want to show that $\policy[][][*]$ is a generalized Markov perfect equilibrium (\MPGNE{}) of $\mgame$.  Since $\policy[][][*] \in \brmap(\policy[][][*]) = \bigtimes_{\player \in \players} \brmap[\player] (\policy[][][*])$, we know that for all $\player \in \players$, $\policy[\player][][*] (\state) \in \brmap[\player][\state] (\policy[][][*])=\argmax_{\action[\player] \in \actions[\player] (\state, \policy[-\player][][*] (\state))}
        \qfunc[\player][{\policy[][][*]}] (\state, \action[\player], \policy[-\player][][*] (\state))$. 
    We now show that for any $\player \in \players$, for any $\policy[\player] \in \fpolicies[\player] (\policy[-\player][][*])$, $\vfunc[\player][{\policy[][][*]}] (\state) \geq \vfunc[\player][{(\policy[\player], \policy[-\player][][*])}] (\state)$ for all $\state \in \states$. Take any policy $\policy[\player] \in \fpolicies[\player] (\policy[-\player][][*])$. Note that $\policy[\player]$ may not be Markov, so we denote $\{ \policy[\player] (\hist[: \numhorizon]) \}_{\numhorizon\in \N}=\{ \action[\player][][\numhorizon] \}_{\numhorizon\in \N}$. Then,
    for all $\state[0] \in \states$,
    \begin{align}
        &\vfunc[\player][{\policy[][][*]}] (\state[0]) \notag\\
        &= \qfunc[\player][{\policy[][][*]}] (\state[0], \policy[\player][][*] (\state[0]), \policy[-\player][][*] (\state[0])) \notag\\
        &=\max_{\action[\player] \in \actions[\player] (\state[0], \policy[-\player][][*] (\state[0]))}
        \qfunc[\player][{\policy[][][*]}] (\state[0], \action[\player], \policy[-\player][][*] (\state[0])) \notag \\
        &= \max_{\action[\player] \in \actions (\state[0], \policy[-\player][][*] (\state[0]))} 
        \reward[\player] (\state[0], \action[\player], \policy[-\player][][*] (\state[0])) 
        + \Ex_{\state[1] \sim \trans (\cdot\mid \state[0], \action[\player], \policy[-\player][][*] (\state[0]))}[\discount \vfunc[\player][{\policy[][][*]}] (\state[1]) ] \notag \\
        &\geq \reward[\player] (\state[0], \action[\player][][0], \policy[-\player][][*] (\state[0])) + \Ex_{\state[1] \sim \trans (\cdot\mid \state[0], \action[\player][][0], \policy[-\player][][*] (\state[0]))}[\discount \vfunc[\player][{\policy[][][*]}] (\state[1]) ] \label{eq:recursive_form_v_prime}
    \end{align}
    For any $\state[0] \in \states$, define $\vfunc[\player][\prime] (\state[0]) \doteq \reward[\player] (\state[0], \action[\player][][0], \policy[-\player][][*] (\state[0])) 
    + \Ex_{\state[1] \sim \trans (\cdot\mid \state[0], \action[\player][][0], \policy[-\player][][*] (\state[0]))}[\discount \vfunc[\player][{\policy[][][*]}] (\state[1]) ]$ . 
    Since $\vfunc[\player][{\policy[][][*]}] (\state) \geq \vfunc[\player][\prime] (\state)$ for all $\player \in \players$, $\state \in \states$, we have for any $\state[0] \in \states$
    \begin{align}
        &\vfunc[\player][{\policy[][][*]}] (\state[0]) \notag\\ 
        &\geq 
        \reward[\player] (\state[0], \action[\player][][0], \policy[-\player][][*] (\state[0])) +
        \Ex_{\state[1] \sim \trans (\cdot\mid \state[0], \action[\player][][0], \policy[-\player][][*] (\state[0]))} [\discount \vfunc[\player][{\policy[][][*]}] (\state[1]) ] \notag \\
        & \geq \reward[\player] (\state[0], \action[\player][][0], \policy[-\player][][*] (\state[0])) +
        \Ex_{\state[1] \sim \trans (\cdot\mid \state[0], \action[\player][][0], \policy[-\player][][*] (\state[0]))}[ \discount \vfunc[\player][\prime] (\state[1])] \notag\\
        & \geq \reward[\player] (\state[0], \action[\player][][0], \policy[-\player][][*] (\state[0])) \notag \\
        &+
        \Ex_{\state[1] \sim \trans (\cdot\mid \state[0], \action[\player][][0], \policy[-\player][][*] (\state[0]))}
        \bigg[ \discount \bigg (\reward[\player] (\state[1], \action[\player][][1], \policy[-\player][][*] (\state[1])) \notag\\
        &+\Ex_{\state[2] \sim \trans (\cdot\mid \state[1], \action[\player][][1], \policy[-\player][][*] (\state[1])}[ \discount \vfunc[\player][{\policy[][][*]}] (\state[2]) ] \bigg) \bigg] \notag \\
         & \geq \reward[\player] (\state[0], \action[\player][][0], \policy[-\player][][*] (\state[0])) \notag \\
         &+
        \Ex_{\state[1] \sim \trans (\cdot\mid \state[0], \action[\player][][0], \policy[-\player][][*] (\state[0]))}
        \bigg[ \discount \bigg (\reward[\player] (\state[1], \action[\player][][1], \policy[-\player][][*] (\state[1]))  \notag\\
        &+\Ex_{\state[2] \sim \trans (\cdot\mid \state[1], \action[\player][][1], \policy[-\player][][*] (\state[1])}[ \discount \vfunc[\player][\prime] (\state[2]) ] \bigg) \bigg] \notag\\
       & \vdots \label{eq:expand_v_prime}\\
       &\geq \vfunc[\player][{(\policy[\player], \policy[-\player][][*])}] (\state) \notag
    \end{align}
    where in \Cref{eq:expand_v_prime}, we recursively expand $\vfunc[\player][\prime]$ and eliminate $\vfunc[][{\policy[][][*]}]$ using \Cref{eq:recursive_form_v_prime}. 
    We therefore conclude that for all states $\state \in \states$, and for all $\player \in \players$, $$\vfunc[\player][{\policy[][][*]}] (\state) \geq \max_{\policy[\player] \in \fpolicies[\player] (\policy[-\player][][*])} \vfunc[\player][{(\policy[\player], \policy[-\player][][*])}] (\state) .$$ 
\end{proof}

\lemmaexploitGNE*
\begin{proof}[Proof of \Cref{lemma:exploit_GNE}]
We first prove the result for state exploitability.

($\policy[][][*]$ is a \MPGNE{} $\implies$ $\sexploit (\state, \policy[][][*]) =0$ for all $\state \in \states$): Suppose that $\policy[][][*]$ is a \MPGNE{}, i.e., for all players $\player \in \players$, $\vfunc[\player][{\policy[][][*]}] (\state) \geq \max_{\policy[\player] \in \fpolicies[\player] (\policy[-\player][][*])} \vfunc[\player][{(\policy[\player], \policy[-\player][][*])}] (\state)$ for all state $\state \in \states$. Then, for all state $\state \in \states$, we have
\begin{align}
    \forall \player \in \players, \; \; \max_{\policy[\player] \in \fpolicies[\player] (\policy[-\player][][*])} \vfunc[\player][{(\policy[\player], \policy[-\player][][*])}] (\state) - \vfunc[\player][{\policy[][][*]}] (\state) =0
\end{align}
Summing up across all players $\player \in \players$, we get
\begin{align}
    \sexploit (\state, \policy[][][*])
    = \sum_{\player \in \players} \max_{\policy[\player] \in \fpolicies[\player] (\policy[-\player][][*])} \vfunc[\player][{(\policy[\player], \policy[-\player][][*])}] (\state) - \vfunc[\player][{\policy[][][*]}] (\state) =0
\end{align}

($\sexploit (\state, \policy[][][*]) =0$ for all $\state \in \states$ $\implies$ $\policy[][][*]$ is a \MPGNE{}):
Suppose we have $\policy[][][*] \in \fmarkovpolicies (\policy[][][*])$ and $\sexploit (\state, \policy[][][*]) =0$ for all $\state \in \states$. That is, for any $\state \in \states$
\begin{align}
     \sexploit (\state, \policy[][][*])
    = \sum_{\player \in \players} \max_{\policy[\player] \in \fpolicies[\player] (\policy[-\player][][*])} \vfunc[\player][{(\policy[\player], \policy[-\player][][*])}] (\state) - \vfunc[\player][{\policy[][][*]}] (\state) =0.
\end{align}
Since for any $\player \in \players$, $\policy[\player][][*] \in \fmarkovpolicies[\player] (\policy[-\player][][*])$,  $\max_{\policy[\player] \in \fmarkovpolicies[\player] (\policy[-\player][][*])} \vfunc[\player][{(\policy[\player], \policy[-\player][][*])}] (\state) - \vfunc[\player][{\policy[][][*]}] \geq \vfunc[\player][{\policy[][][*]}] (\state)-\vfunc[\player][{\policy[][][*]}] (\state) =0$. As a result, we must have for all player $\player \in \players$,
\begin{align}
    \vfunc[\player][{\policy[][][*]}] (\state) = \max_{\policy[\player] \in \fpolicies (\policy[-\player][][*])} \vfunc[\player][{(\policy[\player], \policy[-\player][][*])}] (\state), \; \; \forall \state \in \states
\end{align}
Thus, we can conclude that $\policy[][][*]$ is a \MPGNE{}.

Then, we can prove results for exploitability in an analogous way. 

($\policy[][][*]$ is a GNE $\implies$ $\gexploit (\policy[][][*]) =0$ ): Suppose that $\policy[][][*]$ is a GNE, i.e., for all players $\player \in \players$, $\payoff[\player] (\policy[][][*]) \geq \max_{\policy[\player] \in \fpolicies[\player] (\policy[-\player][][*])} \payoff[\player] (\policy[\player], \policy[-\player][][*])$. Then, we have
\begin{align}
    \forall \player \in \players, \; \; \max_{\policy[\player] \in \fpolicies[\player] (\policy[-\player][][*])} \payoff[\player] (\policy[\player], \policy[-\player][][*]) - \payoff[\player] (\policy[][][*]) =0
\end{align}
Summing up across all players $\player \in \players$, we get
\begin{align}
    \gexploit (\policy[][][*])
    = \sum_{\player \in \players} \max_{\policy[\player] \in \fpolicies[\player] (\policy[-\player][][*])} \payoff[\player] (\policy[\player], \policy[-\player][][*]) - \payoff[\player] (\policy[][][*]) =0
\end{align}

($\gexploit (\state, \policy[][][*]) =0$ $\implies$ $\policy[][][*]$ is a GNE):
Suppose we have $\policy[][][*] \in \fpolicies (\policy[][][*])$ and $\gexploit (\policy[][][*]) =0$. That is, 
\begin{align}
     \gexploit (\policy[][][*])
    = \sum_{\player \in \players} \max_{\policy[\player] \in \fpolicies[\player] (\policy[-\player][][*])} \payoff[\player] (\policy[\player], \policy[-\player][][*]) - \payoff[\player] (\policy[][][*]) =0.
\end{align}
Since for any $\player \in \players$, $\policy[\player][][*] \in \fpolicies[\player] (\policy[-\player][][*])$,  $\max_{\policy[\player] \in \fpolicies[\player] (\policy[-\player][][*])} \payoff[\player] (\policy[\player], \policy[-\player][][*]) - \payoff[\player] (\policy[][][*]) \geq \payoff[\player] (\policy[][][*]) - \payoff[\player] (\policy[][][*]) =0$. As a result, we must have for all player $\player \in \players$,
\begin{align}
    \payoff[\player] (\policy[][][*]) = \max_{\policy[\player] \in \fpolicies (\policy[-\player][][*])} \payoff[\player] (\policy[\player], \policy[-\player][][*])
\end{align}
Thus, we can conclude that $\policy[][][*]$ is a GNE.
\end{proof}

\obsminmax*
\begin{proof}
    The per-player maximum operator can be pulled out of the sum in the definition of state-exploitability, because the $\player$th player's best-response policy is independent of the other players' best-response policies, given a fixed policy profile $\policy$:
    \begin{align}
        \forall\; \state \in \states, 
        \; \; \sexploit (\state, \policy)
        &= \sum_{\player \in \players}
        \max_{\policy[\player][][\prime] \in \fmarkovpolicies[\player] (\policy[-\player])} \vfunc[\player][{(\policy[\player][][\prime], \policy[-\player])}] (\state) - \vfunc[\player][\policy] (\state) \\
        &= \max_{\policy[][][\prime] \in \fmarkovpolicies (\policy)} \sum_{\player \in \players} \vfunc[\player][{(\policy[\player][][\prime], \policy[-\player])}] (\state) - \vfunc[\player][\policy] (\state) \\
        &= \max_{\policy[][][\prime] \in \fmarkovpolicies (\policy)} \scumulreg (\state, \policy, \policy[][][\prime])
    \end{align}
The argument is analogous for exploitability:
    \begin{align}
        \gexploit (\policy)
        &= \sum_{\player \in \players}
        \max_{\policy[\player][][\prime] \in \fmarkovpolicies[\player] (\policy[-\player])} \payoff[\player] (\policy[\player][][\prime], \policy[-\player]) - \payoff[\player] (\policy) \\
        &= \max_{\policy[][][\prime] \in \fmarkovpolicies (\policy)} \sum_{\player \in \players} \payoff[\player] (\policy[\player][][\prime], \policy[-\player]) - \payoff[\player] (\policy) \\
        &= \max_{\policy[][][\prime] \in \fpolicies (\policy)} \gcumulreg (\policy, \policy[][][\prime])
    \end{align}
\end{proof}

\lemmafullsupport*
\begin{proof}

    First, using Jensen's inequality, by the convexity of the $2$-norm $\| \cdot \|$, we have: 
    \begin{align*}
    \Ex_{\state \sim \initstates} \left[ \| \grad[\param] \sexploit (\state, \param) \| \right] 
    &\leq \left\| \Ex_{\state \sim \initstates} \left[ \grad[\param] \sexploit (\state, \param) \right] \right\| \, \\
    &= \left\|   \grad[\param] \Ex_{\state \sim \initstates} \left[\sexploit (\state, \param) \right] \right\| \, \\
    &= \| \grad[\param] \exploit (\param) \|  \enspace .   
    \end{align*}
    
    The first claim follows directly from the fact that for all $\state \in \states$, $\|\grad[\param] \gexploit (\state, \param)\| \geq 0$, and hence for the expectation $ \Ex_{\state \sim \initstates} \left[ \| \grad[\param] \gexploit (\state, \param) \| \right]$ to be equal to $0$, its value should be equal to zero on a set of measure 1.

    Now, for the second part, by Markov's inequality, we have:
    $\Pr \left( \| \grad[\param] \sexploit (\state, \param) \| \, \geq \nicefrac{\varepsilon}{\delta} \right) \leq \frac{\Ex_{\state \sim \initstates} \left[ \| \grad[\param] \sexploit (\state, \policy) \| \right]} {\nicefrac{\varepsilon}{\delta}} \leq \frac{\varepsilon} {\nicefrac{\varepsilon}{\delta}} = \delta$.
\end{proof}

\lemmabrmismatch*
\begin{proof}
In this proof, for any $\player\in \players$,
we define $\deparam[\player](\param)=\depolicy[\player](\cdot, \policy(\cdot;\param); \deparam)$ as player $\player$'s policy in the policy profile $\deparam(\param)=\depolicy(\cdot, \policy(\cdot;\param); \deparam)$. Similarly, we define $\param[\player]=\policy[\player](\cdot;\param)$ as player $\player$'s policy in the policy profile $\param=\policy(\cdot; \param)$.

Given a policy parametrization scheme $(\policy, \depolicy, \params, \deparams)$,  consider any two parameters $\param \in \params, \deparam \in \deparams$, and any two initial state distributions  $\initstates, \diststates \in \simplex(\states)$, we know that
\begin{align}
    &\left\|\grad[\param] \scumulreg(\diststates, \param, \deparam) \right\|\\
    &=\left\|\grad[\param] \sum_{\player\in \players} \payoff[\player]( \deparam[\player](\param), \param[-\player])
    -\payoff[\player](\param)\right\|\\
    &=  \left\|\sum_{\player\in \players} \grad[\param] (\payoff[\player](\deparam[\player](\param), \param[-\player])-\payoff[\player](\param))\right\|\\
    &=\left\|\sum_{\player\in \players} \grad[\param]\left[
    \E_{\substack{\state[][][\prime]\sim \statedist[\diststates][{(\deparam[\player](\param), \param[-\player])}]\\
    \state \sim \statedist[\diststates][\param]}}
    \left[ \reward[\player](\state[][][\prime], \depolicy[\player](\state[][][\prime], \policy(\state;\param); \deparam), 
    \policy[-\player](\state[][][\prime]; \param) 
    - \reward[\player](\state, \policy(\state; \param))
    \right] \right]\right\|\\
    &= \bigg\|\sum_{\player\in \players}
    \E_{\substack{\state[][][\prime]\sim \statedist[\diststates][{(\deparam[\player](\param), \param[-\player])}]\\
    \state \sim \statedist[\diststates][\param]}}
    \bigg[
    \grad[{\action[-\player]}] \qfunc[\player][{\deparam[\player](\param), \param[-\player]}]    (\state[][][\prime], 
    \depolicy[\player](\state[][][\prime], \policy(\state[][][\prime];\param); \deparam), \policy[-\player](\state[][][\prime];\param)) 
    \grad[{\param}] \left( \depolicy[\player](\state[][][\prime], \policy[-\player](\state[][][\prime]; \param); \param), \policy(\state[][][\prime]; \param) \right)
    \nonumber\\
    &- \grad[\action] \qfunc[\player][\param](\state, \policy(\state; \param)) \grad[\param]\policy(\state; \param)
    \bigg] \label{eq:first_dpg} \bigg\|\\
    &\leq \max_{\player\in \players}
    \max_{\statep, \state\in \states}
    \frac{\statedist[\diststates][{(\deparam[\player](\param), \param[-\player])}] (\statep)  
    \statedist[\diststates][\param](\state)}{\statedist[\initstates][{(\deparam[\player](\param), \param[-\player])}](\statep) \statedist[\initstates][\param](\state)} 
    \bigg\|\E_{\substack{\state[][][\prime]\sim \statedist[\initstates][{(\deparam[\player](\param), \param[-\player])}]\\
    \state\sim \statedist[\initstates][\param]}}
    \bigg[
    \grad[{\action[-\player]}] \qfunc[\player][{\deparam[\player](\param), \param[-\player]}]    (\state[][][\prime], 
    \depolicy[\player](\state[][][\prime], \policy(\state[][][\prime];\param); \deparam), \policy[-\player](\state[][][\prime];\param)) 
    \nonumber\\
    &\quad\quad 
    \grad[{\param}] \left( \depolicy[\player](\state[][][\prime], \policy[-\player](\state[][][\prime]; \param); \param), \policy(\state[][][\prime]; \param) \right)
    -  \grad[\action] \qfunc[\player][\param](\state, \policy(\state; \param)) \grad[\param]\policy(\state; \param)
    \bigg] \bigg\|\\
    & \leq \max_{\player\in \players}\max_{\statep, \state\in \states}
    \frac{\statedist[\diststates][{(\deparam[\player](\param), \param[-\player])}](\statep) \statedist[\diststates][\param](\state)}{\statedist[\initstates][{(\deparam[\player](\param), \param[-\player])}](\statep) \statedist[\initstates][\param](\state)}
    \left\|\grad[\param] \left[
    \vfunc[\player][{\deparam[\player](\param), \param[-\player]}](\initstates)- \vfunc[\player][\param](\initstates)\right] \right\|
    \label{eq:second_dpg}\\
    & \leq \left(\frac{1}{1-\discount}\right)^2 
    \max_{\player\in \players}\max_{\statep, \state\in \states}\frac{\statedist[\diststates][{(\deparam[\player](\param), \param[-\player])}](\statep) \statedist[\diststates][\param](\state)}{\initstates(\statep)\initstates(\state)} 
    \left\|\grad[\param] \scumulreg(\initstates, \param, \deparam) \right\|
    \label{eq:back_to_initial_dist}\\
    &= \left(\frac{1}{1-\discount}\right)^2 
    \max_{\player\in \players}
    \left\Vert\frac{\statedist[\diststates][{(\deparam[\player](\param), \param[-\player])}]}{\initstates}\right\Vert_{\infty}
    \left\Vert\frac{\statedist[\initstates][\param]}{\initstates}\right\Vert_\infty
    \left\|\grad[\param] \scumulreg(\initstates, \param, \deparam)\right\|
\end{align}
where \Cref{eq:first_dpg} and \Cref{eq:second_dpg} are obtained by deterministic policy gradient theorem \cite{silver2014deterministic}, and \Cref{eq:back_to_initial_dist} is due to the fact that $\statedist[\initstates][\param](\state)\geq (1-\discount)\initstates(\state)$ for any $\policy\in \policies$, $\state\in \states$. 

Given condition (1) of \Cref{assum:param_gradient_dominance}, fix any $\param\in\params$, there exists $\deparam[][][*]\in \deparams$ s.t. for all $\player\in \players$, $\state\in \states$: $$\qfunc[\player][\param](\state, \depolicy[\player](\state, \policy(\state;\param);\deparam[][][*]), \policy[-\player](\state;\param)) 
= \max_{\policy[\player][][\prime]\in \fpolicies[\player](\policy(\cdot;\param))} 
\qfunc[\player][\param](\state, \policy[\player][][\prime](\state), \policy[-\player](\state;\param)) \enspace .
$$  Thus, $\sexploit(\state, \param)=\scumulreg(\state, \param, \deparam[][][*])$ for all $\state\in \states$. Hence, plugging in the optimal best-response policy $\deparam = \deparam[][][*]$, we obtain that \begin{align}
    \|\grad[\param] \sexploit(\diststates, \param)\|
    &\leq  \left(\frac{1}{1-\discount}\right)^2 
    \max_{\player\in \players}
    \left\Vert\frac{\statedist[\diststates][{(\deparam[\player][][*](\param), \param[-\player])}]}{\initstates}\right\Vert_{\infty}
    \left\Vert\frac{\statedist[\initstates][\param]}{\initstates}\right\Vert_\infty
    \|\grad[\param]  \sexploit(\initstates, \param)\|\\
    &\leq \left(\frac{1}{1-\discount}\right)^2 
    \max_{\player\in \players}
    \max_{\policyp[\player] \in \brmap[\player](\policy[-\player](\cdot;\param) )}
    \left\Vert\frac{\statedist[\diststates][{(\policy[\player][][\prime], \policy[-\player](\cdot;\param))}]}{\initstates}\right\Vert_{\infty}
    \left\Vert\frac{\statedist[\initstates][\param]}{\initstates}\right\Vert_\infty
     \|\grad[\param] \sexploit(\initstates, \param)\| \label{eq:max_over_best_response}
\end{align}
where \cref{eq:max_over_best_response} is due to the fact that $\deparam[\player][][*](\param)\in \brmap[\player](\policy[-\player](\cdot;\param) )$.

\end{proof}

\lemmauncoupledminmax*
\begin{proof}
    Fix $\policy[][][*] \in \fmarkovpolicies(\policy[][][*])$.
    We want to show that $$\max_{\policy[][][\prime] \in \fmarkovpolicies (\policy[][][*])} \gexploit (\policy[][][*], \policy[][][\prime]) = \max_{\depolicy \in \depolicies} \gexploit (\policy[][][*], \depolicy (\cdot, \policy (\cdot))) \enspace .$$ 
    Define $\policies[][{\depolicies, \policy[][][*]}] \doteq \{ \policy: \state\mapsto \depolicy (\state, \policy[][][*] (\state)) \mid \depolicy \in \depolicies \} \subseteq \markovpolicies$. 
    
    First, for all $\policy[][][\prime] \in \policies[][{\depolicies, \policy[][][*]}]$, $\policy[][][\prime] (\state) = \depolicy (\state, \policy[][][*] (\state)) \in \actions (\state, \policy[][][*] (\state))$, for all $\state \in \states$, by the definition of $\depolicies$.
    Thus, $\policy[][][\prime] \in \fmarkovpolicies (\policy[][][*]) = \{ \policy \in \markovpolicies\mid \forall\state \in \states, \policy (\state) \in \actions (\state, \policy[][][*] (\state)) \}$.
    Therefore, $\policies[][{\depolicies, \policy[][][*]}] \subseteq \fmarkovpolicies (\policy[][][*])$, which implies that $\max_{\policy[][][\prime] \in \fmarkovpolicies (\policy[][][*])} \gexploit (\policy[][][*], \policy[][][\prime])
    \geq \max_{\policy[][][\prime] \in \policies[][{\depolicies, \policy[][][*]}]} \gexploit (\policy[][][*], \policy[][][\prime])
    = \max_{\depolicy \in \depolicies} \gexploit (\policy[][][*], \depolicy (\cdot, \policy (\cdot)))$. 

    Moreover, for all $\policy[][][\prime] \in \fmarkovpolicies (\policy[][][*])$, $\policy[][][\prime] (\state) \in \actions (\state, \policy[][][*] (\state))$, for all $\state \in \states$, by the definition of $\fmarkovpolicies$. 
    Define $\depolicy[][][\prime]$ such that for all $\state \in \states$, $\depolicy[][][\prime] (\state, \action) = \policy[][][\prime] (\state)$ if $\action = \policy[][][*] (\state)$, and $\depolicy[][][\prime] (\state, \action) = \action[][][][\prime]$ for some $\action[][][][\prime] \in \actions (\state, \action)$ otherwise.
    Note that $\depolicy[][][\prime] \in \depolicies$, since $\forall (\state, \action) \in \states\times \actionspace, \; \depolicy (\state, \action) \in \actions (\state, \action)$. Thus, as $\policy[][][\prime] (\state) = \depolicy[][][\prime] (\state, \policy[][][*] (\state))$, for all $\state \in \states$, it follows that $\policy[][][\prime] \in \policies[][{\depolicies, \policy[][][*]}]$. 
    Therefore, $\fmarkovpolicies (\policy[][][*]) \subseteq \policies[][{\depolicies, \policy[][][*]}]$, which implies that $\max_{\policy[][][\prime] \in \fmarkovpolicies (\policy[][][*])} \gexploit (\policy[][][*], \policy[][][\prime])
    \leq \max_{\policy[][][\prime] \in \policies[][{\depolicies, \policy[][][*]}]} \gexploit (\policy[][][*], \policy[][][\prime])
    = \max_{\depolicy \in \depolicies} \gexploit (\policy[][][*], \depolicy (\cdot, \policy (\cdot)))$. 

    Finally, we conclude that $\max_{\policy[][][\prime] \in \fmarkovpolicies (\policy[][][*])} \gexploit (\policy[][][*], \policy[][][\prime]) = \max_{\depolicy \in \depolicies} \gexploit (\policy[][][*], \depolicy (\cdot, \policy (\cdot)))$. 
\end{proof}

\thmconvergence*
\begin{proof}
As is common in the optimization literature (see, for instance, \citet{davis2018subgradient}), 
we consider the Moreau envelope of the exploitability, which we simply call the \mydef{Moreau exploitability}, i.e., $$\regexploit (\param) \doteq \min_{\param[][][\prime] \in \params} \left\{ \gexploit (\param[][][\prime]) + \lipschitz[{\grad \scumulreg}] \left\| \param - \param[][][\prime] \right\|^2\right\} \enspace .$$ 
Similarly, we also consider the \mydef{state Moreau exploitability}, i.e., the Moreau envelope of the state exploitability: $$\regsexploit (\state, \param) \doteq \min_{\param[][][\prime] \in \params} \left\{ \sexploit (\state, \param[][][\prime]) + \lipschitz[{\grad \scumulreg}] \left\| \param - \param[][][\prime] \right\|^2\right\} \enspace .$$
We recall that in these definitions, by our notational convention, $\lipschitz[{\grad \scumulreg}] \geq 0$, refers to the Lipschitz-smoothness constants of the state exploitability which in this case we take to be the largest across all states, i.e., for all $\state \in \states$, $(\param, \deparam) \mapsto \scumulreg(\state, \param, \deparam)$ is $\lipschitz[{\grad \scumulreg}]$-Lipschitz-smooth, respectively, and which we note is guaranteed to exist under \Cref{assum:param_lipschitz}. Further, we note that since $\gcumulreg(\param, \deparam) = \Ex_{\state \sim \initstates} \left[\scumulreg(\state, \param, \deparam) \right]$ is a weighted average of $\scumulreg$, $(\param, \deparam) \mapsto \gcumulreg(\param, \deparam)$ is also $\lipschitz[{\grad \scumulreg}]$-Lipschitz-smooth.

We invoke Theorem 2 of \citeauthor{daskalakis2020independent}
(\citeyear{daskalakis2020independent}).
Although their result is stated for gradient-dominated-gradient-dominated functions, their proof applies in the more general case of non-convex-gradient-dominated functions.
    
First, \Cref{assum:param_lipschitz} guarantees that the cumulative regret $\gcumulreg$ is Lipschitz-smooth w.r.t.\@ $(\param, \deparam)$.
    Moreover, under \Cref{assum:param_lipschitz}, which guarantees that $\deparam \mapsto \qfunc[\player][{\param[][][\prime]}] (\state, \depolicy[\player] (\state, \policy[-\player] (\state; \param); \deparam), 
    \policy[-\player] (\state; \param) 
    )$ is continuously differentiable for all $\state \in \states$ and $\param, \param[][][\prime] \in \params$, and \Cref{assum:param_gradient_dominance}, we have that $\gcumulreg$ is $\left(\nicefrac{\left\|\nicefrac{\partial\statedist[\initstates][{\policy[][][*]}]}{\partial \initstates} \right\|_\infty}{1-\discount} \right)$-gradient-dominated in $\deparam$, for all $\param \in \params$, by Theorems 2 and 4 of \citeauthor{bhandari2019global} (\citeyear{bhandari2019global}). 
    Finally, under \Cref{assum:param_lipschitz}, since the policy, the reward function, and the transition probability function are all Lipschitz-continuous, $\estpayoff$, $\estgcumulreg$, and hence $\estG$ are also Lipschitz-continuous, since $\states$ and $\actionspace$ are compact. 
    Their variance must therefore be bounded, i.e., there exists $\varconst[\param], \varconst[\deparam] \in \R$ s.t.\@ $\Ex_{\hist, \hist[][\prime]}[\estG[\param] (\param, \deparam; \hist, \hist[][\prime])- \grad[\param] \gcumulreg(\param, \deparam; \hist, \hist[][\prime])] \leq \varconst[\param]$ and $\Ex_{\hist, \hist[][\prime]}[\estG[\deparam] (\param, \deparam; \hist, \hist[][\prime])- \grad[\deparam] \gcumulreg(\param, \deparam; \hist, \hist[][\prime])] \leq \varconst[\deparam]$. 

    Hence, under our assumptions, the assumptions of Theorem 2 of \citeauthor{daskalakis2020independent} are satisfied.
    Therefore,
        $\nicefrac{1}{\numiters  + 1} \sum_{\numhorizon = 0}^\numiters \|\grad \regulexploit(\param[][][(\numhorizon)]) \|\leq \varepsilon$.
    Taking a minimum across all $\numhorizon \in [\numiters]$, 
    we conclude $\left\|\grad \regulexploit (\bestiter[{\param}][\numiters]) \right\| \leq \varepsilon$.
    
    Then, by the Lemma 3.7 of \cite{lin2020gradient}, there exists some $\param[][][*]\in \params$ such that $\|\bestiter[{\param}][\numiters]-\param[][][*]\|\leq \frac{\varepsilon}{2\lipschitz[\gcumulreg]}$ and $\param[][][*]\in \params_\varepsilon\doteq\{\param\in \params\mid \exists \alpha\in \subdiff\gexploit(\param), \|\alpha\|\leq \varepsilon\}$. 
    That is, $\bestiter[{\param}][\numiters]$ is a $(\varepsilon, \frac{\varepsilon}{2\lipschitz[\gcumulreg]})$-stationary point of $\gexploit$.

    Furthermore, if we assume that $\sexploit(\statedist, \cdot)$ is differentiable at $\param[][][*]$ for any state distribution $\statedist\in \Delta(\states)$, $\gexploit$ is also differentiable at $\param[][][*]$.
    Hence, by the proof of \Cref{lemma:br_mismatch_coef}, we know that for any state distribution $\diststates\in \simplex(\states)$,
    \begin{align}
        \|\grad[\param]\sexploit(\diststates, \param)\|
        &\leq 
        \max_{\deparam[][][*]\in \argmax_{\deparam\in \deparams}\scumulreg(\diststates, \param,\deparam)}\|\grad[\param]\scumulreg(\diststates, \param, \deparam[][][*])\|\\
        &\leq \max_{\player\in \players}\max_{\deparam[][][*]\in \argmax_{\deparam\in \deparams}\scumulreg(\diststates, \param,\deparam)}\\
        &\left(\frac{1}{1-\discount}\right)^2 \left\Vert\frac{\statedist[\diststates][{\deparam[\player][][*](\param), \param[-\player]}]}{\initstates}\right\Vert_\infty
        \left\Vert\frac{\statedist[\diststates][{\param}]}{\initstates}\right\Vert_\infty \|\grad[\param]\gcumulreg(\param, \deparam[][][*])\|\\
       &=\brmismatch(\param, \initstates, \diststates)\|\grad[\param]\gcumulreg(\param, \deparam[][][*])\|
       \end{align}
       \begin{align}
       \frac{1}{\brmismatch(\param, \initstates, \diststates)} \|\grad[\param]\sexploit(\diststates, \param)\|
       \leq \|\grad[\param]\gcumulreg(\param, \deparam[][][*])\| 
    \end{align}
Therefore, 
\begin{align}
        \param[][][*]\in \params_\varepsilon
        &\doteq\{\param\in \params\mid \exists \alpha\in \subdiff\gexploit(\param), \|\alpha\|\leq \varepsilon\}\\
        &\supseteq\{\param\in \params\mid \exists \deparam[][][*]\in \argmax_{\param\in \params}\gcumulreg(\param, \deparam) s.t. \|\grad[\param]\gcumulreg(\param, \deparam[][][*])\|\leq \varepsilon\}\\
        &\supseteq \{\param\in \params\mid \nicefrac{1}{\brmismatch(\param, \initstates, \diststates)} \|\grad[\param]\sexploit(\diststates,\param)\| 
        \leq \varepsilon\}\\
        &=\{\param\in \params\mid \|\grad[\param]\sexploit(\diststates,\param)\| 
        \leq \delta\}
    \end{align}
    Therefore, we can conclude that there exists $\param[][][*]$ such that $\|\bestiter[\param][T]-\param[][][*]\|\leq \frac{\varepsilon}{2\lipschitz[\gcumulreg]}$ and $\|\grad[\param]\sexploit(\diststates, \param)\|\leq \delta$ for any $\diststates$. Thus, $\bestiter[\param][T]$ is a $(\varepsilon, \delta)$-stationary point of $\sexploit(\diststates, \cdot)$ for any $\diststates\in \simplex(\states)$.
    
\end{proof}

\subsection{Omitted Results and Proofs from \Cref{sec:inf_eqa}}\label{sec_app:stochastc_exchange_economy}

\thmexistRRE*
\begin{proof}
    Let $\policy[][][*]=(\consumption[][][][*], \portfolio[][][][*], \price[][][*], \assetprice[][][*]): \states \to \consumptions \times \portfoliospace \times \pricespace \times \assetpricespace $ be an \MPGNE{} of the Radner Markov pseudo-game $\mgame$ associated with $\economy$. 
    We want to show that it is also an RRE of $\economy$.

    First, we want to show that $\policy[][][*]$ is Markov perfect for all consumers. 
    We can make some easy observations: the state value for the player $\buyer\in \buyers$ in the Radner Markov pseudo-game at state $\state\in \states$ induced by the policy $\policy[][][*]$
    \begin{align}
        \vfunc[\buyer][{\policy[][][*]}] (\state)&=
        \Ex_{\histrv\sim \histdistrib[][{\policy[][][*]}]} \left[
        \sum_{\numhorizon=0}^{\infty}\discount^\numhorizon \newreward(\staterv[\numhorizon], \actionrv[][][\numhorizon]) \mid \staterv[0]=\state)
        \right]
        \\&= 
        \Ex_{\histrv\sim \histdistrib[][{\policy[][][*]}]} \left[
        \sum_{\numhorizon=0}^{\infty}\discount^\numhorizon \util[\buyer] (\consumption[\buyer][][][*] (\staterv[\numhorizon]); \typerv[\buyer][][\numhorizon]) \mid \staterv[0]=\state)
        \right]
    \end{align}
    is equal to the consumption state value induced by $(\consumption[][][][*], \portfolio[][][][*], \price[][][*], \assetprice[][][*])$
    \begin{align}
        \vfunc[\buyer][{(\consumption[][][][*], \portfolio[][][][*], \price[][][*], \assetprice[][][*])}] (\state) \doteq \Ex_{\histrv \sim \histdistrib[][{( \consumption[][][][*], \portfolio[][][][*], \price[][][*], \assetprice[][][*])}]} \left[ \sum_{\numhorizon = 0}^\infty \discount^\numhorizon \util[\buyer] \left( \allocation[\buyer][][][*] (\histrv[:\numhorizon][][]); \typerv[][][\numhorizon] \right) \mid \staterv[0] = \state \right]  \enspace .
    \end{align}
    as $\consumption[\player][][][*]$ is Markov. Since $\policy[][][*]$ is a \MPGNE{}, we know that for any $\buyer\in \buyers$:
    $$(\allocation[\buyer][][][*], \portfolio[\buyer][][][*]) \in \argmax_{\substack{(\allocation[\buyer], \portfolio[\buyer]): \states \to \consumptions[\buyer] \times \portfoliospace[\buyer]: \forall \state \in \states, \\(\allocation[\buyer], \portfolio[\buyer])(\state) \in \budgetset[\buyer] (\consendow[\buyer], \price[][][*] (\state), \assetprice[][][*] (\state))
    }} \left\{ \vfunc[\buyer][{(\allocation[\buyer][][][], \allocation[-\buyer][][][*], 
    \portfolio[\buyer][][], \portfolio[-\buyer][][][*], \price[][][*], \assetprice[][][*])}] (\state)  \right\}$$ for all $\state\in \states$, so $(\consumption[][][][*], \portfolio[][][][*], \price[][][*], \assetprice[][][*])$ is Markov perfect. 

    Next, we want to show that $(\consumption[][][][*], \portfolio[][][][*], \price[][][*], \assetprice[][][*])$ satisfies the Walras's law. First, we show that for any $\buyer\in \buyers$, $\state\in \states$, $\consumption[\buyer][][][*] (\state) \cdot\price[][][*] (\state)+ \portfolio[\buyer][][][*] (\state) \cdot \assetprice[][][*] (\state)- \consendow[\buyer] \cdot \price[][][*] (\state)=0$. By way of contradiction, assume that there exists some $\buyer\in \buyers$, $\state\in \states$ such that $\consumption[\buyer][][][*] (\state) \cdot\price[][][*] (\state)+ \portfolio[\buyer][][][*] (\state) \cdot \assetprice[][][*] (\state)- \consendow[\buyer] \cdot \price[][][*] (\state) \neq 0$. Note that $(\consumption[\buyer][][][*] (\state), \portfolio[\buyer][][][*] (\state)) \in \newbudgetset(\state, \action[-\buyer])=\budgetset(\consendow[\buyer], \price[][][*] (\state), \assetprice[][][*] (\state))=\{(\consumption[\buyer], \portfolio[\buyer]) \in \consumptions[\buyer] \times \portfoliospace[\buyer] \mid  \consumption[\buyer] \cdot \price[][][*] (\state) + \portfolio[\buyer] \cdot \assetprice[][][*] (\state)  \leq \consendow[\buyer] \cdot \price[][][*] (\state) \}$, so we must have  $\consumption[\buyer][][][*] (\state) \cdot\price[][][*] (\state)+ \portfolio[\buyer][][][*] (\state) \cdot \assetprice[][][*] (\state)- \consendow[\buyer] \cdot \price[][][*] (\state)< 0$. By the (no saturation) condition of \Cref{assum:existence_RRE}, there exists $\consumption[\buyer][][][+] \in \consumptions[\buyer]$ s.t. $\util[\buyer] (\consumption[\buyer][][][+]; \type[\buyer])>\util[\buyer] (\consumption[\buyer][][][*] (\state); \type[\buyer])$. Moreover, since $\consumption[\buyer] \mapsto \util[\buyer] (\consumption[\buyer]; \type[\buyer])$ is concave, for any $0<t<1$, $\util[\buyer] (t\consumption[\buyer][][][+]+(1-t) \consumption[\buyer][][][*] (\state); \type[\buyer])>\util[\buyer] (\consumption[\buyer][][][*] (\state); \type[\buyer])$. Since $\consumption[\buyer][][][*] (\state) \cdot\price[][][*] (\state)+ \portfolio[\buyer][][][*] (\state) \cdot \assetprice[][][*] (\state)- \consendow[\buyer] \cdot \price[][][*] (\state)< 0$, we can pick $t$ small enough such that $\consumption[\buyer][][][\prime]=t\consumption[\buyer][][][+]+(1-t) \consumption[\buyer][][][*] (\state)$ satisfies $\consumption[\buyer][][]['] (\state) \cdot\price[][][*] (\state)+ \portfolio[\buyer][][][*] (\state) \cdot \assetprice[][][*] (\state)- \consendow[\buyer] \cdot \price[][][*] (\state) \leq 0$ but $\consumption[\buyer][][][\prime] \in \consumptions[\buyer]$ s.t. $\util[\buyer] (\consumption[\buyer][][][+]; \type[\buyer])>\util[\buyer] (\consumption[\buyer][][][*] (\state); \type[\buyer])$. Thus,
    \begin{align}
        &\qfunc[\buyer][{\policy[][][*]}](\state, \consumption[\buyer][][][\prime], \consumption[-\buyer][][][*](\state), \portfolio[][][][*](\state), \price[][][*](\state), \assetprice[][][*](\state))\\
        &= \newreward[\buyer](\state, \consumption[\buyer][][][\prime], \consumption[-\buyer][][][*](\state), \portfolio[][][][*](\state), \price[][][*](\state), \assetprice[][][*](\state)) + \Ex_{\staterv[][][\prime] \sim \trans(\staterv[][][\prime] \mid \state, \portfolio[][][][*](\state))} [\discount \vfunc[\buyer][{\policy[][][*]}](\staterv[][][\prime])]\\
        &=\util[\buyer](\consumption[\buyer][][][\prime];\type[\buyer]) + \Ex_{\staterv[][][\prime] \sim \trans(\staterv[][][\prime] \mid \state, \portfolio[][][][*](\state))} [\discount \vfunc[\buyer][{\policy[][][*]}](\staterv[][][\prime])]\\
        &> \util[\buyer](\consumption[\buyer][][][*](\state);\type[\buyer]) + \Ex_{\staterv[][][\prime] \sim \trans(\staterv[][][\prime] \mid \state, \portfolio[][][][*](\state))} [\discount \vfunc[\buyer][{\policy[][][*]}](\staterv[][][\prime])]\\
        &=\qfunc[\buyer][{\policy[][][*]}](\state, \consumption[][][][*](\state), \portfolio[][][][*](\state), \price[][][*](\state), \assetprice[][][*](\state))
    \end{align}
    This contradicts that fact that $\policy[][][*]$ is a \MPGNE{} since an optimal policy is supposed to be greedy optimal (i.e., maximize the action-value function of each player over its action space at all states) respect to optimal action value function. Thus, we know that for all $\buyer\in \buyers$, $\state\in \states$, $\consumption[\buyer][][][*] (\state) \cdot\price[][][*] (\state)+ \portfolio[\buyer][][][*] (\state) \cdot \assetprice[][][*] (\state)- \consendow[\buyer] \cdot \price[][][*] (\state)=0$. Summing across the buyers, we get $\price[][][*](\state) \cdot \left( \sum_{\buyer \in \buyers} \consumption[\buyer][][][*] (\state) - \sum_{\buyer \in \buyers} \consendow[\buyer] \right)  +  \assetprice[][][*] (\state) \cdot \left(\sum_{\buyer \in \buyers} \portfolio[\buyer][][][*] (\state) \right)=0 $ for any $\state\in \states$, which is the Walras' law.

    Finally, we want to show that $(\consumption[][][][*], \portfolio[][][][*], \price[][][*], \assetprice[][][*])$ is feasible. We first show that $\sum_{\buyer \in \buyers} \consumption[\buyer][][][*] (\state) - \sum_{\buyer \in \buyers} \consendow[\buyer][][] \leq \zeros[\numcommods]$ for any $\state\in \states$. 
    We proved that for any state $\state\in \states$, $\newreward[\numbuyers+1](\state, \consumption[][][][*](\state), \portfolio[][][][*](\state), \price[][][*](\state), \assetprice[][][*](\state))=\price[][][*](\state) \cdot \left( \sum_{\buyer \in \buyers} \consumption[\buyer][][][*] (\state) - \sum_{\buyer \in \buyers} \consendow[\buyer] \right)  +  \assetprice[][][*] (\state) \cdot \left(\sum_{\buyer \in \buyers} \portfolio[\buyer][][][*] (\state) \right)=0$, which implies $\vfunc[\numbuyers+1][{\policy[][][*]}](\state)=0$. For any $\good\in \goods$, consider a $\price:\states\to \pricespace$ defined by $\price(\state)=\j_{\good}$ for all $\state\in \states$ and a $\assetprice: \state\to \assetpricespace$ defined by $\assetprice(\state)=\zeros[\numassets]$ for all $\state\in \states$. Then, we know that
    \begin{align}
        0 &= \vfunc[\numbuyers+1][{\policy[][][*]}]\\
        &= \qfunc[\numbuyers+1][{\policy[][][*]}](\state, \consumption[][][][*](\state), \portfolio[][][][*](\state), \price[][][*](\state), \assetprice[][][*](\state))\\
        &\geq \qfunc[\numbuyers+1][{\policy[][][*]}](\state, \consumption[][][][*](\state), \portfolio[][][][*](\state), \price(\state), \assetprice(\state))\\
        &=\newreward[\numbuyers+1](\state,  \consumption[][][][*](\state), \portfolio[][][][*](\state), \price(\state), \assetprice(\state)) + \Ex_{\staterv[][][\prime] \sim \trans(\staterv[][][\prime] \mid \state, \portfolio[][][][*](\state))} [\discount \vfunc[\buyer][{\policy[][][*]}](\staterv[][][\prime])]\\
        &= \j_\good\cdot \left( \sum_{\buyer \in \buyers} \consumption[\buyer][][][*] (\state) - \sum_{\buyer \in \buyers} \consendow[\buyer] \right) && \forall \good \in \goods\\
        &= \sum_{\buyer \in \buyers} \consumption[\buyer][\good][][*] (\state) - \sum_{\buyer \in \buyers} \consendow[\buyer][\good] && \forall \good \in \goods
    \end{align}
Thus, we know that $\sum_{\buyer \in \buyers} \consumption[\buyer][][][*] (\state) - \sum_{\buyer \in \buyers} \consendow[\buyer][][] \leq \zeros[\numcommods]$ for any $\state\in \states$. 
Finally, we show that $\sum_{\buyer\in \buyers}\portfolio[\buyer][][][*](\state)\leq \zeros[\numassets]$ for all $\state\in\states$. By way of contradiction, suppose that for some asset $\asset \in [\numassets]$, and some state $\state \in \states$, $\sum_{\buyer\in \buyers}\portfolio[\buyer][\asset][][*](\state)> 0$. Then,  
the auctioneer can increase its cumulative payoff by increasing $\assetprice[\asset][][*](\state)$, which contradicts the definition of a \MPGNE{}.  
    
Therefore, we can conclude that $\policy[][][*]=(\consumption[][][][*], \portfolio[][][][*], \price[][][*], \assetprice[][][*]): \states \to \consumptions \times \portfoliospace \times \pricespace
\times \assetpricespace $ is a RRE of $\economy$.
\end{proof}

\corexistRRE
\begin{proof}
    For any infinite horizon Markov exchange economy $\economy$ for which \Cref{assum:existence_RRE} holds, consider the associated exchange economy Markov pseudo-game $\mgame$. 
    By the definition of exchange economy Markov pseudo game, we can see that the transition functions set in the game are all stochastically concave and as such give rise action-value functions which are concave in the actions each of player \cite{atakan2003stochastic}, and it is easy to verify that the game also satisfies all conditions that guarantee the existence of a \MPGNE{} (see Section 4 of \cite{atakan2003stochastic} for detailed proofs). Hence, by \Cref{thm:existence_of_mpgne} which guarantees the existence of \MPGNE{} in Markov pseudo-game, we can conclude that there exists an RRE $(\consumption[][][][*], \portfolio[][][][*], \price[][][*], \assetprice[][][*])$ in any Radner economy $\economy$.
\end{proof}

\subsection{Omitted Results and Proofs from \Cref{sec:computation}}

\thmcomputeSRE*
\begin{proof}
     In the proof of \Cref{thm:existence_RRE}, we can observe that, for any infinite horizon Markov exchange economy $\economy$ and its associated exchange economy Markov pseudo-game $\mgame$, 
    the exploitability of an outcome $(\consumption, \portfolio, \price, \assetprice)$ in $\economy$ is equivalent to the exploitability of a policy $\policy=(\consumption, \portfolio, \price, \assetprice)$ in $\mgame$. 
    Similarly,
    the state exploitability of an outcome $(\consumption, 
    \portfolio, \price, \assetprice)$ in $\economy$ is equivalent to the state exploitability of a policy $\policy=(\consumption, \portfolio, \price, \assetprice)$ in $\mgame$ given any state $\state\in \state$. 

    Therefore, this results follows readily from \Cref{thm:convergence_GNE}.
\end{proof}
\section{Experiments}\label{sec_app:experiments}

\subsection{Neural Projection Method}
\label{sec_app:npm}

The projection method \cite{judd_projection_1992}, also known as the weighted residual methods, is a numerical technique often used to approximate solutions to complex economic models, particularly those involving dynamic programming and dynamic stochastic general equilibrium (DSGE) models. 
These models are common in macroeconomics and often don't have analytical solutions due to their non-linear, dynamic, and high-dimensional nature. The projection method helps approximate these solutions by projecting the problem into a more manageable, lower-dimensional space.

The main idea of the projection method is to express equilibrium of the dynamic economic model as a solution to a functional equation $\operator(\functional)=\zeros$, where 
$\functional: \states \to \R^\numfactions$ is a function that represent some unknown policy, 
$\operator: (\states\to \R^\numfactions) \to (\states \to \R^\numplayers)$, 
and
$\zeros$ is the constant zero function. Some classic examples of the operator $\operator$ includes Euler equations and Bellman equations.
A canonical project method consists of four steps:~1) Define a set of basis functions $\{\basisfunc[i]: \states \to \R^\numfactions\}_{i\in [\numbasis]}$ and approximate each each function $\functional\in \functionals$
through a linear combination of basis functions: $\approxfunc(\cdot;\fparam)=\sum_{i=1}^\numbasis \fparam[i]\basisfunc[i](\cdot)$;
~2) Define a residual equation as a functional equation evaluated at the approximation: $\residual(\cdot;\fparam)\doteq \operator(\approxfunc(\cdot; \fparam))$;
~3) Choose some weight functions $\{\weightfunc[i]: \states \to \R\}_{i\in [\numweights]}$ over the states and
find $\fparam$ that solves $\weightedr(\fparam)\doteq\int_{\states} \weightfunc[i](\state)\residual(\state; \fparam) d\state=0$ for all $i\in [\numweights]$. This gets the residual ``close" to zero in the weighted integral sense;
~4) Simulate the optimal decision rule based on the chosen parameter $\fparam$ and basis functions.

Recently, the neural projection method was developed to extend the traditional projection method \cite{maliar_deep_2021, azinovic2022deep, sauzet_projection_2021}. In the neural projection method, neural networks are used as the functional approximators for policy functions instead of traditional basis function approximations. 
In this section, we show how we can approximate generalized Markov perfect equilibrium of Markov pseudo-game, and consequently Recursive Radner Equilibrium of infinite-horizon Markov exchange economies, through the neural projection method.

\begin{assumption}\label{assum:neural_proj}
Given a Markov pseudo-game $\mgame$, assume that 1.~for any $\player\in \players$, $\state\in\states$, $\action[-\player]\in \actionspace[-\player]$,
$\actions[\player](\state, \action[-\player])\doteq \{\action[\player]\in \actionspace[\player]\mid \actionconstr[\player][\numconstr](\state, \action[\player], \action[-\player])\geq 0\text{ for all $\numconstr\in [\numconstrs]$}\}$ for a collection of \mydef{constraint functions} $\{\actionconstr[\player][\numconstr]: \states\times \actionspace \mid \numconstr\in [\numconstrs]\}$, where $\action[\player]\mapsto \actionconstr[\player][\numconstr](\state, \action[\player], \action[-\player])$ is concave\sadie{is this a reasonable assumption?} for every $\numconstr\in [\numconstrs]$. 
\end{assumption}

\begin{restatable}{theorem}{thmbellman_firstorder}
\label{thm:firstorder+bellman}
    Let $\mgame$ be a Markov pseudo-game that satisfies \Cref{assum:neural_proj}. For any policy profile $\policy\in \fmarkovpolicies$, $\policy$ is a \MPGNE{} if and only if there exists Lagrange multiplier policy $\langmult: \states\to \R_+^{\numplayers\times \numconstrs}$ such that $(\policy, \langmult)$ solves the following functional equation: for all $\player\in \players$, $\state\in \states$,
    \begin{align}
         &0\in \partial_{\action[\player]} \qfunc[\player][\policy](\state, \policy[\player](\state), \policy[-\player](\state))+ \sum_{\numconstr\in [\numconstrs]} \langmult[\player\numconstr](\state) \partial_{\action[\player]}\actionconstr[\player][\numconstr](\state, \policy[\player](\state), \policy[-\player](\state))\\
    &\forall \numconstr\in [\numconstrs], \;\;0=\langmult[\player\numconstr](\state)\actionconstr[\player][\numconstr](\state, \policy[\player](\state), \policy[-\player](\state))\\
    &\forall \numconstr\in [\numconstrs],\;\;0\leq \actionconstr[\player][\numconstr](\state, \policy[\player](\state), \policy[-\player](\state)) \amy{what is $\action[\player][][][*]$?}\sadie{It's a typo, I think it should be $\policy[\player](\state)$}
    \end{align}
    and for all $\player\in \players$, $\state\in \states$,
    \begin{align}
        \vfunc[\player][\policy](\state) =
    \qfunc[\player][\policy](\state, \policy[\player](\state), \policy[-\player](\state))
    \end{align}
\end{restatable}

\begin{proof}
    First, we know that a policy profile $\policy\in \fmarkovpolicies$ is a \MPGNE{} if and only if it satisfies the following generalized Bellman Optimality equations, i.e., for all $\player\in \players$, $\state\in \states$,
    \begin{align}
        \vfunc[\player][\policy](\state)
        &=\max_{\action[\player]\in \actions[\player](\state, \policy[-\player](\state))}
        \reward[\player](\state, \action[\player], \policy[-\player](\state))
        + \discount \E_{\state[][][\prime]\sim \trans(\cdot\mid \state, \action[\player], \policy[-\player](\state))}[\vfunc[\player][\policy](\state[][][\prime])]\label{eq:bellman1}\\
        &=\max_{\action[\player]\in \actions[\player](\state, \policy[-\player](\state))}
        \qfunc[\player][\policy](\state, \action[\player], \policy[-\player](\state))\label{eq:bellman2}
    \end{align}
    Then since $\action[\player]\mapsto \qfunc[\player][\policy](\state, \action[\player], \policy[-\player](\state))$ is concave over $\actions[\player](\state, \policy[-\player](\state))$ by \Cref{assum:existence_of_mpgne}, the KKT conditions provides sufficient and necessary optimality conditions for the constrained maximization problem \begin{align}
        \max_{\action[\player]\in \actions[\player](\state, \policy[-\player](\state))}
        \qfunc[\player][\policy](\state, \action[\player], \policy[-\player](\state)) \label{eq:bellman_opt}
    \end{align}
    That is, $\action[\player][][][*]\in \actions[\player](\state, \policy[-\player](\state))$ is a solution to \cref{eq:bellman_opt} if and only if there exists  $\{\langmult[\player\numconstr][*]:\states\to \R_+\}_{\numconstr\in [\numconstrs]}$ s.t.
    \begin{align}
        &0\in \partial_{\action[\player]} \qfunc[\player][\policy](\state, \action[\player][][][*], \policy[-\player](\state))+ \sum_{\numconstr\in [\numconstrs]} \langmult[\player\numconstr][*](\state) \partial_{\action[\player]}\actionconstr[\player][\numconstr](\state, \action[\player][][][*], \policy[-\player](\state)) \label{eq:bellman_first_order}\\
        &\forall \numconstr\in [\numconstrs], \;\;0=\langmult[\player\numconstr][*](\state)\actionconstr[\player][\numconstr](\state, \action[\player][][][*], \policy[-\player](\state))\\
        &\forall \numconstr\in [\numconstrs],\;\;0\leq \actionconstr[\player][\numconstr](\state, \action[\player][][][*], \policy[-\player](\state))
    \end{align}
    Therefore, we can conclude that $\policy\in \fmarkovpolicies$ is a \MPGNE{} if and only if there exists $\{\langmult[\player\numconstr]:\states\to \R_+\}_{\player\in \players,\numconstr\in [\numconstrs]}$ s.t. for all $\player\in \players$, $\state\in \states$,
    \begin{align}
         &0\in \partial_{\action[\player]} \qfunc[\player][\policy](\state, \policy[\player](\state), \policy[-\player](\state))+ \sum_{\numconstr\in [\numconstrs]} \langmult[\player,\numconstr](\state) \partial_{\action[\player]}\actionconstr[\player][\numconstr](\state, \policy[\player](\state), \policy[\player](\state))\\
    &\forall \numconstr\in [\numconstrs], \;\;0=\langmult[\player\numconstr](\state)\actionconstr[\player][\numconstr](\state, \policy[\player](\state), \policy[\player](\state))\\
    &\forall \numconstr\in [\numconstrs],\;\;0\leq \actionconstr[\player][\numconstr](\state, \policy[\player](\state), \policy[-\player](\state)) \amy{what is $\action[\player][][][*]$?}\sadie{Same here, I changed it}
    \end{align}
        and for all $\player\in \players$, $\state\in \states$,
    \begin{align}
        \vfunc[\player][\policy](\state) =
    \qfunc[\player][\policy](\state, \policy[\player](\state), \policy[-\player](\state))
    \end{align}
\end{proof}

Therefore, for a policy profile $\policy\in \fmarkovpolicies$ and a Lagrange multiplier policy $\langmult: \states\to \R_+^{\numplayers\times \numconstrs}$, 
consider the \mydef{total first-order violation} \begin{align}
    \firsterror(\policy, \langmult)
    &=\sum_{\player\in \players}
\left\|\int_{\state\in \states}
\partial_{\action[\player]} \qfunc[\player][\policy](\state, \policy[\player](\state), \policy[-\player](\state))+ \sum_{\numconstr\in [\numconstrs]} \langmult[\player,\numconstr](\state) \partial_{\action[\player]}\actionconstr[\player][\numconstr](\state, \policy[\player](\state), \policy[-\player](\state)) ds\right\|_2^2 
\end{align} 
and the \mydef{average Bellman error}
\begin{align}
    \bellmanerror(\policy, \langmult)
    &=\sum_{\player\in \players}
\left\|\int_{\state\in \states}
\vfunc[\player][\policy](\state) -
    \qfunc[\player][\policy](\state, \policy[\player](\state), \policy[-\player](\state))
 ds\right\|_2^2.
\end{align} 
We can directly approximate the \MPGNE{} through minimizing the sum of these two errors. 

Typically, approximating the \MPGNE{} using the neural projection method requires optimizing both the policy profile and the Lagrange multiplier policy. However, in exchange economy Markov pseudo-games, we derive a closed-form solution for the optimal Lagrange multiplier, allowing us to focus solely on optimizing the policy profile. 
\subsection{More Results}

\begin{figure}
    \begin{subfigure}{\textwidth}
        \centering
        \includegraphics[width=0.6\textwidth]{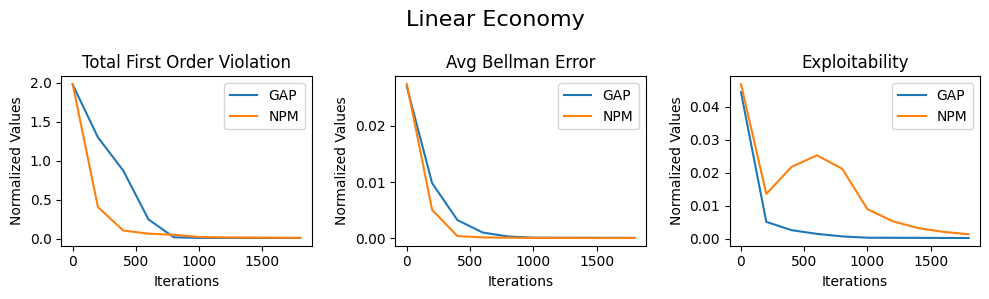}
        \label{fig:linear_nonstochastic}
    \end{subfigure}
    
    
    \begin{subfigure}{\textwidth}
        \centering
        \includegraphics[width=0.6\textwidth]{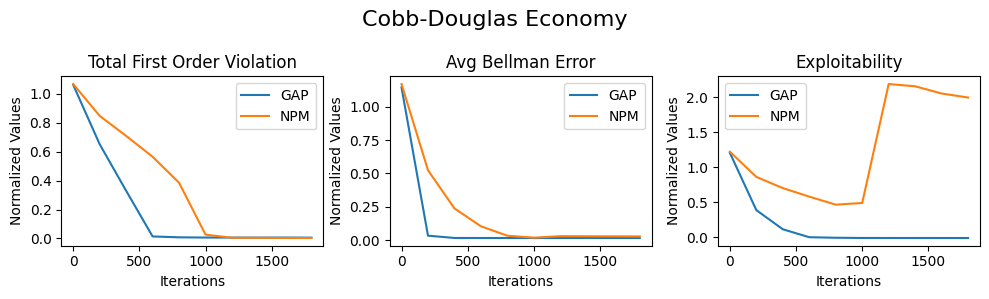}
        \label{fig:cd_nonstochastic}
    \end{subfigure}
    
    
    \begin{subfigure}{\textwidth}
        \centering
        \includegraphics[width=0.6\textwidth]{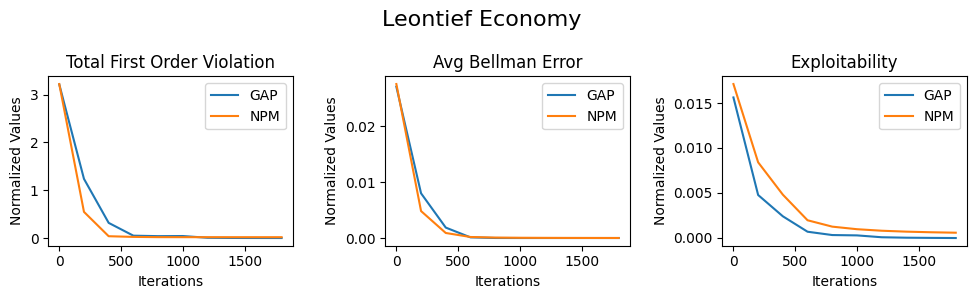}
        \label{fig:leontief_nonstochastic}
    \end{subfigure}
    \caption{Normalized Metrics for Economies with Deterministic Transition Probability Function}
    \label{fig:nonstochastic}
\end{figure}
\subsection{Implementation Details}
\paragraph{Deterministic Case Training Details}
For deterministic transition probability case, for each reward function class
we randomly sampled one economy with 10 consumers, 10 commodities, 1 asset, and 5 world state. The asset return matrix $\returns$
is sampled in a way such that $\returns[\worldstate][\asset][\commod]\sim \unif([0.5, 1.1])$ for all $\worldstate$, $\asset$, and $\commod$. Moreover, we set the length of the stochastic process to be 30.
For the initial state, we sample each consumer's endowment $\consendow[\player]\sim \unif([0.01, 0.1])^\numcommods$ and normalized so that the total endowment of each commodity add up to 1. We also sample each consumer's type $\type[\player]\sim \unif([1.0, 5.0])^\numcommods$, and set the world state to be 0. 
The transition probability function $\trans$ is defined as $\trans(\state[][][\prime]\mid \state, \portfolio)=1$ for all $\state(\worldstate, \consendow, \type)$ where $\state[][][\prime]=(\worldstate[][][\prime], \consendow[][][][\prime], \type[][][][\prime])$ is defined as $\worldstate[][][\prime]=0$, $\consendow[][][][\prime]=0.01\cdot\ones[\numbuyers\times \numcommods]$, and $\type[][][][\prime]=\type$. 

Then, for both GAPNets method and neural projection method, we run 1000 episodes for each learning rate candidate in a grid search manner and measure the performance in terms of minimizing total first-order violation and average Bellman error. Finally, we pick the best hyperparameter for the final experiments. 

In the final experiments, we run GAPNets for 2000 episodes using learning rates $\learnrate[\param]=1\times 10^{-5}, \learnrate[\deparam]=1\times 10^{-5}$ for the linear economy, 
$\learnrate[\param]=1\times 10^{-5}, \learnrate[\deparam]=1\times 10^{-5}$ for the Cobb-Douglas economy, 
and $\learnrate[\param]=1\times 10^{-5}, \learnrate[\deparam]=1\times 10^{-5}$ for the Leontief economy. 
Similarly, we ran neural projection method for 2000 episodes using learning rates
$\learnrate[\param]=1\times 10^{-4}$ for the linear economy,
$\learnrate[\param]=2.5\times 10^{-5}$ for the Cobb-Douglas economy,
and $\learnrate[\param]=1\times 10^{-4}$ for the Leontief economy.
In this process, we compute the exploitability of computed policy profile through gradient ascent of the adversarial network. In specific, we ran 1000 episodes of gradient ascent with learning rate $\learnrate[\deparam]=5\times 10^{-5}$ for the linear economy, 
$\learnrate[\deparam]=1\times 10^{-4}$ for the Cobb-Douglas economy, 
and $\learnrate[\deparam]=1\times 10^{-4}$ for the Leontief economy. 

Next, for each economy, we randomly sample 50 policy profiles and record their total first-order violations, average Bellman errors, and exploitabilities. Finally, we normalize the results by the average of the sampled values.

\paragraph{Stochastic Case Training Details}
For stochastic transition probability case, for each reward function class
we randomly sampled one economy with 10 consumers, 10 commodities, 1 asset, and 5 world state. The asset return matrix $\returns$
is sampled in a way such that $\returns[\worldstate][\asset][\commod]\sim \unif([0.5, 1.1])$ for all $\worldstate$, $\asset$, and $\commod$. Moreover, we set the length of the stochastic process to be 30.
For the initial state, we sample each consumer's endowment $\consendow[\player]\sim \unif([0.01, 0.1])^\numcommods$ and normalized so that the total endowment of each commodity add up to 1. We also sample each consumer's type $\type[\player]\sim \unif([1.0, 5.0])^\numcommods$, and set the world state to be 0. 
The transition probability function will stochastically transition from state $\state(\worldstate, \consendow, \type)$
to state $\state[][][\prime]=(\worldstate[][][\prime], \consendow[][][][\prime], \type[][][][\prime])$ where $\worldstate[][][\prime]\sim \unif(\{0,1,2,3,4\})$, $\consendow[][][][\prime]\sim 0.002+\unif([0.01, 0.1])^{\numbuyers\times \numcommods}$, and $\type[][][][\prime]=\type$. 

Then, for both GAPNets method and neural projection method, we run 1000 episodes for each learning rate candidate in a grid search manner and measure the performance in terms of minimizing total first-order violation and average Bellman error. Finally, we pick the best hyperparameter for the final experiments. 

In the final experiments, we run GAPNets for 2000 episodes using learning rates $\learnrate[\param]=1\times 10^{-5}, \learnrate[\deparam]=1\times 10^{-5}$ for the linear economy, 
$\learnrate[\param]=2.5\times 10^{-5}, \learnrate[\deparam]=2.5\times 10^{-5}$ for the Cobb-Douglas economy, 
and $\learnrate[\param]=5\times 10^{-5}, \learnrate[\deparam]=5\times 10^{-5}$ for the Leontief economy. 
Similarly, we ran neural projection method for 2000 episodes using learning rates
$\learnrate[\param]=5\times 10^{-5}$ for the linear economy,
$\learnrate[\param]=2.5\times 10^{-5}$ for the Cobb-Douglas economy,
and $\learnrate[\param]=5\times 10^{-4}$ for the Leontief economy.
In this process, we compute the exploitability of computed policy profile through gradient ascent of the adversarial network. In specific, we ran 1000 episodes of gradient ascent with learning rate $\learnrate[\deparam]=7.5\times 10^{-4}$ for the linear economy, 
$\learnrate[\deparam]=1\times 10^{-4}$ for the Cobb-Douglas economy, 
and $\learnrate[\deparam]=1\times 10^{-4}$ for the Leontief economy. When estimating the neural loss function—cumulative regret for the GAPNets method and total first-order violations and average Bellman error for the neural projection method—we use 100 samples for GAPNets and 10 samples for the neural projection method. The primary reason for this difference is the high memory consumption of the neural projection method, which makes larger sample sizes infeasible.

Next, for each economy, we randomly sample 50 policy profiles and record their total first-order violations, average Bellman errors, and exploitabilities. Finally, we normalize the results by the average of the sampled values.

\subsection{Other Details}

\paragraph{Programming Languages, Packages, and Licensing}
We ran our experiments in Python 3.7 \cite{van1995python}, using NumPy \cite{numpy},  , CVXPY \cite{diamond2016cvxpy}, Jax \cite{jax2018github}, OPTAX \cite{jax2018github}, Haiku \cite{haiku2020github}, and  JaxOPT \cite{jaxopt_implicit_diff}.
All figures were graphed using Matplotlib \cite{matplotlib}. 

Python software and documentation are licensed under the PSF License Agreement. Numpy is distributed under a liberal BSD license. Pandas is distributed under a new BSD license. Matplotlib only uses BSD compatible code, and its license is based on the PSF license. CVXPY is licensed under an APACHE license. 




\paragraph{Computational Resources}
The experiments were conducted using Google Colab, which provides cloud-based computational resources. Specifically, we utilized an NVIDIA T4 GPU with the following specifications: GPU: NVIDIA T4 (16GB GDDR6), CPU: Intel Xeon (2 vCPUs), RAM: 12GB, Storage: Colab-provided ephemeral storage.

\paragraph{Code Repository}
the
full details of our experiments, including hyperparameter search, final experiment configurations, and visualization code,
can be found in our code repository ({\color{blue}\rawcoderepo}).

\end{document}